\documentclass[aps,prd,preprint,longbibliography,tighten,nofootinbib,eqsecnum,amsmath,amssymb,superscriptaddress]{revtex4-2}
\usepackage{epsf,epsfig,graphics,graphicx}
\usepackage{verbatim,color}%,ulem}
\usepackage{enumitem}
\usepackage{etoolbox}
\usepackage[colorlinks=true, citecolor=blue, linkcolor=blue, urlcolor=blue]{hyperref}
\newcommand{\be}{\begin{equation}}
\newcommand{\ee}{\end{equation}}
\newcommand{\bea}{\begin{eqnarray}}
\newcommand{\eea}{\end{eqnarray}}
\newcommand{\ba}{\begin{array}}
\newcommand{\ea}{\end{array}}
\usepackage{amssymb,amsmath,amsthm,graphicx}
\usepackage[mathscr]{eucal}
\usepackage{enumerate,color,verbatim,multirow,comment}
\definecolor{purple}{rgb}{0.5,0,0.5}

\usepackage{orcidlink}

\begin{document}
\title{Dynamical Love numbers of analogue rotating black and white holes}

\author{Satadal Datta \orcidlink{0000-0002-4687-5254}}
\email{satadaldatta1@gmail.com}
\affiliation{Department of Physics, National Dong-Hwa University, Hualien 974301, Taiwan, R.O.C.}

\author{Wei-Can Syu \orcidlink{0000-0001-8359-4219}}
\email{syuweican@gmail.com}
\affiliation{Center of General Education, Wenzao Ursuline University of Languages, Kaohsiung 80793, Taiwan, R.O.C.}

\author{Da-Shin Lee \orcidlink{0000-0003-3187-8863}}
\email{dslee@gms.ndhu.edu.tw}
\affiliation{Department of Physics, National Dong-Hwa University, Hualien 974301, Taiwan, R.O.C.}

\begin{abstract}
We calculate the dynamical tidal response coefficients (TRCs) of 2+1D analogue black and white holes generated by draining and fountaining bathtub flows, respectively. The parameter space is characterized by the frequency and azimuthal number of the Fourier modes and the rotation of the analogue black or white hole. In general, the TRC is a complex-valued function of these parameters. Its real and imaginary parts are defined as the tidal Love number (TLN) and the tidal dissipation coefficient (TDC), respectively. The TRC of the analogue black hole (ABH) exhibits several interesting features. At certain points in the parameter space, including the static case of vanishing perturbation frequency, the TLN exhibits logarithmic running, while the TDC vanishes. Unlike in Einstein gravity, fluid dynamics allows for the physical existence of an analogue white hole (AWH) event horizon. Time-independent, torque-free, barotropic, inviscid, axisymmetric fluid dynamical equations can yield a pair of transonic background flow solutions. The solutions in the pair, {corresponding to an ABH-AWH pair}, {share} the same angular momentum and Bernoulli constant but have opposite mass flow rates, all of which are conserved quantities. For such an ABH-AWH pair, the TRC of the AWH is the complex conjugate of the TRC of the ABH.
\end{abstract}
\maketitle
\newpage

\section{Introduction}
When a massive object is subjected to a tidal field,  the tidal field induces changes in the object’s mass distribution, which in turn alter its gravitational field. Naturally, tidal effects are prevalent in binary systems in the Universe. The dynamical tidal response coefficients (TRCs) of a massive object are in general dimensionless complex proportionality coefficients that characterizes its {\it response}, which is proportional to the tidal force field according to the linear response theory. The real part of a TRC that encodes tidal deformity of the object, is the tidal Love number (TLN). It is named after the English mathematical physicist Augustus Edward Hough Love, who, within the framework of Newtonian gravity, first theorized a linear {response} of  Earth's surface height and tidal response potential to the tide generating potential by the Sun and the Moon \cite{AEHLove}. The imaginary part of the TRC, capturing dissipative effects, is the tidal dissipation coefficient (TDC). Extending theoretical reach to the relativistic regime, static TLNs have been calculated for compact objects (such as neutron stars), and black holes (BHs) \cite{Hinderer_2008,Flanagan,Damour,Binnington,
Kol2012,hui2021static,Chia,Goldberger2021,Charalambous2021,TiecKerr}. For gravitational wave (GW) signals from neutron stars in binary systems, the signature of tidal effects, and their corresponding TLNs, have been proposed to reveal information about the nuclear equation of state of neutron stars. The works of  \cite{GW170817,Chaves_2019,Yunes2022} have provided the first estimation of the static TLN following the event GW170817 \cite{GW170817}. Static TLNs of asymptotically flat BHs in Einstein gravity are identically zero  \cite{Damour,Binnington,FangLovalace, Kol2012,hui2021static,Chia,Goldberger2021,Charalambous2021,TiecKerr, hiddensymmetryl0, Achour2022,RSVitor,Lovesymmetry,valinishingtlnKS,PhysRevLett.130.091403,Rai2024,Sharma}. However, static TDCs of BHs in Einstein gravity may not be zero. For example, the static TDC of a Kerr BH is nonzero due to the frame dragging \cite{Chia,Goldberger2021,Charalambous2021}. Since a tidal field in general can be time dependent, in recent  years, studies have been made to incorporate the time dependence \cite{Bonelli,Scattering2,Sumanta,perry,chakraborty2025dynamicallovenumbersblack,Maria,
chakraborty2026dynamical}. Calculations by the inclusion of the time dependence produce nonzero corrections in TLNs, and TDCs of BHs in Einstein gravity \cite{Bonelli,Scattering2,perry}.

One of the useful methodologies in the calculation of TRCs is the Teukolsky formalism \cite{Teukolsky}, which provides a separable master equation for perturbations of different spins ($s = 0, ~\pm1, ~\pm2$) in BH spacetimes. In the literature for calculating TRCs, the perturbation field is originally considered to be the spin-2 gravitational field, a  natural choice in the context of  BH perturbation theory. Later, following the footsteps of Teukolsky's paper \cite{Teukolsky}, researchers have included all sorts of perturbation fields to calculate TRCs, e.g., the spin 0 Klein-Gordon perturbation, spin 1 electromagnetic perturbation \cite{hui2021static,Charalambous2021,chakraborty2026dynamical}. The motivation is to see how a BH tidally responds to different types of tidal fields, as they have similar form of Teukolsky equations. There can indeed be notable differences in TRCs for different perturbation fields; for instance, static fermionic TLNs of Schwarzschild, Kerr, and Reissner-Nordstr\"om  BHs are found to be nonzero \cite{2yr1-9ymw,pang2026}. For BHs, in the case of time dependent perturbations in general, there exist analytical solutions of the Teukolsky equation in the near zone. The near zone can be defined by the condition $\omega r \ll 1$  \cite{hiddensymmetryl0,Lovesymmetry,Castro,perry,rodriguez2026lovenumbersblackholes} , in which $\omega$ is dimensionless perturbation's frequency, and $r$ is a dimensionless radius, usually defined in units of the BH exterior event horizon radius. TRCs are calculated from the near zone analytical solutions. This is why, measurements of the TRC from GW events are important where the non vanishing static TLN from a BH merger GW signal could imply  the deviation from Einstein gravity, especially near and potentially inside a BH event horizon \cite{Vitor2,Vitor3,Maselli_2019}. With recent progress in the GW astronomy, upcoming GW events are expected to shed more light about tidal effects relating theories of BHs and neutron stars in general. In this regard, it is interesting to analyze tidal effects in laboratory BH analogues where the observers in the laboratory have full access to all regions of an ABH spacetime such as the interior of the ABH, the near-horizon region, and the far-asymptotic region etc.

Unruh's 1981 pioneering work \cite{unruh} gave birth to the field of {\it analogue gravity}. For an inviscid barotropic fluid with zero vorticity, the linear  perturbation field over a background flow %(a known solution of the fluid dynamical equations)
satisfies a minimally coupled massless Klein-Gordon (KG) scalar field equation in an effective spacetime with a pseudo-Riemannian {\em acoustic} metric, which is determined by the background flow density and velocity \cite{visser1993,Unruh95,BLV}.
When a time-independent background flow transitions from subsonic to supersonic, linear perturbations in the supersonic region cannot propagate upstream. This boundary acts as an analogue black hole (ABH) event horizon. Unlike our current understanding of gravity based on Einstein's general relativity and post-Einsteinian theories, fluid dynamical analogue gravity allows for the physical existence of a white hole event horizon. When a time-independent background flow transitions from supersonic to subsonic in the flow's direction, linear perturbations from the subsonic region cannot enter the supersonic region. In this case, the boundary between the two regions acts as an analogue white hole (AWH) event horizon.
This motivates us to go even beyond the {\it real} gravity framework.  We calculate TRCs of the AWH generated by our model flow. Fluid dynamical ABHs and AWHs have been studied extensively
\cite{unruh,visser1993,Unruh95,MattCQG,Visser,BLV,StefanoLiberati_2000,nvmv,Barcelo_2004,Visser_2005,Rousseaux2008,Weinfurtner,Unruh2014,Euve,PhysRevLett85.4643,
barcelo2001analogue,Basak_2003,Volovik_1999,Volovik2005,Volovik2006,Marino,Nguyen,PhysRevB.86.144505,Carusotto_2008,RalfBill,ros2011,PhysRevD.87.124038, Steinhauer16,Munoz,analogueWormholes,PhysRevD.91.124018,Torres2017,Datta_2018,EuveII,JKBRoyWH, Kolobov2021,ESangshinCaio2022,PhysRevD.105.085022,SyuLee2022,SyuLee2023,PhysRevD.110.044017,bossard2023createanalogueblackhole,Datta_2025}, with huge experimental successes such as the observation of analogue quantum Hawking emission in Bose-Einstein Condensate (BEC) \cite{Munoz},  the observation of superradiance in a draining bathtub rotating hydrodynamical ABH \cite{Torres2017}, and the realization of an analogue wormhole in a hydrodynamical setup \cite{analogueWormholes}.
%Other systems enabling Hawking radiation studies are, e.g., fiber optical systems \cite{Philbin1367,Rosenberg}, superfluid helium , photon fluids  \cite{,Nguyen,PhysRevB.86.144505}, and Weyl semimetals \cite{VolovikWSM_2016,WilczekWSM,VolovikWSM}.\colb{advantage, experiments in BEC, water, and others; existence of white hole event horizon, experiment, motivation w.r.t observation in general relativity, and w.r.t fluid dynamics of waves}\colp{Advantage of wave scattering experiments by an analogue black hole is that, }\colp{motivation as white hole event horizon, beyond the scope of general relativity and near region access}

{\it Methodology:-} we consider a barotropic, irrotational, inviscid, constant density two spatial dimensional background flow model in a bathtub. Therefore, the corresponding analogue spacetime is 2+1 dimensional. For example, surface waves of water (in inviscid regime), as massless KG scalar fields, experience 2+1 D analogue spacetime \cite{RalfBill}. The Teukolsky equation for a massless KG field in our analogue spacetime, can be expressed as an expansion in its frequency $\omega$, involving only whole number  powers up to quadratic order. In the near zone limit, the quadratic term in $\omega$ can be dropped. The transition of the TRC from the static to the dynamical regime is highly nontrivial. To gain a systematic understanding of this transition, we first derive analytical solutions of the Teukolsky equation at zeroth order in the frequency parameter $\omega$. We then refine the analysis by obtaining analytical solutions of the Teukolsky equation truncated at linear order in $\omega$. These solutions enable us to compute the corresponding zeroth- and first-order TRCs. The zeroth-order solution reproduces the static TLNs obtained in the previous study \cite{PRDLuca}. Furthermore, we derive explicit analytical expressions of the real (TLN) and the imaginary part (TDC) of the TRCs.
%We would see, the transformation of TRC from the static to the dynamic regime is nontrivial. This is why, to carefully understand the static to dynamic transition, first we seek analytical solutions of the Teukolsky equation approximated in the zeroth power of $\omega$, and then we improve our analysis by finding analytical solution of the Teukolsky equation truncated at linear order in $\omega$. Thus, we find out zeroth and first order TRCs.

The content of the paper is arranged as follows. We briefly review TRC, and analogue gravity in Sec. \ref{DTLND}, and \ref{FD}, respectively. In Sec. \ref{AWHBH}, we introduce 2+1 dimensional ABHs and AWHs generated by a time-independent {\it general} axisymmetric background flow. This construction is necessary because, to the best of our knowledge, an ABH-AWH pair suitable for our analysis has not yet been discussed in existing literature. In Sec. \ref{TSE}, we then derive the radial wave equation using a Teukolsky-inspired separation of variables technique.  The TRC of our ABH by the aforementioned methodology is computed in Sec. \ref{m0}. In Sec. \ref{WTD},  the TRC of our AWH is also obtained. We conclude our results in Sec. \ref{CSEC}.

\section{A Brief review on TRC}\label{DTLND}
In Newtonian  gravity, when a massive spherical  body of mass $M$ and radius $R$, is subject to an external tidal field, its mass multipole moments change.  The resulting gravitational potential outside the body, takes the following form \cite{Binnington,poisson2014gravity}:
\begin{equation}\label{tidald}
	U=-\frac{GM}{r}+\frac{(\ell-2)!}{\ell!}\sum_{\ell=2}^{\infty}\sum_{m=-\ell}^{m=\ell} Y_{\ell m}(\theta,\phi)\xi_{\ell m}r^\ell\left[1+k_{\ell m}\left(\frac{r}{R}\right)^{-2\ell -1}\right]
\end{equation}
in terms of the  spherical harmonics $Y_{\ell m}$ with $\ell\in\mathbb{Z}_{\geq 0}$  orbital angular momentum umber and $m\in \mathbb{Z}$ azimuthal number,
where $\xi_{\ell m}$ correspond to the tidal moment.
In Eq. \eqref{tidald}, multiplying the factor outside the brackets reveals that the first term inside the brackets, which grows with radial distance r from the centre of the spherical body, is the tidal force term. The second term, which decays with r, is the change in the gravitational potential of the body due to the tidally induced effect on its mass multipole moments.  This results in a dimensionless proportionality coefficient, namely, the TRC $k_{\ell m}$. {Following Eq. \eqref{tidald}, in the relativistic regime, using the weak gravitational field approximation, the metric perturbation $h_{tt}$ has the following expression \cite{perry,rodriguez2026lovenumbersblackholes,chakraborty2026tidalresponsecompactobjects}
\begin{equation}\label{tidald2}
\frac{h_{tt}}{2}=-\frac{(\ell-2)!}{\ell!}\sum_{\ell=2}^{\infty}\sum_{m=-\ell}^{m=\ell} Y_{\ell m}(\theta,\phi)\xi_{\ell m}r^\ell\left[1+k_{\ell m}\left(\frac{r}{R}\right)^{-2\ell -1}+{\rm subleading~terms}\right].
\end{equation} }
$k_{\ell m}$, in general is a complex number depending on the tidal perturbation's frequency $\omega$, and can be separated into real and imaginary part \cite{perry,rodriguez2026lovenumbersblackholes},
\begin{equation}\label{TLNDF}
	k_{\ell m}(\omega)=\kappa_{\ell m}(\omega)+i\nu_{ \ell m}(\omega).
\end{equation}
$\kappa_{\ell m}(\omega)$  and $\nu_{ \ell m}(\omega)$ are defined as the TLN and the TDC, respectively.
 Note that $k_{\ell m}$ can also be a slowly varying function of $r$, e.g., natural logarithm of $r$ \cite{Charalambous2021,PRDLuca}.
\section{A Brief review on analogue gravity}\label{FD}
The continuity and Euler equations of an ideal fluid \cite{Landau1987Fluid} in the nonrelativistic limit are given, respectively, by:
\begin{equation}\label{cont}
	\partial_t \rho+\nabla\cdot (\rho {\bf v})=0~,
\end{equation}
\begin{equation}\label{euler}
	\partial_t {\bf v}+{\bf v}\cdot\nabla{\bf v}=-\frac{\nabla p}{\rho}-\nabla V_{\rm ext}~,
\end{equation}
where $\rho(\bf{x},t)$ is the density field, $p(\bf{x},t)$ is the fluid pressure, and ${\bf v}({\bf x}, t)$ is the velocity vector field. We assume the presence of an external conservative force field with the potential $V_{\rm ext}({\bf x},t)$.
For a barotropic flow, the pressure $p$ is related to the density $\rho$ by  a barotropic relation,
\begin{equation}\label{barotropic}
	p=p(\rho)~,
\end{equation}
so that we now have two independent flow variables $\rho(\bf{x},t)$ and ${\bf v}({\bf x}, t)$ satisfying two equations, Eq. \eqref{cont}, Eq. \eqref{euler}.
The sound speed $c_s$ is defined by,
\begin{equation}
	c_s^2=\frac{dp}{d\rho}.
\end{equation}
For an irrotational fluid, i.e, $\nabla\times{\bf v}=0$, we can introduce the velocity scalar field  $\Phi$  as
\begin{equation}
{\bf v}=\nabla \Phi \,.
\end{equation}
 {The Euler equation, upon integration, can be rewritten as
 \begin{equation}\label{Phi}
\partial_t \Phi + \left( \frac{1}{2} \nabla\Phi \right)^2+ h+V_{\rm ext}({\bf x},t)=0\, ,
\end{equation}}
where $h(p)$ is the specific enthalpy of the barotropic fluid, defined by
\begin{equation}\label{enthalpy}
	h(p)=\int \frac{dp}{\rho}\,
\end{equation}
with $\nabla h=\frac{1}{\rho} \nabla p$.

{We represent the background quantities with subscript $0$, and denote the linear perturbation fields by a preceding $\delta$.}  Linearization of Eqs.(\ref{cont}) and (\ref{Phi}) leads to
\begin{equation}
{\partial_t}\delta\rho+ \nabla (\delta\rho {\bf v}_0 +\rho_0 \delta{\bf v})=0\, ,
\end{equation}
\begin{equation}
{\partial_t}\delta \Phi +{\bf v}_0 \cdot \nabla \delta \Phi+ \frac{\delta p}{\rho_0}=0 \,
\end{equation}
with $\delta h=\delta p/\rho_0$ from { Eq. \eqref{enthalpy}}. Given the fact that {$ \delta \rho=\frac{\partial \rho}{\partial p} \delta p= \frac{\delta p}{c_{s0}^2}$}, and combining the two equations above, the equation of motion for the perturbed potential $\delta \Phi$ is obtained as
\begin{equation}\label{waveE}
-\partial_t \left(\frac{\rho_0}{c_{s0}^2} \left( \partial_t \delta \Phi+ {\bf v}_0 \cdot \nabla \delta \Phi \right) \right)+ \nabla \cdot \left( \rho_0 \nabla \delta \Phi -\frac{\rho_0 {\bf v}_0}{c_{s0}^2} \left(\partial_t \delta \Phi+ {\bf v}_0 \cdot \nabla \delta \Phi \right)\right)=0\, .
\end{equation}
The wave equation \eqref{waveE} can be written as a massless KG field equation minimally coupled to an effective spacetime \cite{unruh,BLV}: 
  \begin{equation}\label{we}
	\partial_\mu (\sqrt{-g}g^{\mu\nu}\partial_\nu\delta\Phi)=0 \,.
\end{equation}
In the context of our topic, we are interested in two spatial dimensional flows. The metric in 2+1 D takes the following form \cite{BLV}:
\begin{equation}\label{theM}
ds^2=g_{\mu\nu}dx^\mu dx^\nu=\left(\frac{\rho_{0}}{c_{s0}}\right)^2\left[-\left(c_{s0}^2-v_0^2\right)d{t}^2-2{\bf v}_0\cdot dtd{\bf x} + d{\bf x}^2\right]~,
\end{equation}
where $v_0$ is the magnitude of ${\bf v}_0$.
\section{ 2+1 D ABH and AWH spacetimes}\label{AWHBH}
In this  {section}, we introduce a time independent torque free axially symmetric barotropic flow, such that $\rho_0=\rho_0(r)$, giving  $p_0=p_0(r)$ and ${\bf v}_0=(v_0^r (r), v_0^\phi (r), v_0^z=0)$ in cylindrical coordinate $(r,\phi,z)$.
The continuity equation leads to the conserved mass flow rate \cite{clarke2007principles},
\begin{equation}\label{fr}
	\rho_0 v_0^r r={\rm constant}=C_1.
\end{equation}
Since $\rho_0 >0$,  Eq. \eqref{fr} implies that $v_0^r$ cannot  change sign.

When no torque  {is exerted} by the conservative force field, i.e., $V_{\rm ext}= V_{\rm ext}(r)$, {we consider} the simplest possible axially symmetric flow. The Euler equation (\ref{euler}) along the $\phi$ direction  gives us the conserved specific angular momentum,
\begin{equation}\label{l}
	rv_0^\phi={\rm constant}=l.
\end{equation}
{Eq. \eqref{l} leads to $\nabla\times {\bf v}_0=0$.}
In addition, Eq. (\ref{euler}) along $r$ direction can be expressed as
\begin{equation}\label{eulerr}
	v_0^r\frac{d v_0^r}{d r}-\frac{(v_0^\phi)^2}{r}=-\frac{ dh_0}{d r}-\frac{d V_{ext}}{d r},
\end{equation}
which gives us a Bernoulli's constant,
\begin{equation}\label{b}
	\frac{1}{2}(v_0^r)^2+\frac{l^2}{2r^2}+h_0 (r)+V_{ext}={\rm constant}=C_2\,.
\end{equation}
If any of the followings is a solution of the fluid dynamical equations (Eq. \eqref{fr}, Eq. \eqref{l}, Eq. \eqref{eulerr})\\
	(i) $\left(\rho_0(r),~v_0^r(r),~v_0^\phi =\frac{l}{r}\right)$,\\
	(ii) $\left(\rho_0(r),~ v_0^r(r),~v_0^\phi =-\frac{l}{r}\right)$,\\
	(iii) $\left(\rho_0(r),~-v_0^r(r),~v_0^\phi =\frac{l}{r}\right)$,\\
	(iv) $\left(\rho_0(r),~ -v_0^r(r),~ v_0^\phi=-\frac{l}{r}\right)$,\\
then the other three are also solutions with same $\vert l\vert$,  $\vert C_1 \vert$, $C_2$ and under the same $V_{ext}(r)$.

For simplicity, we restrict the dynamics of both the background and the waves to effectively two spatial dimensions. The corresponding $2+1$ D acoustic spacetime metric \cite{BLV} in $(t,~r,~\phi)$ coordinates of the laboratory frame, is given by
\begin{equation}\label{agmn}
	ds^2=\left(\frac{\rho_{0}}{c_{s0}}\right)^2\left[-\left(c_{s0}^2-(v_0^r)^2-\frac{l^2}{r^2}\right)d{t}^2-2v_0^r dtdr -2ldtd\phi + dr^2+r^2d{\phi}^2\right].
\end{equation}
The radius $r_h$ of the event horizon {satisfies} \cite{carroll_2019}
\begin{equation}\label{EH}
	g^{rr}(r=r_h)=0\Rightarrow \left({v}_0^r(r=r_h)\right)^2 = \left(c_{s0}(r=r_h)\right)^2,
\end{equation}
and the radius $r_s$ of the {stationary limit surface}\footnote{{Note that, for a general analogue spacetime \eqref{agmn}, there can be multiple event horizons and multiple stationary limit surfaces, corresponding to multiple roots of equations \eqref{EH} and \eqref{ES}, respectively.}} {is given by}, \begin{equation}\label{ES}
	g_{tt}(r=r_s)=0\Rightarrow \left(\left({v}_0^r(r=r_s)\right)^2 +\frac{l^2}{r_s^2}\right)= \left(c_{s0}(r=r_s)\right)^2.
\end{equation}
For such a general 2+1 D ABH/AWH, $r_s\geq r_h$ in \cite{Datta_2025}.
A null ray satisfies $ds^2=0$, giving $\frac{d{\bf x}}{dt}=c_{s0}\hat{e}+{\bf v}_0$ where the null ray points out in the direction of a unit vector $\hat{e}$. For $\hat{e}=\pm\hat{r}$, $\frac{dr}{dt}=\pm c_{s0}+v_0^r$ and  $r\frac{d \phi}{dt}=v_0^\phi=\frac{l}{r}$. Hence, if a radially inward flow ($v_0^r<0$) transitions from a subsonic  region ($|v_0^r(r)|<c_{s0}(r)$) to a supersonic  region ($|v_0^r(r)|>c_{s0}(r)$) along the flow's direction, then the geometry \eqref{agmn} corresponds to an ABH spacetime with a event horizon at the subsonic-supersonic transition boundary (Eq. \eqref{EH}). {This follows because, inside the event horizon (i.e., in the supersonic  region), every future directed timelike curve is necessarily directed towards decreasing $r$.} If we refer to such a background flow as the aforementioned solution (i) in this section, then the solution (ii) corresponds to an ABH with counter-rotation. Furthermore, the solution (iii) corresponds to an AWH with the same rotation, and the solution (iv) corresponds to an AWH with counter-rotation. Eq. \eqref{cont} and \eqref{euler} are invariant in form under time reversal for $\nabla V_{ext}({\bf x},t)= \nabla V_{ext}({\bf x},-t)$, because under $t\to -t$, ${\bf v}$ changes sign, and  $p$ and $\rho$ remain unchanged. For the time independent background considered in this section, the solution (iv) corresponds to the time reversed flow of the solution (i). {Under $t\to -t$, using the transformation property of a rank 2 covariant tensor $g_{\mu\nu}$ in the $(t,~r,~\phi)$ coordinate, $g_{tt}\to g_{tt}$, $g_{tr}\to-g_{tr}$, $g_{t\phi}\to-g_{t\phi}$, $g_{rr}\to g_{rr}$, $g_{r\phi}\to g_{r\phi}$ and $g_{\phi\phi}\to g_{\phi\phi}$.   {According to Eq. \eqref{agmn},} time reversal of the metric \eqref{agmn} is equivalent to reversing the background flow's direction. The same argument also holds true for the more general 2+1 D acoustic metric \eqref{theM} as for the case of the background solution (iv) relative to the background solution (i). Following discussion above, reversal of the background flow's direction is equivalent to time reversal of the background flow solution. If the solution (i) corresponds to an ABH then the solution (iv) corresponds to an AWH with $t\to -t$ of the ABH.

For a regular $V_{\rm ext}(r)$, there exists a regular background solution $(\rho_0(r), v_0^r(r))$ according to Eq. \eqref{eulerr}.  We usually do not expect {singularity} in the metric \eqref{agmn} at finite nonzero $r$. The spacetime metric \eqref{agmn} is of the Painlev\^e-Gullstrand form \cite{BLV} with no coordinate singularity at the event horizon for a regular background flow. However, around/at $r=0$, a fluid dynamical singularity must exist for a nonzero radial flow. Thus, a drain or a fountain with a finite radius is needed around $r=0$ to maintain a steady-state flow, where the continuity equation  \eqref{cont} with nonzero radial velocity is satisfied. For certain radial flow profiles, $r = 0$ can correspond to a physical singularity that is geometrically similar to a real BH singularity \cite{Datta_2025} as described by the famous Penrose singularity theorem \cite{Penrose65PRL}.

We introduce the  coordinate transformation, which diagonalizes the $t-r$ block of the $g_{\mu\nu}$ matrix \eqref{agmn}, and will be useful in the subsequent analysis. The transformation is given by
\begin{eqnarray}
	& dt=d\tilde{t}-\frac{v_0^r}{\left(c_{s0}^2-(v_0^r)^2\right)}dr,~~d\phi =d\tilde{\phi}-\frac{l v_0^r}{r^2\left(c_{s0}^2-(v_0^r)^2\right)}dr. \label{ct2}
\end{eqnarray}
Eq. \eqref{ct2} {represents} a valid coordinate transformation given that $c_{s0}=c_{s0}(r)$, and $v_0^r=v_0^r(r)$, because they are integrable: $\frac{\partial^2 t}{\partial \tilde{t}\partial r}=\frac{\partial^2 t}{\partial r\partial \tilde{t}}=0$, and $\frac{\partial^2 \phi}{\partial r\partial \tilde{\phi}}=\frac{\partial^2 \phi}{\partial \tilde{\phi}\partial r}=0$.
The line element now becomes
\begin{equation}\label{agmn2}
	 ds^2=\left(\frac{\rho_{0}}{c_{s0}}\right)^2\left[-\left(c_{s0}^2-(v_0^r)^2-\frac{l^2}{r^2}\right)d\tilde{t}^2+\frac{c_{s0}^2}{\left(c_{s0}^2-(v_0^r)^2\right)}dr^2-2ld\tilde{t}d\tilde{\phi}+r^2d\tilde{\phi}^2\right].
\end{equation}
{We observe from Eq. \eqref{agmn2} that the coordinate transformation \eqref{ct2} however introduces a coordinate singularity at the event horizon. Nevertheless,, the coordinate transformation \eqref{ct2} is invertible. For a given metric of the form \eqref{agmn2}, it is in fact possible to construct two distinct coordinate systems that remove the coordinate singularity at the event horizon (see further detail in the Appendix \ref{RCS}). In one coordinate system with the $`-$' sign in (\ref{ct4}), the metric takes the form \eqref{agmn}, and is identified in the laboratory frame as an ABH spacetime, whereas in the other  with the $`+$' sign in (\ref{ct4}), it is identified as an AWH spacetime. Such a removal of the coordinate singularity at the event horizon of the analogue spacetime \eqref{agmn2} is comparable to the removal of the coordinate singularity at the event horizon of the Schwarzschild solution by coordinate transformations from Schwarzschild coordinates to the pair of ingoing and outgoing Eddington-Finkelstein coordinate systems (or ingoing and outgoing Painlev\'e –Gullstrand coordinates), where in the ingoing Eddington-Finkelstein frame, the spacetime appears as a BH spacetime, and in the outgoing Eddington-Finkelstein frame, the spacetime appears as a WH spacetime.}
\section{Sound waves in a bathtub}\label{TSE}
In the bathtub model, for constant condensate density and zero torque on the fluid,  the background medium flows with the velocity,
\begin{equation}\label{db}
	{\bf v}_0=\left(-\frac{d}{r}{\bf e}_r +\frac{l}{r}{\bf e}_\phi\right),
\end{equation}
respecting the time independent fluid dynamical equations,  Eq. \eqref{fr} and Eq. \eqref{eulerr}.  Therefore, $-d=\frac{C_1}{\rho_0}$ and {$V_{ext}(r)=C_2-h_0-\frac{(d^2+l^2)}{2 r^2}$.}
Such a background flow with a constant background density $\rho_0$ (hence a constant barotropic pressure $p_0$ and constant enthalpy $h_0$), gives rise to a 2+1 D spacetime with the metric of the form \eqref{agmn}. Note that the  angular momentum  can take any real value, i.e, $l\in\mathbb{R}$ in  Eq. \eqref{l}. {The line element \eqref{agmn2} takes the following form  \cite{PhysRevD.87.124038},} as
\begin{equation}\label{bathtub} ds^2=\left(\frac{\rho_{0}}{c_{s0}}\right)^2\left[-\left(c_{s0}^2-\frac{d^2+l^2}{r^2}\right)d\tilde{t}^2+\frac{c_{s0}^2}{\left(c_{s0}^2-\frac{d^2}{r^2}\right)}dr^2-2ld\tilde{t}d\tilde{\phi}+r^2d\tilde{\phi}^2\right].
\end{equation}
The event horizon radius is given by, $r_h = \frac{d}{c_{s0}}$. The radius of the stationary limit surface is given by, $r_s=\frac{\sqrt{d^2+l^2}}{c_{s0}}$.
 {Hereafter, we express all relevant quantities in a dimensionless form where $r$ is scaled by $r_h$, and speeds by $c_{s0}$. From Eq. \eqref{fr} and from the expression of $d$, $|d|=\frac{|C_1|}{\rho_0}=\frac{\rho_0 c_{s0} r_h}{\rho_0}=1$.}

 We now solve the wave equation \eqref{waveE} in the bathtub spacetime using the method of separation of variables. Henceforth, we denote $\tilde{\phi}$ by $\phi$ and $\tilde{t}$ by $t$ for simplicity. The perturbation $\delta\Phi$ can be decomposed into
\begin{equation}\label{deltaPhi}
	\delta\Phi =\sum_{m\in \mathbb{Z}}\int d \omega  \, \delta\Phi_{\omega,m} =\sum_{m\in \mathbb{Z}}\int d \omega  R_{\omega, m} (r) e^{i m \phi} e^{-i\omega t}.
\end{equation}
 The radial part of the scalar field in the draining bathtub spacetime satisfies
\begin{equation}\label{Teukolsky}
	\Delta (r)\frac{d}{dr}\left(\Delta (r) \frac{dR_{\omega, m}}{dr}\right)+\left[\left(\frac{l^2}{r^2}-\frac{\Delta (r)}{r}\right) m^2 + r^2 \omega ^2 -2\omega m l\right]R_{\omega, m}=0,
\end{equation}
where $\Delta (r) = r\left(1-\frac{1}{r^2}\right)$. We observe that the {Teukolsky} equation \eqref{Teukolsky} is identical for a pair of the analogue black and  white hole with the same spin, generated by the background flow solution  pair ((i),(iii)), or ((ii), (iv)) in Sec. \ref{AWHBH}.
\section{Dynamical TRCs of ABHs in the draining bathtub model}\label{m0}
We find the analytical solution of the  radial wave equation \eqref{Teukolsky} in the near-zone regime ({ the {aforementioned $\omega r\ll 1$ condition}}) in powers of $\omega$. {This can be used to analyze} how the behavior of the perturbations  and their corresponding TRC transitions from the static to dynamic regime. Rewriting Eq. \eqref{Teukolsky} in powers of $\omega$, we have
\begin{equation}\label{Teukolsky3}
	\Delta (r)\frac{d}{dr}\left(\Delta (r) \frac{dR_{\omega, m}}{dr}\right)+f(m,l,\omega,r)R_{\omega,m}=0,
\end{equation}
where
\begin{eqnarray}\label{f}
	& f(m,l,\omega,r)=f_0(m,l,r)+f_1(m,l,\omega,r)+f_2(m,l,\omega,r),&\label{f}\\
	& f_0(m,l,r)= m^2\left(\frac{l^2+1}{r^2}-1\right),~ f_1(m,l,\omega,r)=-2\omega m l;
	~f_2(m,l,\omega,r)=\omega ^2 r^2.& \label{f012}
\end{eqnarray}
Therefore, in this terminology the near-zone condition means
\begin{equation}\label{NZ}
f_2\ll 1.
\end{equation}
As a consequence of causality, perturbation modes must be purely ingoing at {the}  BH and  ABH {horizons} \cite{Hawking1975,Teukolsky}.
{Translational symmetry of the spacetime in $t$ and $\phi$ provides a null Killing  vector $K^\mu=(1,0,\Omega_H)$ at the horizon, where $\Omega _H$ is the angular velocity of the event horizon \cite{carroll_2019}. In our ABH, $\Omega_H=l$. Therefore, the corresponding conserved frequency of the mode is $\tilde{\omega}=-k_\mu K^\mu=(\omega-ml)$, where $k_t=-\omega$, and $k_\phi=m$ in Eq. \eqref{deltaPhi}.}  The frequency $\tilde{\omega}$ is the natural frequency at the event horizon with respect to the ingoing and outgoing Eddington-Finkelstein coordinate $v$ and $u$, where $v=t+r_*$, and $u=t-r_*$. The tortoise coordinate $r_*$ is defined as
$r_*=r +\frac 12\ln\left|\frac{r-1}{r+1}\right| $, so that
$r_*\to -\infty$  at the horizon, and $r_*\to +\infty$  as $r$ approaches infinity.
 Therefore, $\delta\Phi_{\omega,m}$ in Eq. \eqref{deltaPhi} at the ABH event horizon must be of the form ({also see Appendix \ref{hofr}}):
 \begin{equation}\label{ingoing}
	\delta\Phi_{\omega,m}  \propto e^{-i\tilde{\omega}v}e^{im(\phi+lt)}.
\end{equation}
Such a mode has a radially inward group velocity at the event horizon \cite{Starobinskii:1973vzb}.

\subsection{\bf Analytical solution of the radial wave equation to zeroth order in $\omega$} \label{zero}
{At zeroth order in $\omega$},  the radial wave equation \eqref{Teukolsky3} is approximated to its static form:
\begin{equation}\label{Teukolsky0}
	\Delta (r)\frac{d}{dr}\left(\Delta (r) \frac{dR^{(0)}_{\omega, m}}{dr}\right)+f_0(m,l,r)R^{(0)}_{\omega, m}=0\, .
\end{equation}
Eq. \eqref{Teukolsky0} is  analytically solvable $\forall r\geq 1$. The analytical solution for the static ($\omega=0$) case exists everywhere exterior to the event horizon \cite{PRDLuca}, where {$R_{\omega=0,m}(r)=R^{(0)}_{\omega,m}(r)$ $\forall r\geq 1$. For the dynamical case, i.e., for nonzero $\omega$,  $R^{(0)}_{\omega, m}$ can be regarded as the zeroth-order approximation of $R_{\omega,m}$. Far away from the near zone where $f_2$ in  Eq. \eqref{Teukolsky3} dominates, $R_{\omega,m}$ deviates significantly from $R^{(0)}_{\omega, m}$.}

We introduce a new variable $z=\frac{1}{r^2}$, and define {$R^{(0)}_{\omega, m}(r):=z^\alpha (1-z)^\beta F(z)$} in terms of $z$, where $\alpha=\frac{|m|}{2}$, and $\beta=\frac{i m l}{2}$.   Note that such a choice of $\beta$ can be understood by studying the inclusion of the linear $\omega$  term later in Sec. \ref{1storderT} where
$\beta\simeq -\frac{i}{2}\tilde{\omega}=\frac{i}{2}(\omega-lm)$ {satisfies the ingoing boundary condition} at the horizon {{\footnote{The value of our $\beta=\frac{iml}{2}$  is different from  $\beta=\frac{il \vert m\vert}{2 }$ in  Ref. \cite{PRDLuca}. }}}. In  the case of $l=0$, i.e., for $\beta=0$, the factor $(1-z)^\beta$ in $R^{(0)}_{\omega,m}$  is undefined at $z=1$, and so is the solution of $R^{(0)}_{\omega,m}$. Applying Riemann's theorem on removable singularities \cite{stein2010complex},  $R^{(0)}_{\omega, m}(r)(l=0,r)$ can be obtained by taking the limit $\lim _{l\to 0}R^{(0)}_{\omega, m}(l, r)$.
After substituting the aforementioned variable change in \eqref{Teukolsky0}, $F(z)$ satisfies
\begin{equation}\label{odehy}
	z(1-z) F''(z)+(c-(1+a+b)z)F'(z)-abF(z)=0,
\end{equation}
where
\begin{equation}\label{a,b,c}
	a=1+\alpha+\beta, \qquad b=\alpha+\beta,\qquad c=1+2\alpha.
\end{equation}
Eq. \eqref{odehy}, which has  three regular singular points at $z=0$, $z=1$ and $z=\infty$, is the ordinary differential equation (ODE) for the Gauss hypergeometric function $_2 F_1(a,b;c;z)$ for $m\neq 0$. The analytic solution takes the following form \cite{PRDLuca,Beals_Wong_2010}:
\begin{align}\label{Theformula2}
	F(z)=&C_{\omega,m}\frac{\Gamma |m|}{\Gamma (a)\Gamma (b)}z^{-|m|}\sum_{n=0}^{|m|-1}\frac{(a-|m|)_n (b-|m|)_n}{n!(1-|m|)_n}z^n\nonumber\\
	&-C_{\omega,m}\frac{(-1)^{|m|}}{\Gamma (a-|m|)\Gamma (b-|m|)}\sum _{n=0}^{\infty}\frac{(a)_n(b)_n}{n!(n+|m|)!}z^n\nonumber\\
	&\qquad\times[\ln z-\psi (n+1)-\psi (n+|m|+1)+\psi (a+n) +\psi (b+n)],
\end{align}
where $\psi$ is the digamma function given by $\psi (z)=\frac{\Gamma '(z)}{\Gamma (z)}$ and $( )_n$ denotes the Pochhammer symbol. The series expression of Eq. \eqref{Theformula2} is absolute convergent for $|z|<1$, and $|\arg (z)|<\pi$ \cite{Beals_Wong_2010}. The tidal effect manifests as the zeroth-order term in $R^{(0)}_{\omega,m}(r)$ in the near-zone region.  We find it out by evaluating the asymptotic limit of $z\to 0$ ($r \to \infty$) of the expression \eqref{Theformula2}:
\begin{align}\label{rmasymp}
	& \lim _{r\to\infty} R^{(0)}_{\omega,m}(r) =z^\alpha F(z)=\frac{C_{\omega,m}~ \Gamma |m|}{\Gamma (a)\Gamma (b)}\left(\frac{r}{r_h}\right)^{|m|}\left[1+k^{(0)}_{m}(l,r) \left(\frac{r}{r_h}\right)^{-2|m|}\right]\,,
\end{align}
where the $n=0$ term in \eqref{Theformula2} is considered. In \eqref{rmasymp},  when multiplying the factor outside the brackets, the first term  inside the bracket grows with $r$ ($\propto |r|^m$) corresponding to the tidal field, and the 2nd term  falls with $r$ ($\propto r^{-|m|}$) corresponding to the tidal response field by the ABH. By comparing the  form of  Eq. \eqref{rmasymp} with  Eq. \eqref{tidald}, we reproduce the zeroth-order TRC in \cite{PRDLuca},
\begin{equation}\label{sTLN2}
	k^{(0)}_{m}(l,r) =2\times(-1)^{|m|}\times\frac{\Gamma \left[1+\frac{|m|}{2}+i \frac{lm}{2}\right]\Gamma\left[\frac{|m|}{2} + i \frac{lm}{2}\right]}{\Gamma (|m|)\Gamma (|m|+1) \Gamma \left[1-\frac{|m|}{2}+i \frac{lm}{2}\right] \Gamma\left[ -\frac{|m|}{2}+i \frac{lm}{2} \right]} \ln \left(r\right).
\end{equation}
Moreover, we write $k^{(0)}_{m}(l,r)$  in terms of  its real and imaginary parts, $\kappa ^{(0)}_m$ and $\nu^{(0)}_m$, respectively.  They are (for derivation, see Appendix \ref{td})
\begin{equation} \label{skapa}
	\kappa ^{(0)}_m(l,r)=\frac{(-1)^{|m|+1}}{\pi^2}\times\frac{1}{\Gamma(|m|)\Gamma(|m|+1)}\times |\delta |^2|\Gamma(-\delta)|^4 \times \left(1+(-1)^{|m|+1}\cosh\pi lm\right)\ln r,
\end{equation}
and
\begin{equation}\label{snu}
	\nu^{(0)}_m=0\, 
\end{equation}
{with $\delta=\beta-\alpha$.}
\begin{figure}[hbt]
	\centering
	\includegraphics[scale=0.25]{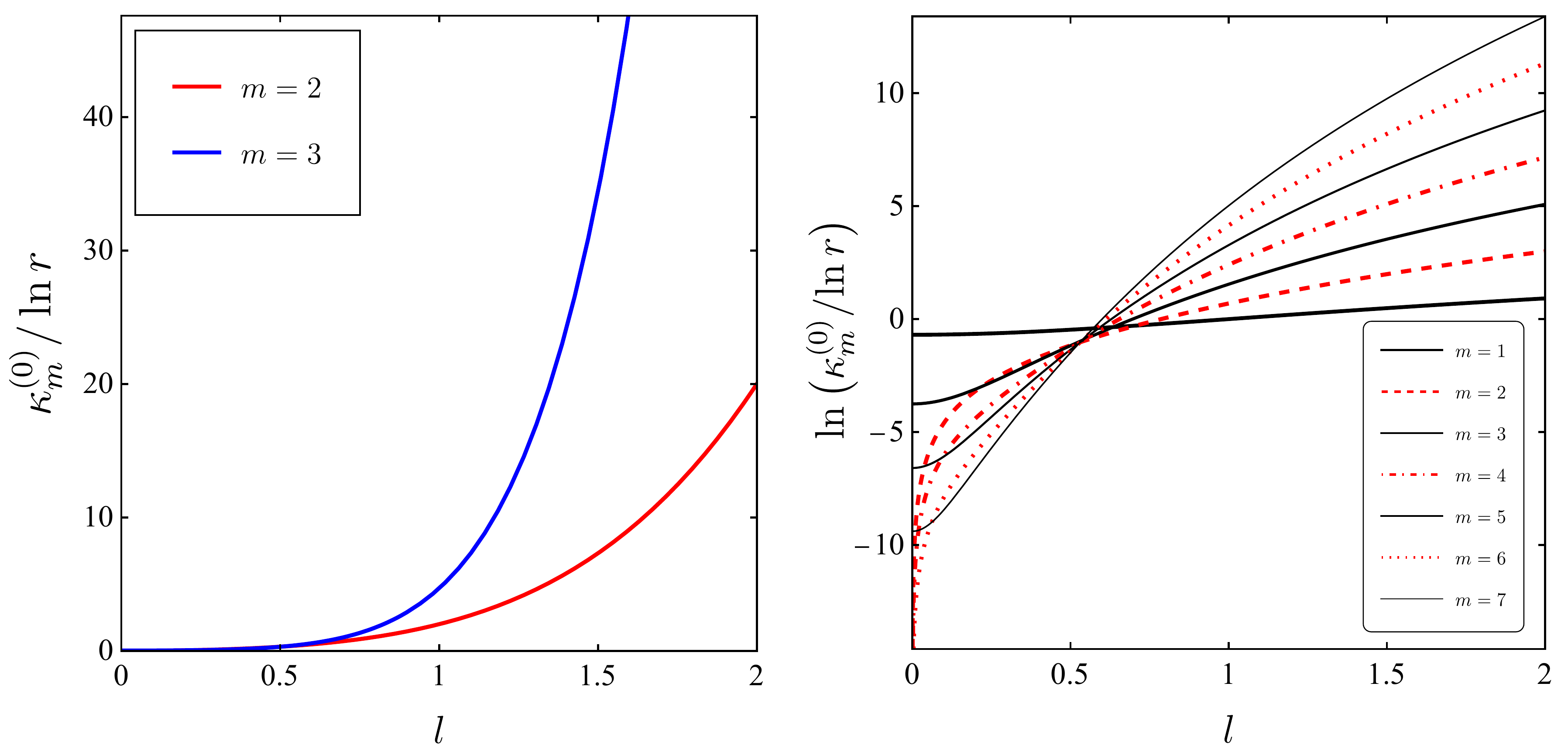}
	\caption{{ Variation of $\kappa^{(0)}_m/(\ln r)$ with $m$ and $l$ [Eq. \eqref{skapa}].} (Left) $\kappa^{(0)}_m$ monotonically increases with $|l|$ in the domain. (Right) For even $m$ (dashed curves), $\lim_{l\to 0} \ln(\kappa^{(0)}_m/\ln r)=-\infty$, since  $\kappa^{(0)}_m =0$ at $l=0$. For odd $m$ (solid curves), $\ln (\kappa^{(0)}_m/\ln r)$ is finite-valued for finite $l$. See the properties of $\kappa^{(0)}_m$ in the main text for reference.}
	\label{fig2}
\end{figure}

{The TLN  $\kappa_m^{(0)}(r)$ has nice mathematical properties for $r > 1$, which are listed  below (for proof, see Appendix \ref{properties})}:
\begin{itemize}[itemsep=3pt, topsep=4pt, parsep=0pt]
\item  $\kappa^{(0)} _m (l,r)=\kappa^{(0)}_{-m}(-l,r)=\kappa^{(0)} _{-m}(l,r)=\kappa^{(0)} _m (-l,r)$.
\item {\it $\kappa^{(0)}_m =0$  if and only if $m$ is an even number, and $l=0$.}
\item $\kappa^{(0)}_m\geq 0$.
\item {\it $\frac{\partial \kappa^{(0)} _m}{\partial l}=0$ at $l=0$.}
\item {\it $\kappa^{(0)}_m(l,r)$ is a smooth ($C^\infty$) function of $l$.}
\end{itemize}
The variation of $\kappa^{(0)}_m/(\ln r)$ with $m$ and $l$ is illustrated in Fig. \ref{fig2}.

\subsection{\bf Analytical solution of the radial wave equation to  {first order in $\omega$}  }\label{1storderT}

We now write Eq. \eqref{Teukolsky3} up to linear order in $\omega$, and {employ the aforementioned near zone approximation} to obtain
\begin{equation}\label{Teukolsky2}
	\Delta (r)\frac{d}{dr}\left(\Delta (r) \frac{dR^{(1)}_{\omega, m}}{dr}\right)+\left[f_0(m,l,r)+f_1(m,l,\omega, r)\right]R^{(1)}_{\omega, m}=0\, .
\end{equation}
{Therefore, $R^{(1)}_{\omega,m}(r)$, capturing the dynamical effect, can be regarded as the first-order approximation of $R_{\omega,m}(r)$. Like $R^{(0)}_{\omega,m}(r)$, $R^{(1)}_{\omega,m}(r)$ also naturally differs from $R_{\omega,m}(r)$ {when $r$ is far from the near zone.}}
 As in the study of the static TRC, after substituting $z=\frac{1}{r^2}$ and {$R^{(1)}_{\omega,m}(r):=z^\alpha (1-z)^\beta F(z)$},  Eq. \eqref{Teukolsky2} reduces to the differential equation for the hypergeometric function \eqref{odehy} {for $m\neq 0$} with
\begin{align}
	& (\alpha,\beta)=\left(\frac{|m|}{2}\sqrt{1+\frac{2\omega l}{m}},\frac{iml}{2}\sqrt{1-\frac{2\omega}{lm}}\right),\label{alphabetaT}&\\
	&(a~,b~,c)=\left(1+\alpha+\beta,\alpha+\beta, 1+2\alpha\right)\label{abcT}.&
\end{align}
 Note that there is a sign ambiguity in the choice of $\alpha$ and $\beta$. None of the choices change the general solution of the {radial wave equation} in the hypergeometric form, which is a second-order differential equation \eqref{Teukolsky2}, and consequently, has two linearly independent solutions (For further detail see  the Appendix \ref{hy1}). {We observe that}, for a non-rotating ABH ($l=0$), the time dependence does not introduce any correction over zeroth-order (static) approximation to the solution at first order in $\omega$. {Secondly, for $\omega=0$, the values of $\alpha$ and $\beta$ in \eqref{alphabetaT} reduce to their respective zeroth-order values in Sec. \ref{zero}, as expected.}

 From the expression in \eqref{alphabetaT}, for $ml\neq 0$, $\beta$ is is either purely imaginary or purely real. If $\beta$ is real and nonzero, {then the overall multiplicative factor $(1-z)^\beta$ in $R^{(1)}_{\omega,m}$ at the event horizon at $z=1$ either diverges for $\beta <0$  or becomes zero for $\beta>0$.} For $\beta=0$, $(1-z)^\beta$ at $z=1$ is undefined. For the existence of a sensible solution of  Eq. \eqref{Teukolsky2}, we require $\beta$ to be purely imaginary. When $\beta$ is purely imaginary, {the limit of the multiplication factor at $z=1$, giving $\lim_{z\to 1}(1-z)^\beta=e^{\beta \ln (1-z)}$,} corresponds to an infinite blue shift at the event horizon.
{Since $\beta$ is purely imaginary, $(c-a-b)= -2\beta\notin \mathbb{Z}$. For $c-a-b\notin\mathbb{Z}$, we can write down any general solution of the hypergeometric function's ODE as a linear combination of two linearly independent solutions \cite{Beals_Wong_2010}.}
\begin{align}
	F(z)=&C_{\omega,m} ~_{2}F_{1}(a,b;a+b+1-c;1-z)\nonumber\\
	&\hspace{2cm}+D_{\omega,m}(1-z)^{c-a-b}~_{2}F_{1}(c-b,c-a;1+c-a-b;1-z)\nonumber\\
	=&C_{\omega,m}~_{2}F_{1}(1+\alpha+\beta,\alpha+\beta;1+2\beta;1-z)\nonumber\\
	&\hspace{2cm}+D_{\omega,m}(1-z)^{-2\beta}~_{2}F_{1}(1+\alpha-\beta,\alpha-\beta;1 -2\beta;1-z).\nonumber \\
\label{CDom2}
\end{align}
  Now, the solution of (\ref{Teukolsky2}) becomes
\begin{eqnarray}\label{Ralphabeta2}
	R^{(1)}_{\omega,m}(r)&&\equiv z^\alpha\left[\left(1-z\right)^\beta C_{\omega,m}~_{2}F_{1}(1+\alpha+\beta,\alpha+\beta;1+2\beta;1-z)\right.\nonumber\\
&&\quad\quad\left.+D_{\omega,m}(1-z)^{-\beta}~_{2}F_{1}(1+\alpha-\beta,\alpha-\beta;1 -2\beta;1-z)\right].
\end{eqnarray}
{For $\omega=0$, Eq. \eqref{Ralphabeta2} gives the zeroth-order analytical solution, since $c-a-b\notin\mathbb{Z}$ in this case.}

Approximating $\beta$ up to linear order in $\omega$  by assuming
\begin{equation}\label{cbeta2}
\omega\ll \frac{|lm|}{2} \, ,
\end{equation}
it becomes
\begin{equation}\label{betal}
	\beta\approx -\frac i2 (\omega-lm)+O(\omega^2)=-\frac i2 \tilde{\omega}+O(\omega^2).
\end{equation}
{Note that \eqref{cbeta2} is a sufficient, but not necessary condition for $\beta$ to be purely imaginary.}
{Substituting the approximate expression for $\beta$ in \eqref{betal} into  Eq. \eqref{Ralphabeta2}, we calculate $\delta\Phi^{(1)}_{\omega,m}$ near the event horizon as}
\begin{equation}\label{ioh}
	\lim_{r\to 1}\delta\Phi^{(1)}_{\omega,m}(r)\equiv \tilde {C}_{\omega,m}e^{-i\tilde{\omega}v}e^{im(\phi+lt)}+\tilde {D}_{\omega,m}e^{-i\tilde{\omega}u}e^{im(\phi+lt)}
\end{equation}
where $ \tilde {C}_{\omega,m}=4^\beta e^{-2\beta} {C}_{\omega,m}$, and $\tilde{D}_{\omega,m}=4^{-\beta}e^{2\beta}{D}_{\omega,m}$. By comparing Eq. \eqref{ioh} with \eqref{ingoing}  {under} the ingoing boundary condition at the horizon, $\tilde{D}_{\omega,m}=0$, and thus $D_{\omega,m}=0$ in \eqref{CDom2} and in \eqref{Ralphabeta2}.
Eq. \eqref{cbeta2} naturally satisfies the condition for BH superradiance,  i.e., $0<\omega < lm$ in draining bathtub ABHs \cite{Basak_2003,PhysRevD.91.124018}. Similar situations also occur for rotating BHs \cite{Starobinskii:1973vzb} in general relativity.

{From the expression \eqref{abcT}, $\operatorname{Re}(c)>0$. The transformation of the hypergeometric function in  \eqref{CDom2} (for $D_{\omega,m}=0$) under $1-z\rightarrow z$ depends on the value of $c$, leading to two distinct cases \cite{Beals_Wong_2010}. }

(I) $c\notin\mathbb{Z^+}$:-
Regarding the the term with the coefficient $C_{\omega,m}$ in  \eqref{CDom2}, we employ the transformation rule for a noninteger $c$ \cite{Beals_Wong_2010} as
\begin{multline}\label{te2}
	_{2}F_{1}(a,b;a+b+1-c;1-z)=\frac{\Gamma(a+b+1-c)\Gamma (1-c)}{\Gamma(a+1-c)\Gamma(b+1-c)}~_{2}F_{1}(a,b;c;z)\\
	+z^{1-c}~\frac{\Gamma(a+b+1-c)\Gamma (c-1)}{\Gamma(a)\Gamma(b)}~_{2}F_{1}(a+1-c,b+1-c;2-c;z).
\end{multline}
Thus, the term with the coefficient $C_{\omega,m}$ in  \eqref{Ralphabeta2} in terms of $\alpha$, $\beta$ becomes
\begin{multline}\label{Coma2}
	\frac{\Gamma(1+2\beta)\Gamma (-2\alpha)}{\Gamma(1+\beta -\alpha)\Gamma(\beta-\alpha)}~_{2}F_{1}(1+\alpha+\beta,\alpha +\beta;1+2\alpha;z)(1-z)^\beta z^\alpha\\
	+\frac{\Gamma(1+2\beta)\Gamma (2\alpha)}{\Gamma(1+\beta +\alpha)\Gamma(\beta+\alpha)}~_{2}F_{1}(1-\alpha+\beta,-\alpha +\beta;1-2\alpha;z) (1-z)^\beta z^{-\alpha}.
\end{multline}
%The expression \eqref{Coma} is invariant under $\alpha\rightarrow -\alpha $. And similarly, the term with the coefficient $D_{\omega,m}$ is also invariant under the transformation $\alpha\rightarrow -\alpha $. Thus we establish with an explicit example of $c-a-b\notin\mathbb{Z}$, $c\notin \mathbb{Z}$, all four choices in $(\alpha,\beta)$ generate the same general solution for $R_{\omega,m}(r)$. We choose
%For $\omega =0$, $c=(1+|m|)\in \mathbb{Z^+}$, the transformation equation \eqref{te} is not valid anymore; the Eq. \eqref{Theformula2} is the limiting behaviour of the Eq. \eqref{te}. At $\omega= 0$, the transformation equation \eqref{te} possesses singularity of the form $\frac{0}{0}$; using L'H\^opital rule, it is possible to derive the Eq. \eqref{Theformula2} as limiting function\cite{Beals_Wong_2010}.

{Therefore, $R^{(1)}_{\omega,m}(r)$ from\eqref{Ralphabeta2}, respecting the ingoing condition at the event horizon can be expressed as
\begin{equation}\label{match}
\begin{split}
R^{(1)}_{\omega,m}(r)=C^1_{\omega,m}r^{2\alpha}\left(1-\frac{1}{r^2}\right)^\beta\left[_2 F_1\left(a+1-c,b+1-c;2-c;\frac{1}{r^2}\right) \right. \\
    \left. +k^{(1)}_mr^{-4\alpha}~_2F_1\left(a,b;c;\frac{1}{r^2}\right)\right],
\end{split}
\end{equation}
where $C^1_{\omega,m}=C_{\omega,m}\frac{\Gamma(a+b+1-c)\Gamma(c-1)}{\Gamma(a)\Gamma(b)}$, and
\begin{align}
	& k^{(1)}_{m}=\frac{\Gamma (1-c)\Gamma(a)\Gamma(b)}{\Gamma(c-1)\Gamma(a+1-c)\Gamma(b+1-c)}=\frac{\Gamma(-2\alpha)\Gamma(1+\alpha+\beta)\Gamma(\alpha+\beta)}{\Gamma(2\alpha)\Gamma(1-\alpha+\beta)\Gamma(-\alpha+\beta)}& \label{abk}\\
	& =\frac{\Gamma\left(-|m|\sqrt{1+\frac{2\omega l}{m}}\right)\Gamma\left(1+\frac{|m|}{2}\sqrt{1+\frac{2\omega l}{m}}+\frac{iml}{2}\sqrt{1-\frac{2\omega}{lm}}\right)\Gamma\left(\frac{|m|}{2}\sqrt{1+\frac{2\omega l}{m}}+\frac{iml}{2}\sqrt{1-\frac{2\omega}{lm}}\right)}{\Gamma\left(|m|\sqrt{1+\frac{2\omega l}{m}}\right)\Gamma\left(1-\frac{|m|}{2}\sqrt{1+\frac{2\omega l}{m}}+\frac{iml}{2}\sqrt{1-\frac{2\omega l}{m}}\right)\Gamma\left(-\frac{|m|}{2}\sqrt{1+\frac{2\omega l}{m}}+\frac{iml}{2}\sqrt{1-\frac{2\omega }{lm}}\right)}. \label{kom2}
\end{align}
{From the definition of the Gauss hypergeometric series, we rewrite $R^{(1)}_{\omega,m}(r)$ in the series form as}
{\small
\begin{align}
& R^{(1)}_{\omega,m}(r)
= C^1_{\omega,m}\left(1-\frac{1}{r^2}\right)^\beta
\left[
\sum_{n=0}^{\infty}
\frac{(a+1-c)_n(b+1-c)_n}{(2-c)_n}
\frac{r^{2(\alpha-n)}}{n!}
+k^{(1)}_m
\sum_{n=0}^{\infty}
\frac{(a)_n(b)_n}{(c)_n}
\frac{r^{-2(\alpha+n)}}{n!}
\right]
\label{id}\\
& = C^1_{\omega,m}\left(1-\frac{1}{r^2}\right)^\beta
\left[
\left(r^{2\alpha}+k^{(1)}_m r^{-2\alpha}\right)
+\left(
\frac{(a+1-c)(b+1-c)}{2-c}r^{2(\alpha-1)}
+k^{(1)}_m\frac{ab}{c}r^{-2(\alpha+1)}
\right)
+\cdots
\right].
\label{id2}
\end{align}
}
Comparing \eqref{id2} with \eqref{tidald} and \eqref{tidald2}, we must have  $\alpha\in\mathbb{R^+}$, so that we have {\it the tidal term}, i.e., $\left(r^{2\alpha}+k^{(1)}_m r^{-2\alpha}\right)$  inside \eqref{id2} from the $n=0$ term, and $k^{(1)}_m$ is thus identified with TRC.  In the series \eqref{id}, the terms with $n\geq 1$ are subleading terms {at $r\gg 1$}. Note that, Eqs. \eqref{tidald} and \eqref{tidald2} describe the tidal effect {in three spatial dimensions} where $\ell (>2) $ takes positive integer values.  In our 2+1 D case, $\alpha$ plays the role of $\ell$, and it can even take positive noninteger values.

{Since $\alpha\in\mathbb{R}^+$}, {the real and imaginary parts of $k^{(1)}_m$ can be calculated (see Appendix \ref{td})} as
\begin{equation}\label{TLN1}
	\kappa^{(1)}_{m}(\omega,l)=-\frac{1}{2\pi^2}\times\frac{\Gamma (-2\alpha)}{\Gamma (2\alpha)}\times|\delta|^2|\Gamma(-\delta)|^4\times\left(1-\cos 2\pi\alpha\cosh\left(2\pi\operatorname{Im}(\beta)\right)\right),
\end{equation}
and
\begin{equation}\label{DC1}
	\nu^{(1)}_{m}(\omega,l)=\frac{1}{2\pi^2}\times\frac{\Gamma (-2\alpha)}{\Gamma (2\alpha)}\times|\delta|^2|\Gamma(-\delta)|^4\times\sin2\pi\alpha\sinh\left(2\pi \operatorname{Im}(\beta)\right),
\end{equation}
where $\delta=\beta-\alpha$. {Following \eqref{TLN1} and \eqref{DC1}, we present Fig. \ref{fig4} describing the $\omega$ dependence of $\kappa^{(1)}_{m}$ and $\nu^{(1)}_{m}$ after setting up parameter ranges in Sec. \ref{Validity}.}

{Using the fact that the Gamma function $\Gamma(x)$ diverges for $x\in\mathbb{Z}_{\leq 0}$ \cite{stein2010complex}, together with the reflection formula $\Gamma (x)\Gamma (1-x)=\frac{\pi}{\sin \pi x}$ applied to \eqref{TLN1} and \eqref{DC1}, { we obtain the limiting values of the TRC and TDC, for  $\alpha\to\frac{N}{2}$, respectively, with $N\in\mathbb{Z^+}$ to be}
\begin{equation}\label{TRCN}
\lim _{\alpha\to\frac{N}{2}}\kappa^{(1)}_m=\pm\infty,\qquad \lim _{\alpha\to\frac{N}{2}} \nu^{(1)}_m=-\frac{1}{2\pi N((N-1)!)^2}\times\times|\delta_N|^2|\Gamma(-\delta_N)|^4\times\sinh\left(2\pi \operatorname{Im}(\beta)\right)
\end{equation}
{with $\delta_N=\beta-\frac{N}{2}$} from the expression of $\delta$ after \eqref{DC1}.
{For $\omega\to 0$, $2 \alpha \to N$ where  $N=|m|$ and $|m|\in\mathbb{Z^+}$,  {Eq. \eqref{odehy} applies}. {In Fig. \ref{fig4},  as $\omega\to 0$, $\kappa^{(1)}_m$ diverge, and $\nu^{(1)}_m$ take a finite value.} However, the case of $2\alpha=N$ turns out to be markedly different from $2\alpha\to N$ in \eqref{TRCN}, {since  $c=1+N \in\mathbb{Z^+}$}. Hence, it falls outside the scope of the case (I), where $c\notin\mathbb{Z^+}$, and requires separate treatment as discussed below.}
%Such a mild time dependence behaviour of the wave is quite distinct than static wave's behaviour as we realize in the Sec. \ref{nH}, and reflected in the expression of TLN because of hypergeometric functions behaviour is quite different for integer $c$, c.f., Sec. \ref{ambiguity} (see \cite{Beals_Wong_2010}).*)

%(vii)\begin{lemma}
	% $\kappa _m$ is a monotonically increasing function of $|l|$ given integer $|m| \geq 1$, and $l\in \mathbb{R}$.
	%\end{lemma}
	%\begin{proof}
	%In the expression \eqref{kapam}, the factor $(-1)^{|m|+1}|\delta|^2\left(1+(-1)^{|m|+1}\cosh\pi lm\right)$ is a monotonically increasing function of $|l|$. From the expression \eqref{identity}, we observe $\frac{\partial |\Gamma(-\delta)|^2}{\partial |l|}>0$ for a given integer $m$ with $|m| \geq 1$, and $|l|>0$. Hence proved.
	%\end{proof}\begin{figure}
(II){  $c\in \mathbb{Z}^+$:-}

{For $2\alpha=N$, and {consequently} $c=1+N$,}  {the transformation equation \eqref{te2} encounters a removable singularity \cite{Beals_Wong_2010}, where the L'H\^opital rule can be applied to find {the limit of $R^{(1)}_{\omega,m}$ as $2 \alpha \to N$}. Thanks to the {Riemann's theorem on removable singularities} \cite{stein2010complex}, the limiting behaviour of $R^{(1)}_{\omega,m}$, i.e., $\lim_{2\alpha\to N}R^{(1)}_{\omega,m}$ captures the case of $2\alpha=N$ corresponding to a positive integer $c$ \cite{Beals_Wong_2010}. The solution of the hypergeometric equation takes the form of  Eq. \eqref{Theformula2} with $|m|$ replaced by $N$ as}
	\begin{multline}\label{Theformula3}
		F(z)=C_{\omega,m}\frac{\Gamma |N|}{\Gamma (a)\Gamma (b)}z^{-N}\sum_{n=0}^{N-1}\frac{(a-N)_n (b-N)_n}{n!(1-N)_n}z^n\\
		-C_{\omega,m}\frac{(-1)^{N}}{\Gamma (a-N)\Gamma (b-N)}\sum _{n=0}^{\infty}\frac{(a)_n(b)_n}{n!(n+N)!}z^n[\ln z-\psi (n+1)-\psi (n+N+1)+\psi (a+n) +\psi (b+n)].
	\end{multline}
It is straightforward to derive the most dominating term as in (\ref{rmasymp}) where the TRC can be obtained as
	\begin{equation}\label{tsar}
		k^{(1)_N}_m(\omega,l,r)=2\times(-1)^{N}\times\frac{\Gamma (a) \Gamma (b)}{\Gamma (N)\Gamma (N+1) \Gamma (a-N)] \Gamma(b-N)} \ln \left(r\right).
	\end{equation}
The	TLN and TDC are given, respectively, by
	\begin{equation}\label{kapastar}
		\kappa^{(1)_N}_m(\omega,l,r)=\frac{(-1)^{N+1}}{\pi^2}\times\frac{1}{\Gamma(N)\Gamma(N+1)}\times |\delta |^2|\Gamma(-\delta)|^4 \times \left(1+(-1)^{N+1}\cosh\left(2\pi \operatorname{Im}(\beta)\right)\right)\ln r,
	\end{equation}
	\begin{equation}\label{nustar}
		\nu_m^{(1)_N}(\omega,l,r)=0.
	\end{equation}
Interestingly, when $\omega = 0$, $N =\vert m\vert$ and $a=1+\frac{|m|}{2}+i \frac{lm}{2}, b=\frac{|m|}{2} + i \frac{lm}{2}$,
the static TRC  in (\ref{skapa}) and (\ref{snu}) can be achieved.  {Comparing Eq. \eqref{kapastar} and Eq. \eqref{nustar} with Eq. \eqref{TRCN}, {we observe that the behaviour of TRCs does not change smoothly at $2\alpha=N$. However, due to the aforementioned Riemann's theorem, the solution $R^{(1)}_{\omega,m}$ is a smooth function of $\alpha$ at $2\alpha=N$, because this point is a removable singularity.} {The bottom-left panel of Fig. \ref{fig3} illustrates} a smooth dynamic to static ($\omega=0$) transition of $R_{\omega,m}$ in the near zone as $R_{\omega,m} (r)\approx R^{(1)}_{\omega,m}(r)$ for the chosen parameter value. The static case (i.e., $\omega=0$) likewise corresponds to a removable singularity of the solution with respect to $\omega$.

\subsection{Validity of the radial wave equation at linear order in $\omega$}\label{Validity}
Unlike the static case ($\omega=0$) in Sec. \ref{zero} with the full analytic solution $\forall r\geq 1$,  the dynamical case, i.e., the  nonzero $\omega$ case in Sec. \ref{1storderT} however requires the assumptions for the analytic computation of the TRC. We have already introduced three conditions in Sec. \ref{1storderT} so that the analytic solution $R^{(1)}_{\omega,m}(r)$ is a good approximation of the full radial part $R_{\omega,m}(r)$ in the near zone. These can be listed as follows: (a) The near zone approximation gives  $f_2\ll 1$ in \eqref{NZ}, which implies that $\omega\ll 1$ at the event horizon and in the near zone, $r\ll R_1=\frac{1}{\omega}$. (b) Condition on $\beta$ gives $\omega\ll\frac{|m||l|}{2}$ in \eqref{cbeta2} so that $\beta\approx-\frac{i\tilde{\omega}}{2}$ in \eqref{betal}, and (c) Condition on $\alpha$ that $\alpha\in\mathbb{R^+}$ for the tidal effect to manifest, thus  $\omega< \frac{|m|}{2|l|} ~~{\rm for}~ lm < 0$.

Additionally, to distinguish the dynamical case from the static one, we require condition (d) $|f_1|\gg f_2$ under the near zone approximation. {This is because  the time dependence} in  Eq. \eqref{Teukolsky2} enters through  the term $f_1$ for nonzero $\omega$. Condition (d) implies that in the near zone, $r\ll R_2$, where $R_2=\sqrt{\frac{ 2 \vert m \vert \vert l \vert}{\omega}}$. {Moreover, we consider condition (e) that $R_1$ and $R_2$ are greater than $r_s=\sqrt{1 + l^2}$, which yields the following constraints, respectively,}
\begin{equation}
\omega<\frac{1}{\sqrt{1+l^2}},~ {\rm and~} \omega < \frac{2 \vert m\vert \vert l\vert }{1+l^2}\,.
\end{equation}

{Combining (a) to (e), the maximum value of $\omega$ is constrained by $ \omega_{\rm max}=\inf (S)$
with
}
\begin{align}
& S=\left\{ \frac{|lm|}{2},\frac{1}{\sqrt{1+l^2}},\frac{2|lm|}{1+l^2}\right\} ~{\rm for}~lm>0,~~ S=\left\{ \frac{|m|}{2|l|}, \frac{|lm|}{2},\frac{1}{\sqrt{1+l^2}},\frac{2|lm|}{1+l^2}\right\} ~{\rm for}~lm<0,&\label{lmS}
\end{align}
where $\inf (S)$ stands for the infimum of the set $S$.}
\begin{figure}
		\centering
		\includegraphics[scale=0.25]{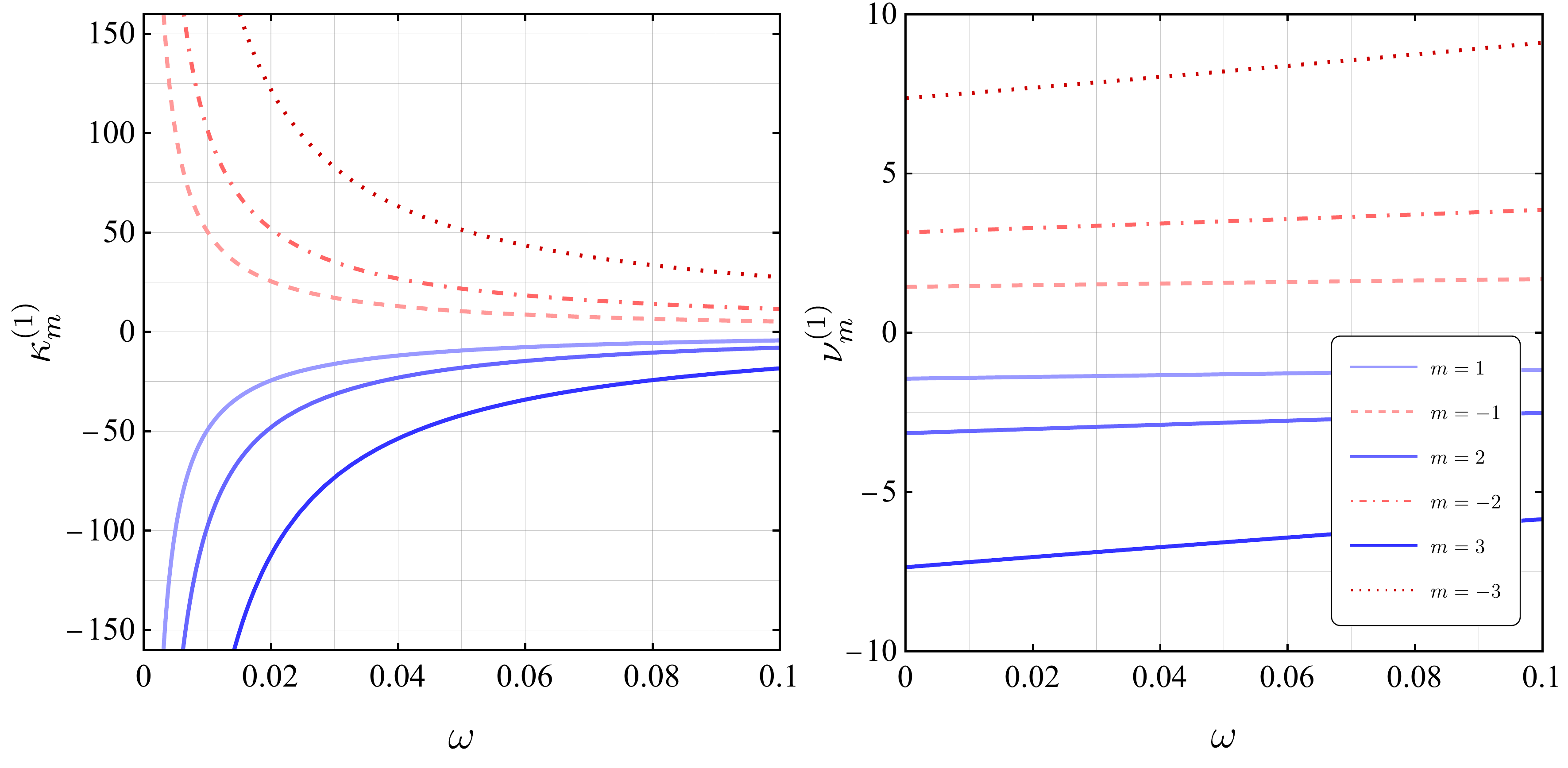}
		\caption{{Variation of TRC $k^{(1)}_m$ with $\omega$.} For all curves, $l=1$ and $\omega_{\max}=0.707$ [Eq. \eqref{lmS}]. In the limit $\omega\to 0_+$: (Left) $\kappa^{(1)}_m$ diverge, whereas (Right) $\nu^{(1)}_m$ take finite values [Eq. \eqref{TRCN}].}
		\label{fig4}
	\end{figure}
{In Fig. \ref{fig4}, for all the chosen parameter values, $\omega_{\rm max}$ comes from $\frac{1}{\sqrt{1+l^2}}$ in the set $S$.}
{Following \eqref{lmS}, the permissible range of $|l|$ for a given value of  $\omega$ and $m$ is $\sup (L)\ll |l|\ll \inf (L')$. The notation $\sup (L)$}  stands for the supremum of the set $L$, and $\inf (L')$ stands for the infimum of the set $L'$, where
\begin{align}
& L=\left\{ \frac{2\omega}{|m|},\frac{|m|}{\omega}\left(1-\sqrt{1-\frac{\omega^2}{m^2}}\right)\right\}, ~L'=\left\{ \sqrt{\frac{1}{\omega^2}-1},\frac{|m|}{\omega}\left(1+\sqrt{1-\frac{\omega^2}{m^2}}\right)\right\}\label{lVV}.
\end{align}

\subsection{Numerical extraction of TRCs of the ABH spacetime}
\begin{figure}
		\centering
		\includegraphics[scale=0.4]{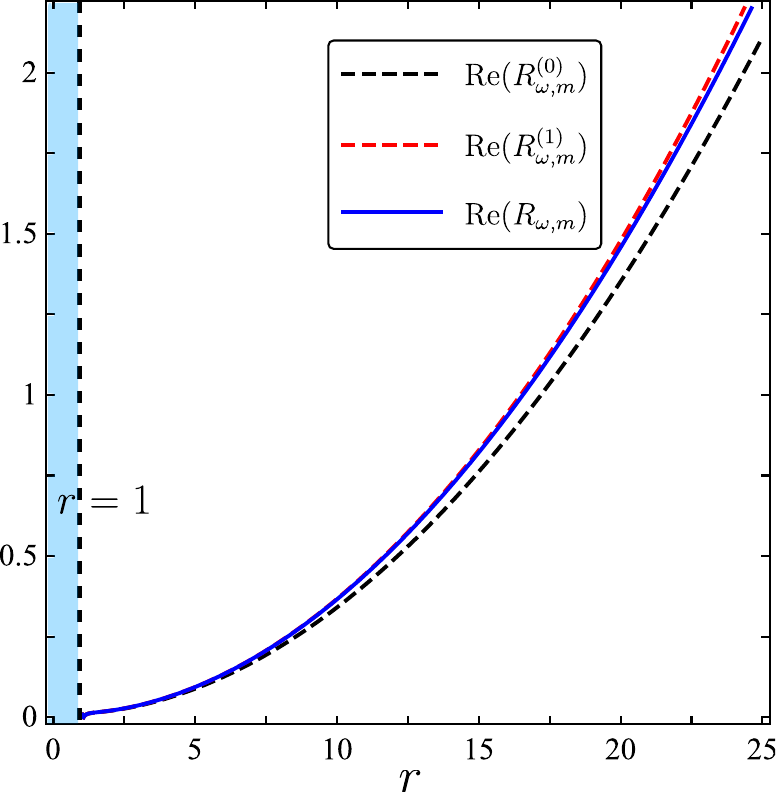}~\includegraphics[scale=0.48]{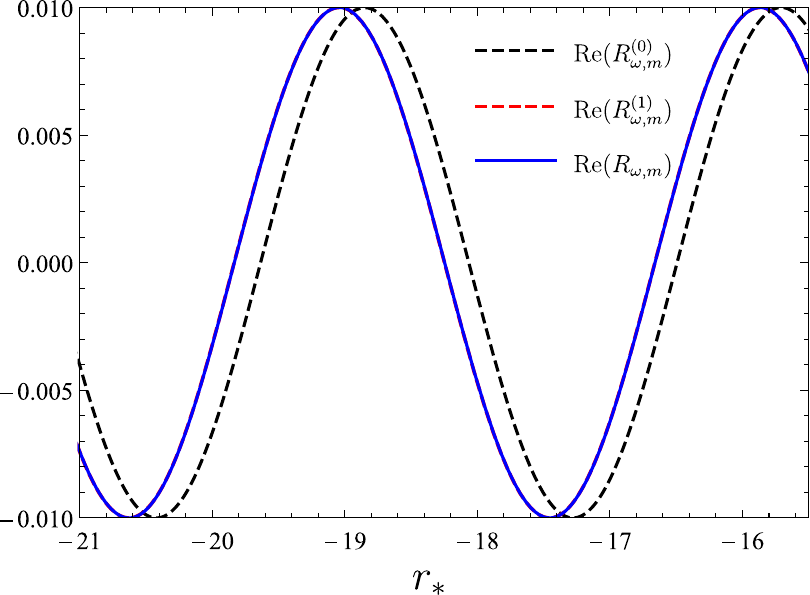} \\
		\includegraphics[scale=0.2]{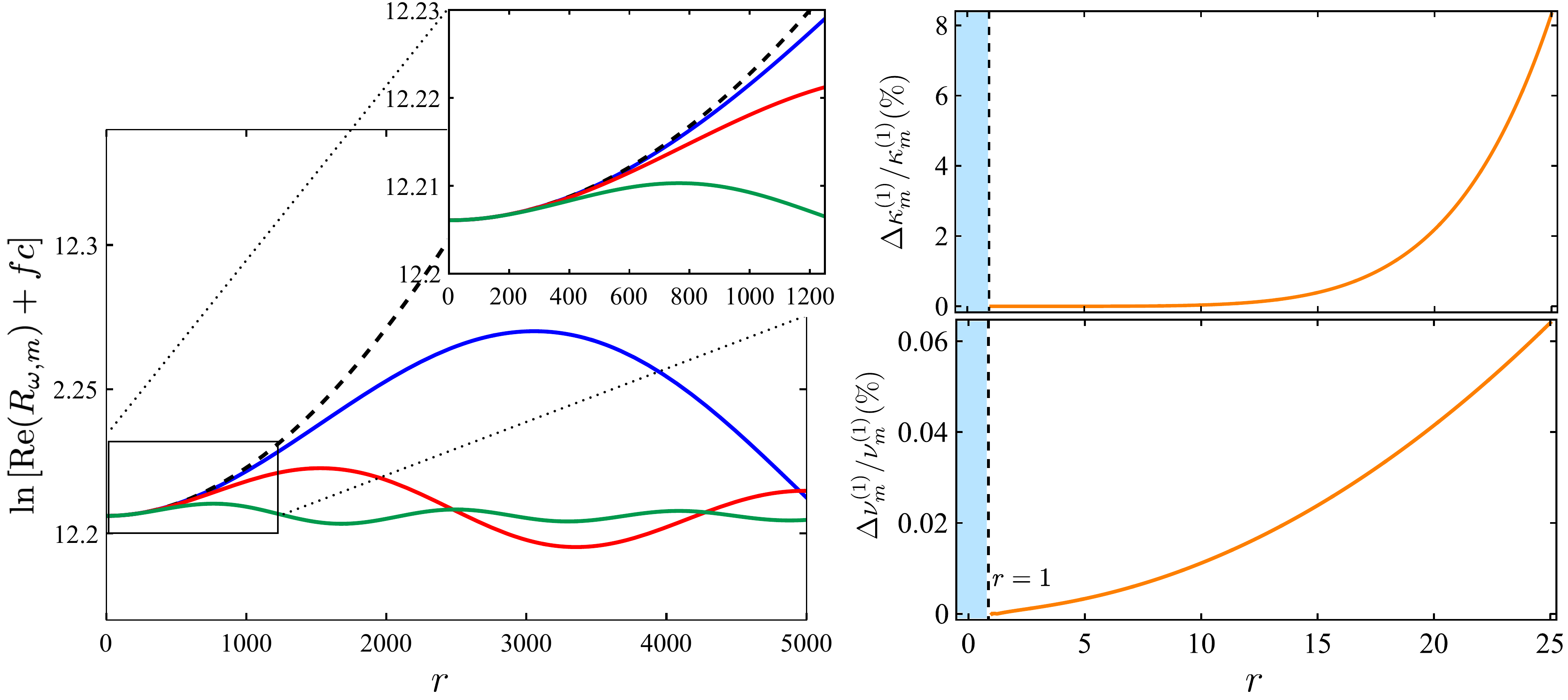}
		\caption{In all plots, $l=1$ (thus $r_s\approx 1.41$), $m=2$, and initial value $h(r=1)=0.01$ for the ingoing modes at the horizon \eqref{ingoing}. The aforementioned parameter values with $\omega=2\times 10^{-2}$ in the top panels yield $R_1=50$, $R_2\approx 14.14$. (Top-left) Comparison of the analytical solutions $R^{(0)}_{\omega,m}$ and $R^{(1)}_{\omega,m}$ obtained from \eqref{Ralphabeta2} ($D_{\omega,m}=0$ in \eqref{Ralphabeta2} to ensure that the solutions are ingoing at the ABH event horizon) with the full numerical solution $R_{\omega,m}$ in the near zone.   (Top-right) When we plot the numerical solution $R_{\omega,m}$ and the analytical solutions $R^{(0)}_{\omega,m}$, $R^{(1)}_{\omega,m}$ in Mathematica with a step size of $10^{-20}$, the near horizon high frequency oscillations caused by large blue shift are captured in the $r_*$ coordinate within the narrow interval $1+10^{-19}< r<1+10^{-14}$ (see Eqs. \eqref{ingoing} and \eqref{blue}). The oscillations of $R^{(1)}_{\omega,m}$ closely match with those of $R_{\omega,m}$. (Bottom-left) The tidal deformation of the full solution $R_{\omega,m}$ at various frequencies. Blue, red and green curves correspond to $\omega=10^{-3}$, $\omega=2\times 10^{-3}$, and $\omega=4\times 10^{-3}$, respectively. The logarithmic scale is used, and the factor $fc=2\times 10^5$ is chosen to fit all the plots in the frame nicely. A zoomed-in near-horizon view depicts a smooth static to dynamic transition. (Bottom-right) The deviation of the numerically extracted TRC $B(r)$ in \eqref{fitting} from its analytic value $k^{(1)}_{m}$ in Eq. \eqref{kom2} is shown, where $\Delta \kappa^{(1)}_m=\operatorname{Re}(B(r))-\kappa^{(1)}_m$, $\Delta \nu^{(1)}_m=\operatorname{Im}(B(r))-\nu^{(1)}_m$. The parameter values with $\omega=2\times 10^{-3}$ in the plot yield $k^{(1)}_{2}\approx -498.09$, $\nu^{(1)}_{2}\approx -3.14$, with corresponding length scales $R_1=500$, $R_2\approx 44.72$.}\label{fig3}
	\end{figure}
We numerically integrate the full  radial equation after factoring out the ingoing horizon behaviour, $ R_{\omega,m}=e^{-i\omega r*} h(r)/\sqrt{r}$, where $h(r)$ is approximated as a polynomial in $(r-1)$ \cite{QNMBertiE}. The initial conditions for the integration can be determined by $h(r=1)$. The details are given in Appendix \ref{hofr}. We then evaluate $R^{(0)}_{\omega,m}(r)$ and $R^{(1)}_{\omega,m}(r)$ using their analytic solutions in \eqref{Ralphabeta2} by setting $D_{\omega,m}=0$, $\omega=0$, and $D_{\omega,m}=0$, $\omega\neq0$, respectively. The constant $C_{\omega,m}$ appearing in these analytic solutions can be obtained from the corresponding initial values of $h(r=1)$ used in the numerical integration  (see Appendix \ref{hofr}), given by
  \begin{equation}\label{Ch}
h(r=1)=h_0=C_{\omega,m}e^{-2\beta}4^\beta.
\end{equation}
 We find that $R^{(1)}_{\omega,m}\simeq R_{\omega,m}$ in the near zone as seen in the top panel of Fig. \ref{fig3}, {whereas} $R^{(1)}_{\omega,m}$  starts to deviate from $R_{\omega,m}$ at large $r$, as expected from our analysis in Sec. \ref{1storderT}. It is also shown that $R^{(1)}_{\omega,m}$ provides a better approximation for $R_{\omega,m}$ than $R^{(0)}_{\omega,m}$.}
  Similar to \cite{tidaldeformabilitydisk}, {we  extract the TRCs {from the numerical solution $R_{\omega,m}$}} by fitting it with the function $R^{fit}_{\omega,m}$. {This function  is motivated by the first-order analytical expression in Eq.~\eqref{match} given by}
 \begin{align}
	R^{fit}_{\omega,m}(r)=A(r) r^{2\alpha}
\left(1-\frac{1}{r^2}\right)^\beta&\Bigg[{}_2 F_1\left(a+1-c,b+1-c;2-c;\frac{1}{r^2}\right)\nonumber\\
&+B(r) r^{-4\alpha}~_2F_1\left(a,b;c;\frac{1}{r^2}\right)\Bigg] \label{fitting}
\end{align}
with two functions $A(r)$ and $B(r)$.
 The numerical  solution of $R_{\omega,m}$ and its first derivative are equated with the expression (\ref{fitting}) and its first derivative, respectively, in order to calculate $A(r)$ and $B(r)$ in the fitting function $R^{fit}_{\omega,m}(r)$.
In the near zone, the real and imaginary parts of $B$, i.e., the numerical TRC, turn out to be close to $\kappa^{(1)}_m$ \eqref{TLN1}) and $\nu^{(1)}_m$ given by (\ref{DC1}), respectively. The value of $B$ then becomes $r$-dependent for large $r$ and departs from the {analytical} value in the near zone shown in the bottom panel of Fig. \ref{fig3}.
\section{Dynamical TRCs of AWHs in a fountaining bathtub}\label{WTD}
According to the background solutions (i) and (iii) in  Sec. \ref{AWHBH}, the perturbations in rotating ABH and AWH spacetimes with the same rotation parameter $l$ satisfy the same radial wave equation in (\ref{Teukolsky}). The AWH metric can be constructed from a fountaining bathtub. For calculating TRCs of the {AWHs}, the outgoing boundary condition at the horizon needs to be imposed. To do so, in  Eq. \eqref{CDom2}, $C_{\omega,m}=0$ but $ D_{\omega,m} \neq 0$  {for a nontrivial solution $R^{(1)}_{\omega,m}$ to exist}.
 Carrying out the similar analysis, we find out  $k_{m}^{(1) WH}$ for the fountaining bathtub AWH in the linear $\omega$ approximation {for $c\notin\mathbb{Z^+}$,}
	\begin{equation}\label{komW}
		k_{m}^{(1) WH}(\omega,l)=\frac{\Gamma (1-c)\Gamma(c-b)\Gamma(c-a)}{\Gamma(c-1)\Gamma(1-b)\Gamma(1-a)}=\frac{\Gamma(-2\alpha)\Gamma(1+\alpha-\beta)\Gamma(\alpha-\beta)}{\Gamma(2\alpha)\Gamma(1-\alpha-\beta)\Gamma(-\alpha-\beta)}=\overline{{k}^{(1)}_m(\omega,l,r)}\,,
	\end{equation}
where ${k}^{(1)}_m(\omega,l,r)$ is the TRC of the ABH in \eqref{kom2} with $\alpha, \beta$ in (\ref{alphabetaT}) or $a,b,c$ in (\ref{abcT}). To obtain the above {relation between} the TRC of AWH {and the TRC of ABH}, we have used the fact that $\beta$ is purely imaginary, and $\alpha$ is purely real. From the definition \eqref{TLNDF},
	\begin{eqnarray}\label{kapanuW}
		\kappa_m^{(1) WH}(\omega,l,r)=\kappa^{(1)}_m(\omega,l,r),\label{rkm}\\
		\nu_m^{(1) WH}(\omega,l,r)=-\nu^{(1)}_m(\omega,l,r) \label{numw}
	\end{eqnarray}
 with $\kappa^{(1)}_m(\omega,l,r)$ in (\ref{TLN1}) and $\nu^{(1)}_m(\omega,l,r)$ in (\ref{DC1})  for $ c {\notin}  \mathbb{Z^+}$.  {Similarly for} {$c=1+N\in\mathbb{Z^+}$,
 \begin{equation}
 \kappa_m^{(1)_N WH}=\overline{{k}^{(1)_N}_m(\omega,l,r)}\,
 \end{equation}
 with  ${k}^{(1)_N}_m(\omega,l,r)$ in(\ref{tsar}).}
In the same manner, one can show
\begin{equation}
k_{m}^{(0) WH}=\overline{k_{m}^{(0)}}={k_{m}^{(0)}}\, ,
\end{equation}
when ${k_{m}^{(0)}}$ in (\ref{sTLN2}) is real.
%Denoting tidal response coefficient of the time reversed analogue white hole as $k_{m}^{'WH}$, therefore we have
%\begin{equation}\label{bhwhk}
%k_{m}^{'WH}=k_{m}^{WH}(\omega,-l,r)=\overline{{k}_m(\omega,-l,r)}.
%\end{equation}
%Following Sec. \ref{AWHBH}, for $t\to -t$ of the ABH spacetime  in Sec. \ref{AWHBH}, the solution (iv) for the AWH relative to the solution (i) for the ABH, the TRC would be specified by $k_{m}^{WH}(\omega,-l,r)$.
According to Sec. \ref{FD} and \ref{AWHBH}, the effects of time reversal of both the background and the linear perturbation, i.e., ${\bf v}={\bf v}_0+\delta {\bf v}\to-{\bf v}_0-\delta{\bf v}$,  lead to the replacement of $(m,l)\rightarrow(-m,-l)$. Therefore,
\begin{equation}\label{bhwhk2}
k_{-m}^{WH}(\omega,-l,r)=k_{m}^{WH}(\omega,l,r)=\overline{{k}_m(\omega,l,r)},
\end{equation}
where the natural symmetry  under $(l,m)\to(-l,-m)$ is adopted.
\section{Conclusions and discussion}\label{CSEC}
{The analytic solutions in the near zone } corresponding to our ABH and AWH involve the Gauss hypergeometric function \cite{Beals_Wong_2010}, a common occurrence for {\it real} BHs as well  \cite{Teukolsky,rodriguez2026lovenumbersblackholes,chakraborty2026tidalresponsecompactobjects}.
  Remarkably, the static TDCs of our ABH and AWH are identically zero for arbitrary $l$ and $m$. From the perspective of TRCs, this result stands in sharp contrast to asymptotically flat BHs in Einstein gravity, where the static TLNs identically vanish, as noted in the Introduction. We also observe that the static-dynamic transition remarkably modifies the feature of the TRC where the TRC, as a function of $\omega$, is not analytic at $\omega= 0$. Although, the static-dynamic transition in terms of the TRC is nontrivial, the corresponding  solution of the radial wave equation undergoes a  smooth transition where $\omega=0$ is simply a removable singularity. In that regard, the static case can be viewed on the same footing as other special parameter values corresponding to $2\alpha\in\mathbb{Z}^+$. The associated TRC reveals logarithmic running for TLN and the vanishing of the TDC. Note, the TRC of the Kerr BH also exhibits logarithmic running, however at second order in $\omega$ \cite{Charalambous2021}. The solution $R^{(1)}_{\omega,m}$ in the linear $\omega$ approximation presented in the main text has one limitation: the oscillation frequency with respect to Eddington-Finkelstein coordinates, given by $ \tilde{\omega}^{(1)}=2 i\beta=-ml\sqrt{1-\frac{2\omega}{ml}}$, does not exactly match the oscillation frequency of the full solution, given by $ \tilde{\omega}=(\omega-ml)$ in (\ref{cbeta2}) and (\ref{betal}). This discrepancy can become important at moderate to high frequencies. This limitation can be overcome by extending the analysis to include the correct horizon oscillation frequency, yielding another analytic solution, $\tilde{R}^{(2)}_{\omega,m}$, in Appendix~\ref{2O}. Since the techniques employed in the analysis are similar to those used in the first-order approximation, we present it in Appendix \ref{2O} for brevity. This modification has two advantages over the first-order approximation. Firstly, the value of $\omega$ can be arbitrarily large. Secondly, it provides the dynamical TRC for a non-rotating $(l=0)$ ABH/AWH through the second-order near-horizon approximation. By contrast, the first-order solution $R^{(1)}_{\omega,m}$ does not capture the dynamical behaviour of the radial waves in a nonrotating analogue spacetime due to Eq. \eqref{lVV}. Consequently, one must rely on the solution $R^{(0)}_{\omega,m}$ derived under the static approximation in Sec. \ref{zero}.

In hydrodynamical analogue gravity experiments involving surface waves \cite{RalfBill}, the flow variables are the fluid surface height and fluid velocity. It is in fact  practically possible to work with long-wavelength and small-amplitude surface waves in a regime where viscosity has a negligible effect (as required for an analogue of gravity and discussed in \cite{Unruh95}), where in most of the hydrodynamical experiments, the viscous layer adjacent to the bottom of the setup is very thin \cite{bionditwolayer}.
We expect to observe near horizon tidal deformity as in Fig. \ref{fig3}, due to the undulation from the wave maker outside the event horizon. This setup can be a hydrodynamical draining bathtub for observation of superradiance effect \cite{PhysRevD.91.124018,Torres2017}, where the flow profiles are qualitatively similar to those in our model. The further scope involves extending our analysis of calculating TRC for a general rotating analogue black (white) hole spacetime in Eq. \eqref{agmn}. 
In addition, the experimental realization of a two-dimensional Bose-Einstein condensate (BEC) flow is a promising development. A recent experiment with BEC \cite{2dsupersonicBEC} has demonstrated that  a supersonic flow region can be created  inside of an acoustic horizon induced by a particle sink in the presence of an attractive potential. For more information, see {also} the earlier proposal in \cite{Datta_2025}. In such a BEC {ABH}, the waves can be created by shining a localized repulsive laser \cite{Andrews} outside the acoustic event horizon, and that would serve as a source of the tidal field. Moreover, there is a short-wavelength correction to the dispersion relation for any analogue gravity medium, known as the {\it analogue} trans-Planckian effect (see, for example, \cite{Unruh95, Jacobson2013, Datta_2023}).  The near-horizon trans-Planckian signature could also be observed in the experimental extraction of TRCs of ABHs and AWHs due to the fact that the wave  oscillates rapidly
with $r$ near the horizon, as seen in the top right panel of Fig. \ref{fig3}}.
\section{Acknowledgments}
DSL is grateful to Yu-Tin Huang, Lam Hui and Bei-Lok Hu for their inspiring discussions. This
work was supported in part by the National Science
and Technology Council (NSTC) of Taiwan, Republic
of China under
grant numbers  114-2112-M-259 -007 -MY3 (DSL) and {114-2112-M-160-001-MY2 (WCS)}.
\appendix %\section{Supplemental Material}
\section{Resolution of coordinate singularity in 2+1 D ABH and AWH spacetimes}\label{RCS}
In the {\it tilde} frame, defined by  Eq. \eqref{ct2}, the corresponding line element is
\begin{equation}\label{agmn3}
	 ds^2=\left(\frac{\rho_{0}}{c_{s0}}\right)^2\left[-\left(c_{s0}^2-(v_0^r)^2-\frac{l^2}{r^2}\right)d\tilde{t}^2+\frac{c_{s0}^2}{\left(c_{s0}^2-(v_0^r)^2\right)}dr^2-2ld\tilde{t}d\tilde{\phi}+r^2d\tilde{\phi}^2\right].
\end{equation}
However, unlike the metric \eqref{agmn} in the laboratory  frame, in this new coordinate system, the metric \eqref{agmn3} has a coordinate singularity at the event horizon, where $c_{s0}^2-(v_0^r)^2=0$. Thus, the component $g_{rr}$ diverges at the event horizon, according to Eq. \eqref{EH}. In this new coordinate system, neither the ingoing nor the outgoing null rays can be connected to those inside the event horizon due to the coordinate singularity. At the event horizon, both the ingoing and outgoing null rays tend to have zero radial coordinate velocity ($\frac{dr}{d\tilde{t}}$). Therefore, in this new coordinate system, there is no way to determine whether the spacetime corresponds to a black or a white hole. This also happens in general relativity, where the Schwarzschild solution, expressed in Schwarzschild coordinates, has a coordinate singularity at the horizon. The metric \eqref{agmn3} is insensitive to the sign of $v_0^r$. Analogue spacetimes in the laboratory frame, governed by solution (i) and (iii) of Sec. \ref{FD} (or solution (ii) and (iv)), have the same form \eqref{agmn3} in this new coordinate system.
Conversely, without having prior knowledge about the sign of $v_0^r$, (substituting $(v_0^r)^2=V_0^2$ in the metric \eqref{agmn3} with $V_0>0$) we can start from the metric \eqref{agmn3} with a coordinate singularity at the event horizon, where the coordinate singularity at the event horizon can be removed by a pair of coordinate transformations below,
\begin{eqnarray}
 & d\tilde{t}=d{t}\pm \frac{V_0}{\left(c_{s0}^2-V_0^2\right)}dr, ~~d\tilde{\phi}=d\phi\pm\frac{l V_0}{r^2\left(c_{s0}^2-V_0^2\right)}dr.\label{ct4}
\end{eqnarray}
The metric \eqref{agmn3} now becomes, respectively,
\begin{equation} \label{whbh}
ds^2=\left(\frac{\rho_{0}}{c_{s0}}\right)^2\left[-\left(c_{s0}^2-(V_0)^2-\frac{l^2}{r^2}\right)d{t}^2\mp 2V_0 dtdr -2ldtd\phi + dr^2+r^2d{\phi}^2\right].
\end{equation}
The metric \eqref{whbh} does not have a coordinate singularity at any event horizon for a regular flow in general. For $V_0(r)>0$, the spacetime can be related to an AWH spacetime in the laboratory frame for the $`-$' sign in the $g_{tr}$ component of the metric \eqref{whbh} and to an ABH spacetime for the $`+$' sign.

\section{Numerical integration of the radial wave equation}\label{hofr}
Our goal is to numerically find $R_{\omega,m}$ for $r\geq 1$, which satisfies the {ingoing} boundary condition \eqref{ingoing} at the horizon.  We first introduce $H_{\omega,m}(r)$ as
\begin{equation}\label{H_R}
 H_{\omega,m}(r)=\sqrt{r}R_{\omega,m}\,.
 \end{equation}
From \eqref{Teukolsky}, the equation of motion for $H_{\omega,m}(r)$,  which is similar in form to  a Schr\"odinger equation \cite{PhysRevD.87.124038}, is derived as
\begin{equation}\label{Sch}
\left[\frac{d^2}{d r_*^2}+V_{eff}(r)\right]H_{\omega, m}(r)=0
\end{equation}
with
\begin{equation}\label{Veff}
V_{eff}(r)= \left(\omega - \frac{ml}{r^2}\right)^2-\left(\frac{r^2 -1}{r^2}\right)\left[\frac{5}{4r^4}+\frac{1}{r^2}\left(m^2-\frac{1}{4}\right)\right].
\end{equation}
Since $\lim_{r\to 1}V_{eff}(r)=(\omega-ml)^2=\tilde{\omega}^2$, and  $\lim_{r\to \infty}V_{eff}(r)=\omega^2$, the solutions {of $H_{\omega,m}$} at the ABH event horizon and in the asymptotic region satisfy the boundary conditions:
\begin{eqnarray}
& \lim_{r\to 1}H_{\omega, m}(r)\approx A_{\omega,m}e^{-i\tilde{\omega} r_*}, \label{HIf}\\
& \lim_{r\to \infty}H_{\omega, m}(r)\approx \tilde{A}^{in}_{\omega,m}e^{-i\omega r}+\tilde{B}^{out}_{\omega,m}e^{i\omega r}~(\because \lim_{r\to\infty}r_*=r). \label{HIf2}
\end{eqnarray}
{In Eq. \eqref{HIf2},} the incoming mode  with a constant amplitude $\tilde{A}^{in}_{\omega,m}$ represents the wave from the past null infinity $\mathcal{I}^-$, and the outgoing mode with a constant amplitude $\tilde{B}^{out}_{\omega,m}$ represents the outgoing wave to the future null infinity $\mathcal{I}^+$ in the Penrose diagram. As an aside, in the case of a fountaining bathtub AWH, the  boundary condition for outgoing modes at the event horizon is such that $-i$ in  \eqref{HIf} is replaced by $+i$. Since Eq. \eqref{Sch} readily provides the behavior of the solution near the horizon and in the asymptotic region, we choose to work with  Eq. \eqref{Sch} instead of the radial wave equation \eqref{Teukolsky} in the main text for the numerical integration.

Eq. \eqref{HIf} expresses an infinite blue shift at the horizon for ingoing modes at the horizon given by the fact that $\lim_{r\to 1} r_*=-\infty$. Consequently, we cannot prescribe a regular ingoing boundary condition directly at the horizon in the numerical integration of \eqref{Sch}. Instead, using Eq. \eqref{HIf}, we {determine} the values of $H_{\omega,m}(1+\epsilon)$ and $\frac{dH_{\omega,m}}{dr}|_{r=1+\epsilon}$ with $0<\epsilon\ll 1$ for the boundary condition of the numerical integration, where $H_{\omega,m}(r)$  satisfies the ingoing boundary condition at the ABH event horizon. Since Eq. \eqref{HIf} is an approximate near-horizon solution,   we introduce the following ansatz to improve the numerical accuracy as
\begin{equation}\label{ansatz}
H_{\omega,m}(r)=e^{-i \tilde{\omega} r_*}h(r)\, ,
\end{equation}
where $h(r)$ is analytical at the horizon ($r=1$). In the neighborhood of $r=1$,  $h(r)$ can be approximated  as a polynomial expansion around $r=1$,
\begin{equation}\label{Polynomial}
h(r)\approx\sum_{i=0}^{n}h_i (r-1)^i.
\end{equation}
The degree $n$ of the polynomial depends on the chosen accuracy. {If we choose $n$ to be $0$, then the expression \eqref{ansatz} reduces to \eqref{HIf} with $h_0=A_{\omega,m}$.}
Inserting the ansatz \eqref{ansatz} into \eqref{Sch} leads to the equation 
\begin{equation}\label{exph}
	A_2(r) h''(r)+A_1(r) h'(r) +A_0(r) h(r)=0,
\end{equation}
where the coefficients of the differential equation \eqref{exph} are functions of $r$, given by
\begin{eqnarray}
& A_2(r) =4r^2(r^2-1)^2,\\
& A_1(r)=8r(r^2-1)(1-i\tilde{\omega}r^3),\\
& A_0(r)=\left[5+\left\{-6+4(1+l^2)m^2\right\}r^2+\left\{1-4m(m+2l\omega)\right\}r^4+4r^6(\omega^2-\tilde{\omega}^2)\right].
\end{eqnarray}
Notice that in the case of a fountaining bathtub AWH,  $\tilde{\omega}$ is replaced by $-\tilde{\omega}$ in the expression of $A_1$ and $A_2$.
% $r=0$, and $r=1$ are regular singular points of the differential equation \eqref{exph}, so that at $r=0$ and at $r=1$, $A_{2}(r)=0$, and $A_1(r)=0$. Moreover, we observe that $A_0(r=1)=0$ as well.
%Given $h_1$ in \eqref{ansatz}, we can calculate $h(1+\Delta r)$, where $\Delta r$ is numerical step size. Given $h_{0}$, and $h_{1}$ in \eqref{exph}, we can also have $\frac{dh}{dr}|_{1+\Delta r}$ from Eq. \eqref{exph}.
%After substituting the polynomial $h(r)$ from \eqref{Polynomial} into \eqref{exph},
%the left hand side of the equation \eqref{exph} can be approximated as a nth degree polynomial expansion around $r=1$, yielding $\sum_{i=0}^{i=n}D_i(r-1)^i\approx 0$. Eq. \eqref{exph} yields that $D_i$s are linear combination of of $h_i$s with coefficients that are algebraic functions of the parameters, i.e., $\omega,~l$, and $m$. The polynomial $\sum_{i=0}^{i=n}D_i(r-1)^i\approx 0$ for any $r$ close to $r=1$. Therefore, we obtain a homogeneous system of $n+1$ linear equations in $n+1$ unknowns $h_i$,
%\begin{equation}
%D_i\approx 0,\qquad {\rm for}{~i=0~}{\rm to}~ n
%\end{equation}
\begin{figure}
\centering
\includegraphics[scale=0.5]{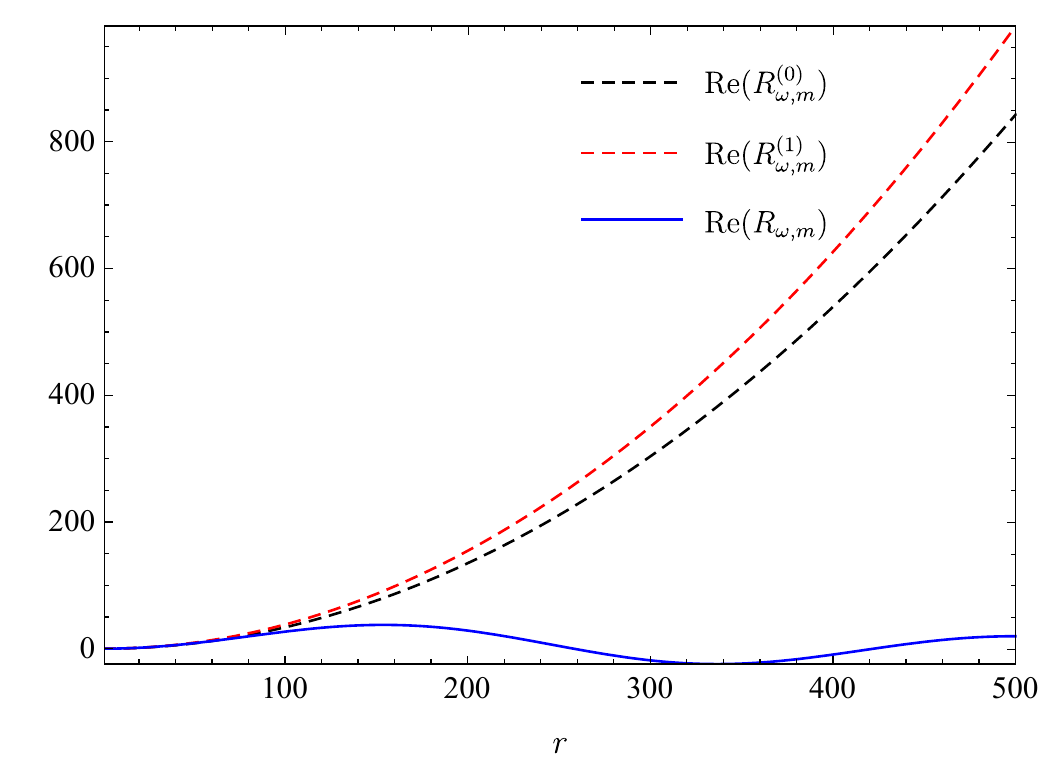}
\caption{{ Real parts of the analytical solutions $R^{(0)}_{\omega,m}(r)$ and $R^{(1)}_{\omega,m}(r)$ in Eq. \eqref{Ralphabeta2} and the numerical solution $R_{\omega,m}(r)$ in the region $r \geq 1$.} Initial value $h_0=0.01$  is assigned to all the curves with the same $m=2$, $l=1$ and $\omega$ is 0.02. Near the horizon, all three curves tend to converge, as also seen in Fig. \ref{fig3}. Far away from the event horizon, $R_{\omega,m}$ oscillates with frequency $\omega$ according to Eq. \eqref{HIf2}, and deviates from the analytical approximations.}\label{NI}
\end{figure}
   {After substituting the polynomial $h(r)$ \eqref{Polynomial} into \eqref{exph}, we derive the left-hand side of the equation \eqref{exph} as a nth degree polynomial expanded around $r=1$, for a chosen $n$. The right hand side of Eq. \eqref{exph} vanishes for any $r>1$. The fact that $A_2(r=1)=A_1(r=1)=A_0(r=1)=0$ renders the coefficient of the zeroth power of $(r-1)$ in the resulting polynomial zero.  {Therefore,} we equate each coefficient of $(r-1)^i$ in this polynomial to zero for $i=1,..n$.  These coefficients are linear combinations of the $h_i$ according to \eqref{exph} and \eqref{Polynomial}.
 Then, we effectively have $n$ homogeneous linear equations of $(n+1)$ coefficients $h_i$. The system can be solved by fixing $h_0$, and the remaining $h_i$ for $i=1$ to $n$ are determined accordingly.} We set $n=3$ for all the numerical integrations presented in this paper.
The differential equation \eqref{Sch} is {a} second order differential equation. By choosing the ingoing boundary condition at the event horizon \eqref{HIf},  one of the two independent initial conditions has been specified. We then set the other initial condition by fixing the value of $h_0$.
 After $h(r)$ is thus determined, we  {calculate} $H_{\omega,m}(1+\epsilon)$ and $\frac{dH_{\omega,m}}{dr}|_{r=1+\epsilon}$  using \eqref{ansatz}.  We then numerically integrate the differential equation \eqref{Sch} for $H_{\omega,m}$, starting from $r=1+\epsilon$ to some $r>1$, and  obtain $R_{\omega,m}$ using \eqref{H_R}. As for {the solutions of} the zeroth-order radial wave equation \eqref{Teukolsky0} and the first-order radial wave equation \eqref{Teukolsky2}, we  {use} the analytical expression  \eqref{Ralphabeta2} by setting $D_{\omega,m}=0$, $\omega=0$, and $D_{\omega,m}=0$, $\omega\neq0$, respectively.  In order to compare with the numerical $R_{\omega,m}$,  the numerical solution and the zeroth- and first-order analytical solutions {must be assigned} the same amplitude at $r=1$. To do so, one must first find the value of the constant $C_{\omega,m}$ from a given $h(r=1)=h_0$.
For $\omega=0$ in \eqref{Sch} and using \eqref{H_R}, the natural frequency of the zeroth-order solution, $R^{(0)}_{\omega,m}(r)$ at the event horizon is $\tilde{\omega}^{(0)}=-ml$. Then, dropping the $\omega ^2$ term in the effective potential  $V_{eff}$ in \eqref{Veff}, leads to the natural frequency at the event horizon for the  first-order solution $R^{(1)}_{\omega,m}(r)$ {as} $\tilde{\omega}^{(1)}=-ml\sqrt{1-\frac{2\omega}{ml}}=2 i \beta$. In  \eqref{Ralphabeta2} , the solution of $R^{(1)}_{\omega,m}$ of the ingoing modes at the horizon with $D_{\omega,m}=0$ is given by
\begin{equation}\label{blue}
R^{(1)}_{\omega,m}=\frac{H_{\omega,m}}{\sqrt{r}}=C_{\omega,m}e^{-i \tilde{\omega}^{(1)} r_*}r^{-2\alpha}e^{-2\beta\left(r+\ln \frac{r}{r+1}\right)} ~_2F_1(a,b;a+b+1-c,1-\frac{1}{r^2}).
\end{equation}
After factoring out the blue-shift term $e^{-i \tilde{\omega}^{(1)}r_*}$, and using (\ref{H_R}) and (\ref{ansatz}), the solution reduces to
\begin{equation}
\frac{h(r)}{\sqrt{r}}=C_{\omega,m}r^{-2\alpha}e^{-2\beta\left(r+\ln \frac{r}{r+1}\right)} ~_2F_1(a,b;a+b+1-c,1-\frac{1}{r^2})\, .
\end{equation}
In the first-order approximation, the coefficient $C_{\omega,m}$ can be obtained from $h(r=1)=h_0$ by
\begin{equation}\label{Ch}
h(r=1)=h_0=C_{\omega,m}e^{-2\beta}4^\beta.
\end{equation}
 {The relation \eqref{Ch} at $\omega=0$  also holds in the case of the zeroth-order approximation.}
In the effective potential $V_{eff}(r)$ \eqref{Veff}, dropping {the term } $\omega^2$ in the first-order approximation, and dropping the terms $\omega^2$ and $-\frac{2\omega m l}{r^2}$  in the zeroth-order approximation cause respective $R^{(1)}_{\omega,m}(r)$  and $R^{(0)}_{\omega,m}(r)$ to be nonoscillatory in the asymptotic region as can be seen in Fig. \ref{NI}.
\section{Properties of $\kappa^{(0)}_m$}\label{properties}
All the following properties are valid for the integer $m$ (as defined) with $|m| \geq 1$, and $l\in \mathbb{R}$ where $l$ is angular momentum of the fluid based on  Eq. \eqref{odehy}. {Additionally,} all the following statements  {hold} for $r>1$.
\newenvironment{ilemma}[1][]
{\refstepcounter{lemma}%
	\textbf{Lemma \thelemma\ifstrempty{#1}{}{ (#1)}.}\ \itshape}
{\upshape}

\begin{itemize}[itemsep=3pt, topsep=4pt, parsep=0pt]
\item{\begin{ilemma}\label{L2}
		$\kappa^{(0)}_m(l,r)=\kappa^{(0)}_{-m}(-l,r)
		=\kappa^{(0)}_{-m}(l,r)=\kappa^{(0)}_m(-l,r)$.
\end{ilemma}}
\begin{proof}
	The first equality  holds from the symmetry of the equations, namely $m\to -m$, and $l\to -l$ in Eqs. \eqref{Veff}, \eqref{alphabetaT},  \eqref{abcT}. The second and the third equality can be proven by showing  that the result of $\kappa^{(0)}_m(l,r)$ is an even function of $m$ and $l$.
	In Eq. \eqref{skapa}, $\cosh (\pi lm)$ and $|\delta|^2= \frac{m^2}{4} (l^2 +1)$ are  even functions of $l$ and $m$.
	 Using the identity \cite{handbook}, $|\Gamma(x+i y)|^2=|\Gamma(x)|^2\prod_{k=0}^{\infty}\frac{1}{1+\frac{y^2}{x^2+k^2}}$ for real $x$ and $y$, {we obtain} 
	\begin{equation}\label{identity} |\Gamma(-\delta)|^2=\left|\Gamma\left(\frac{|m|}{2}\right)\right|^2\prod_{k=0}^{\infty}\frac{1}{1+\frac{m^2l^2}{4\left(\frac{|m|}{2}+k\right)^2}}.
	\end{equation}
	Therefore, the prefactor $|\Gamma(-\delta)|^4$  is also an even function of $l$ and $m$. {Hence, the lemma is proved.}
\end{proof}

\item{\begin{ilemma}
	For $\ln r \neq 0$, $\kappa^{(0)}_m =0$  if and only if $m$ is an even number and $l=0$.
\end{ilemma}}
\begin{proof}
 {$\kappa^{(0)}_m(l=0,r) =0$ for even $m$, because the term $(-1)^{|m|+1}\cosh\pi (lm)+1$ in \eqref{skapa} is zero for even $m$ at $l=0$.} {Since $\delta\notin\mathbb{Z}_{\leq 0}$, $\forall~m (\geq 1)$ and $\forall l\in\mathbb{R}$, the Gamma function prefactor in \eqref{skapa} remains finite and well-defined.} Hence, the lemma is proved.
\end{proof}
\item{\begin{ilemma} $\kappa^{(0)}_m\geq 0$.\end{ilemma}}
\begin{proof}
In \eqref{skapa}, the term $(-1)^{|m|+1} \times \left(1+(-1)^{|m|+1}\cosh\pi lm\right)\geq 0$ $\forall |m|\geq 1$, and $\forall l\in\mathbb{R}$. The other multiplication factors are positive and nonzero. Hence, the lemma is proved.
\end{proof}

\item{\begin{ilemma}\label{lemmasmooth} $\kappa^{(0)}_m(l)$ is a smooth ($C^\infty$) function of $l$.
\end{ilemma}}
\begin{proof}
	The factor $|\delta|^2\times \left(1+(-1)^{|m|+1}\cosh\pi lm\right)$ in  Eq. \eqref{skapa} is a smooth function of $l$. 
{From the definition of $\Gamma(z)$ \cite{stein2010complex}, we have $\overline{\Gamma(z)}=\Gamma(\bar{z})$. Using the identity $|\Gamma(z)|^2=\overline{\Gamma(z)}\Gamma(z)$, we obtain}	
	\begin{equation}\label{pgamma}
		|\Gamma (-\delta)|^4=\left(\Gamma \left(\frac{|m|}{2}-\frac{iml}{2}\right)\right)^2\left(\Gamma \left(\frac{|m|}{2}+\frac{iml}{2}\right)\right)^2.
	\end{equation}
	$(1/\Gamma (z))$ is an entire function with zeros at $z=0,~-1,~-2,~..$ \cite{stein2010complex}. $\Gamma (z)$ has poles at $z\in\mathbb{Z}_{\leq 0}$. Since $|m|\geq 1$, $\Gamma \left(\frac{|m|}{2}+\frac{iml}{2}\right)$ and $\Gamma \left(\frac{|m|}{2}+\frac{iml}{2}\right)$ in the Eq. \eqref{pgamma} are infinitely differentiable w.r.t $l$ $\forall~l$. Hence, the lemma is proved.
\end{proof}
\item{\begin{ilemma}$\frac{\partial \kappa^{(0)} _m}{\partial l}=0$ at $l=0$. \end{ilemma}}
\begin{proof}
 $\kappa^{(0)} _m(l,r)$ is a smooth function of $l$ according to Lemma \ref{lemmasmooth}, and it is also an even function of $l$ according to Lemma \ref{L2}.  Therefore, $\frac{\partial \kappa^{(0)} _m}{\partial l}=0$  vanishes at $l=0$.
\end{proof}
\end{itemize}
\section{Hypergeometric ODE form of the radial wave equation linear in $\omega$}\label{hy1}
The radial wave equation to {first order} in $\omega$ in  Eq. \eqref{Teukolsky2} can be transformed into the form of the ODE for
the hypergeometric function in Eq. \eqref{odehy} with the undetermined $(a$, $b$, $c$) given by the values of $\alpha $ and $\beta$. After substituting, $z=\frac{1}{r^2}$ and {$R^{(1)}_{\omega, m}(r):=z^\alpha (1-z)^\beta F(z)$} in \eqref{Teukolsky2}, we factor out $z^\alpha (1-z)^\beta$, and obtain
\begin{equation}\label{hyp}
	a_{2}(z) F''(z)+ a_1(z) F'(z) +a_0(z) F(z)=0,
\end{equation}
where
\begin{eqnarray}
	& a_{2}(z)=4z^2(1-z)^2, \label{a2}\\
	& a_1(z)= 4z(1-z)\left[(2\alpha +1)-(1+2\alpha+2\beta +1)z \right], \label{a1}\\
	& a_0(z)= 4\alpha ^2 (1-z)^2+4\beta ^2 z^2-4(\alpha +\beta +2\alpha\beta)z(1-z)+l^2m^2z-(1-z) m^2-2\omega ml. \nonumber\\\label{a0}
\end{eqnarray}
In order to compare with Eq. \eqref{hyp},  Eq. \eqref{odehy} is multiplied by $4z(1-z)$,  {resulting in} 
\begin{equation}\label{ODEhy3}
	4z^2(1-z)^2 F''(z)+4z(1-z)(c-(1+a+b)z)F'(z)-4abz(1-z)F(z)=0.
\end{equation}
{Comparing the coefficients of $F'(z)$ in \eqref{hyp} and \eqref{ODEhy3}, we obtain
\begin{eqnarray}
	& c=2\alpha +1,\\
	& a+b=1+2\alpha+2\beta.\label{a+b}
\end{eqnarray}
{Comparing the $z^0$ term in the coefficients of $F(z)$ between Eqs. \eqref{hyp} and \eqref{ODEhy3}, we obtain} 
\begin{equation}
	4\alpha^2-m^2-2\omega m l=0\, ,
\end{equation}
giving $\alpha$, \begin{equation} \label{alpha}
	\alpha =\pm\frac{|m|}{2}\sqrt{1+\frac{2\omega l}{m}}.
\end{equation}
{Further, comparing the coefficients of $F(z)$ for the terms $z^2$ and $z$ between Eqs. \eqref{hyp} and \eqref{ODEhy3} leads to}
\begin{equation}\label{ab}
	ab=(\alpha+\beta+2\alpha\beta)+\alpha ^2 +\beta ^2=(\alpha+\beta+2\alpha\beta)+2\alpha ^2-\frac 14(l^2 +1)m^2\,.
\end{equation}
Then the formulas of $\alpha$ and $\beta$ is obtained as
\begin{equation}\label{beta}
	-\alpha ^2 +\beta ^2=-\frac{(l^2+1)m^2}{4},
\end{equation}
yielding $\beta$ from $\alpha$ in \eqref{alpha}, 
\begin{equation}
 \beta = \pm\frac{iml}{2}\sqrt{1-\frac{2\omega}{lm}}\, .
 \end{equation}
 From the equations of $a$ and $b$ in \eqref{ab} and  \eqref{a+b}, they can be expressed in terms of $\alpha$ and $\beta$ as
\begin{equation}\label{a,b}
	(a,b)=(1+\alpha +\beta,\alpha +\beta),~{\rm Or}~(a,b)=(\alpha +\beta,1+\alpha +\beta).
\end{equation}
Since the differential equation of the hypergeometric function is invariant under the interchange of $a$ and $b$, we can work with either choice in Eq. \eqref{a,b}.  We choose the first equality in Eq.\eqref{a,b}.

% For $\omega=0$, we get back the same values of $\alpha,~\beta,~a,~b,~c$ as in the Sec. \ref{stlne}.
%\begin{lemma}\label{Lemma1}
%{\it If the criteria \eqref{criteria} holds then $\beta$ is purely imaginary for $l\neq 0$, and $m\neq 0$.}
%\end{lemma}
%\begin{proof}
%First of all for $\omega =0$ as we already know $\beta$ is purely imaginary. For $\omega\neq 0$, the expression of $\beta$ in the Eq. \eqref{beta} implies either $\beta $ is purely a real number or purely an imaginary number for a given $l$ and $m$. For $\beta $ to be a real number, we must have
%\begin{equation}
%\frac{\omega}{lm}>0, \qquad {\rm and}\qquad \frac{|\omega|}{|m|} > \frac{|l|}{2}.
%\end{equation}
%Using the second criterion  in the Eq. \eqref{criteria} we find
%\begin{equation}
%r<<\frac{2}{|lm|},
%\end{equation}
%which contradicts the first criterion in the Eq. \eqref{criteria} even for minimum $|lm|$ ($m=1$).
%\end{proof}
{From  \eqref{alpha} and from \eqref{beta},} the possible $(\alpha,\beta)$ choices  for $\omega>0$ are
\begin{align}
	&\left(\frac{|m|}{2}\sqrt{1+\frac{2\omega l}{m}},\frac{iml}{2}\sqrt{1-\frac{2\omega}{lm}} \right),\qquad
	\left(\frac{|m|}{2}\sqrt{1+\frac{2\omega l}{m}},-\frac{iml}{2}\sqrt{1-\frac{2\omega}{lm}}\right),&\nonumber\\
	&\left(-\frac{|m|}{2}\sqrt{1+\frac{2\omega l}{m}},\frac{iml}{2}\sqrt{1-\frac{2\omega}{lm}}\right),\qquad
	\left(-\frac{|m|}{2}\sqrt{1+\frac{2\omega l}{m}},-\frac{iml}{2}\sqrt{1-\frac{2\omega}{lm}}\right).
\end{align}
However,  Eq. \eqref{Teukolsky2} is a second-order ordinary differential equation (ODE).  There can only be two linearly independent solutions for every choice of $\alpha$ and $\beta$ among these four choices. Examining the solution makes this explicit in the following example.

%For the case, $(c-a-b)\notin \mathbb{Z}$ (in our case $c-a-b=-2\beta\notin \mathbb{Z}$ for $l\neq 0$) , we can write down any general solution of the Eq. \eqref{ODEhy3} as linear combination of two linearly independent solutions \cite{Beals_Wong_2010}
%\begin{eqnarray}
%& F(z)=C_{\omega,m} ~_{2}F_{1}(a,b;a+b+1-c;1-z)+D_{\omega,m}(1-z)^{c-a-b}~_{2}F_{1}(c-b,c-a;1+c-a-b;1-z)\label{CDom}\\
%&=C_{\omega,m}~_{2}F_{1}(1+\alpha+\beta,\alpha+\beta;1+2\beta;1-z)+D_{\omega,m}(1-z)^{-2\beta}~_{2}F_{1}(1+\alpha-\beta,\alpha-\beta;1 -2\beta;1-z).
%\end{eqnarray}
%Therefore,
{The solution  in Eq. \eqref{Ralphabeta2} remains {the} same under the exchange of  $\beta\rightarrow -\beta $ and  $C_{\omega,m} \to D_{\omega,m}$. Thus, the form of the general solution remains unchanged under the transformation $\beta\rightarrow -\beta$.}
%Regarding the the term with coefficient $C_{\omega,m}$ in the Eq. \eqref{CDom}, we have the transformation rule for noninteger $c$ \cite{Beals_Wong_2010},
%\begin{multline}\label{te}
%_{2}F_{1}(a,b;a+b+1-c;1-z)=\frac{\Gamma(a+b+1-c)\Gamma (1-c)}{\Gamma(a+1-c)\Gamma(b+1-c)}~_{2}F_{1}(a,b;c;z)\\
%+z^{1-c}~\frac{\Gamma(a+b+1-c)\Gamma (c-1)}{\Gamma(a)\Gamma(b)}~_{2}F_{1}(a+1-c,b+1-c;2-c;z).
%\end{multline}
%Thus the first term with the coefficient $C_{\omega,m}$ in the Eq. \eqref{Ralphabeta} becomes
%\begin{multline}\label{Coma}
%\frac{\Gamma(1+2\beta)\Gamma (-2\alpha)}{\Gamma(1+\beta -\alpha)\Gamma(\beta-\alpha)}~_{2}F_{1}(1+\alpha+\beta,\alpha +\beta;1+2\alpha;z)(1-z)^\beta z^\alpha\\
%+\frac{\Gamma(1+2\beta)\Gamma (2\alpha)}{\Gamma(1+\beta +\alpha)\Gamma(\beta+\alpha)}~_{2}F_{1}(1-\alpha+\beta,-\alpha +\beta;1-2\alpha;z) (1-z)^\beta z^{-\alpha}.
%\end{multline}
The expression \eqref{Coma2} is invariant under $\alpha\rightarrow -\alpha$. {Thus,} we show an explicit example of $c-a-b\notin\mathbb{Z}$, $c\notin \mathbb{Z}$, for which all four choices {of} $(\alpha,\beta)$ generate the same general solution for $R^{(1)}_{\omega,m}(r)$. Here we choose
\begin{align}
	& (\alpha,\beta)=\left(\frac{|m|}{2}\sqrt{1+\frac{2\omega l}{m}},\frac{iml}{2}\sqrt{1-\frac{2\omega}{lm}}\right),\label{alphabeta}&\\
	&(a~,b~,c)=\nonumber\\
	&\quad\left(1+\frac{|m|}{2}\sqrt{1+\frac{2\omega l}{m}}+\frac{iml}{2}\sqrt{1-\frac{2\omega}{lm}},\frac{|m|}{2}\sqrt{1+\frac{2\omega l}{m}}+\frac{iml}{2}\sqrt{1-\frac{2\omega}{lm}}, 1+|m|\sqrt{1+\frac{2\omega l}{m}}\right)\label{abc}.&
\end{align}
\section{Calculation of zeroth- and first-order TLNs and TDCs from TRCs}\label{td}
As mentioned in \eqref{TLN1} and \eqref{DC1}, we introduce a complex number $\delta$,
\begin{equation}
	\delta = \beta -\alpha =a-c=-\frac{|m|}{2}\sqrt{1+\frac{2\omega l}{m}}+\frac{iml}{2}\sqrt{1-\frac{2\omega}{lm}},
\end{equation}
where
\begin{equation}
\operatorname{Re}(\delta) =-\alpha \, ,\quad \operatorname{Im}(\delta)=-i\beta.
\end{equation}
 Since $\alpha$ is purely real, and $\beta$ is purely imaginary,
 the expression of $k^{(1)}_{m}$ in Eq. \eqref{kom2} simplifies to 
\begin{equation}
	k^{(1)}_{m}=\frac{\Gamma ( -2\alpha)}{\Gamma (2\alpha)}\frac{\Gamma(-\bar{\delta})\Gamma(1-\bar{\delta})}{\Gamma(\delta)\Gamma(1+\delta)}.
\end{equation}
{$\bar{\delta}$ is the complex conjugate of $\delta$.} Now, it is possible to find out the real and imaginary parts of $k^{(1)}_{m}$ from the definition \eqref{TLNDF}.  Using the properties of the Gamma function, $\Gamma(\bar{z})=\overline{\Gamma(z)}$, and $\Gamma (\bar{z})\Gamma(z)=\overline{\Gamma(z)}\Gamma(z)\in \mathbb{R}$, we obtain
\begin{equation}
	\kappa^{(1)}_{m}(\omega,l)=-\frac{1}{2\pi^2}\times\frac{\Gamma (-2\alpha)}{\Gamma (2\alpha)}\times|\delta|^2|\Gamma(-\delta)|^4\times\left(1-\cos 2\pi\alpha\cosh\left(2\pi\operatorname{Im}(\beta)\right)\right),
\end{equation}
and
\begin{equation}
	\nu^{(1)}_{m}(\omega,l)=\frac{1}{2\pi^2}\times\frac{\Gamma (-2\alpha)}{\Gamma (2\alpha)}\times|\delta|^2|\Gamma(-\delta)|^4\times\sin2\pi\alpha\sinh\left(2\pi \operatorname{Im}(\beta)\right).
\end{equation}
{Similarly, for the $\omega =0$ in Sec. \ref{zero}, the real and imaginary parts of $k^{(0)}_{m}(l,r)$ in \eqref{sTLN2}) are obtained as }
\begin{equation} \label{kapam}
	\kappa^{(0)} _m(l,r)=\frac{(-1)^{|m|+1}}{\pi^2}\times\frac{1}{\Gamma(|m|)\Gamma(|m|+1)}\times |\delta |^2|\Gamma(-\delta)|^4 \times \left(1+(-1)^{|m|+1}\cosh\pi lm\right)\ln r,
\end{equation}
and
\begin{equation}
	\nu^{(0)}_m=0.
\end{equation}
\section{Analytical solution exhibiting the correct oscillation frequency at the event horizon}\label{2O}
\begin{figure}
\includegraphics[scale=0.5]{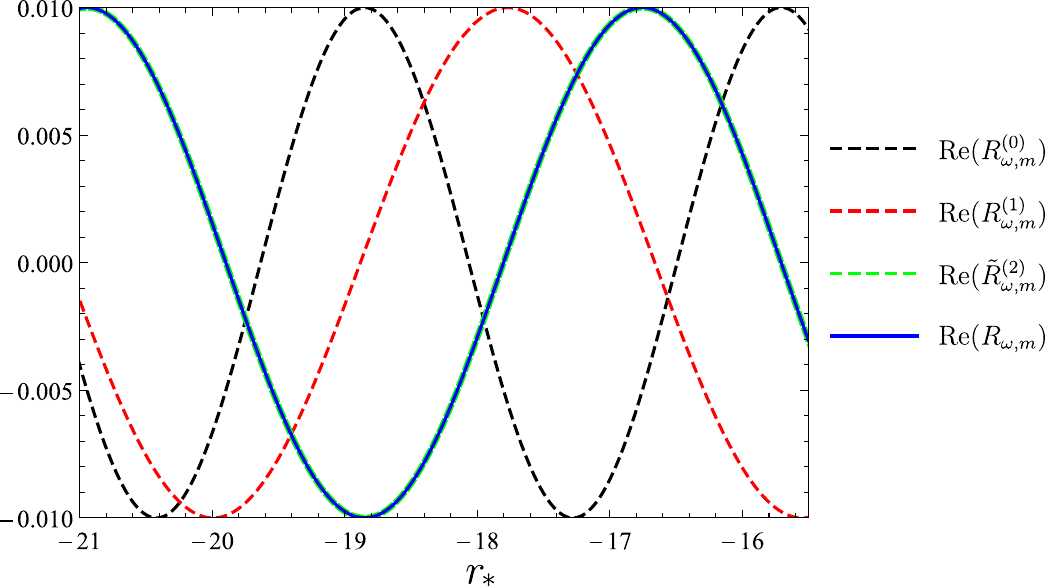} \\
\includegraphics[scale=0.55]{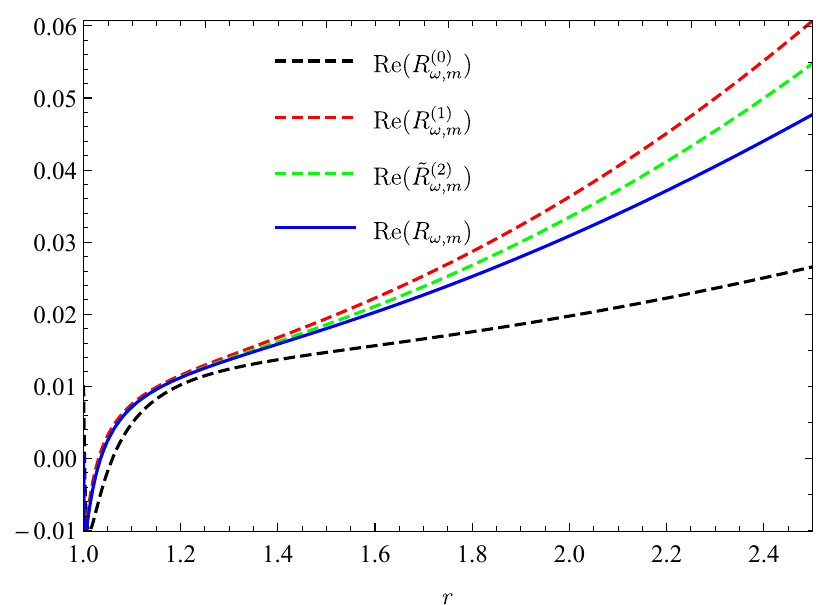}
\includegraphics[scale=0.55]{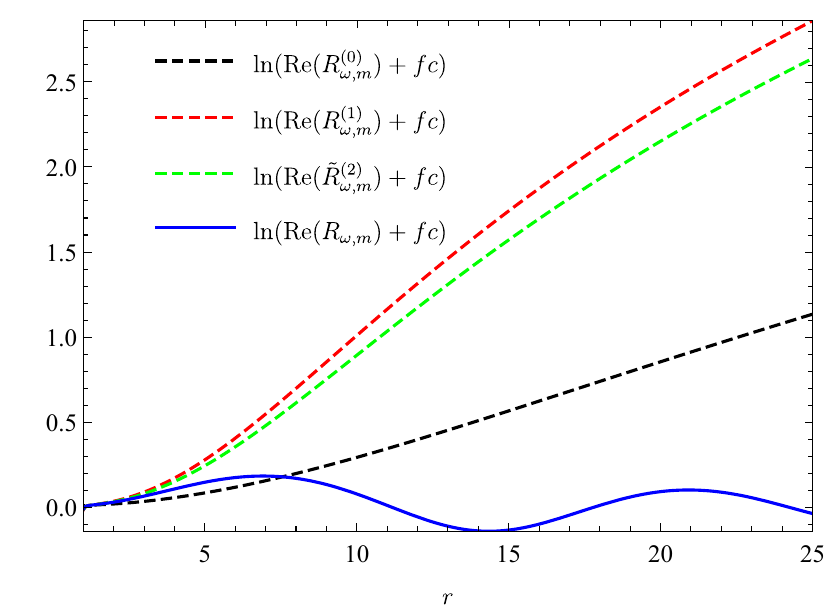}
\caption{ { Comparison of the analytical solutions with the full numerical solution.} The analytical solutions $R^{(0)}_{\omega,m}(r)$, $R^{(1)}_{\omega,m}(r)$ and $R^{(2)}_{\omega,m}(r)$ are obtained from Eq. \eqref{Ralphabeta2} using the $(\alpha,\beta)$ parameter values in Eq. \eqref{alphabetaT} (with $\omega=0$), Eq. \eqref{alphabetaT} (with $\omega\neq 0$) and Eq. \eqref{alphabetaT2}, respectively. $l=1$, $m=2$, $\omega=0.5$ and $h_0=0.01$ for the ingoing modes at the event horizon in all the plots. (Top) Near-horizon oscillations are displayed.  As expected from the analysis, oscillations of $\tilde{R}^{(2)}_{\omega,m}$ match exactly with near-horizon oscillations of $R_{\omega,m}$, viewed in a very narrow range of $r$ (the same as in Fig. \ref{fig3}), whereas the oscillations of $R^{(0)}_{\omega,m}$ and $R^{(1)}_{\omega,m}$ deviate significantly from the oscillations of the full radial solution $R_{\omega,m}$ because their oscillation frequencies with respect to the ingoing Eddington-Finkelstein coordinate $v$ at the horizon are $-m l$ and $-{ml}\sqrt{1-\frac{2\omega}{ml}}$, respectively (see Secs. \ref{zero},  \ref{1storderT} and \ref{hofr}). (Bottom-left) We illustrate how successive analytical approximations of $R_{\omega,m}$ improve the agreement with the full solution in the near zone. (Bottom-right) We see that far away from the near zone, all the analytical approximate solutions deviate from the full solution. The logarithmic scale is used, and $fc=1$ is chosen to fit all the curves nicely in the frame.}
\label{HOS}
\end{figure}
The first-order approximation presented in the main text (also see Eq. \eqref{blue})  corresponds to  {the}  near-horizon oscillation frequency $\tilde{\omega}^{(1)}=2 i \beta=-ml\sqrt{1-\frac{2\omega}{ml}}$ with respect to {the} Eddington-Finkelstein coordinates. {The equality} $\tilde{\omega}^{(1)}=\tilde{\omega}$ in \eqref{betal} {holds} only when $\omega\ll\frac{|lm|}{2}$,  {i.e.,} when the inequality \eqref{cbeta2} is satisfied.  However, following the discussion about the boundary condition at the horizon in the Sec. \ref{m0} and in Appendix \ref{hofr}, the correct oscillation frequency at the event horizon is $\tilde{\omega}$. This  motivates us to {improve upon the first-order approximation {by} equating $\beta$ to $-\frac{i \tilde{\omega}}{2}$}.  To incorporate such a modification, we rewrite the Teukolsky equation \eqref{Teukolsky} as
\begin{equation}\label{Teukolsky4}
\Delta (r)\frac{d}{dr}\left(\Delta (r) \frac{dR_{\omega, m}}{dr}\right)+\left[\left(\frac{l^2}{r^2}-\frac{\Delta (r)}{r}\right) m^2 + \omega ^2 -2\omega m l\right]R_{\omega, m}+(r^2-1)\omega^2=0.
\end{equation}
Now, we approximate  Eq. \eqref{Teukolsky4} by dropping the last term $(r^2-1)\omega^2$, where the near zone is redefined as $(r^2-1)\omega^2\ll 1$. We introduce $\tilde{R}^{(2)}_{\omega,m}$ in the modified {near-zone} limit as
\begin{equation}\label{Teukolsky5}
\Delta (r)\frac{d}{dr}\left(\Delta (r) \frac{d\tilde{R}^{(2)}_{\omega, m}}{dr}\right)+\left[\left(\frac{l^2}{r^2}-\frac{\Delta (r)}{r}\right) m^2 + \omega ^2 -2\omega m l\right]\tilde{R}^{(2)}_{\omega, m}=0.
\end{equation}
\begin{figure}
\centering
\includegraphics[scale=0.25]{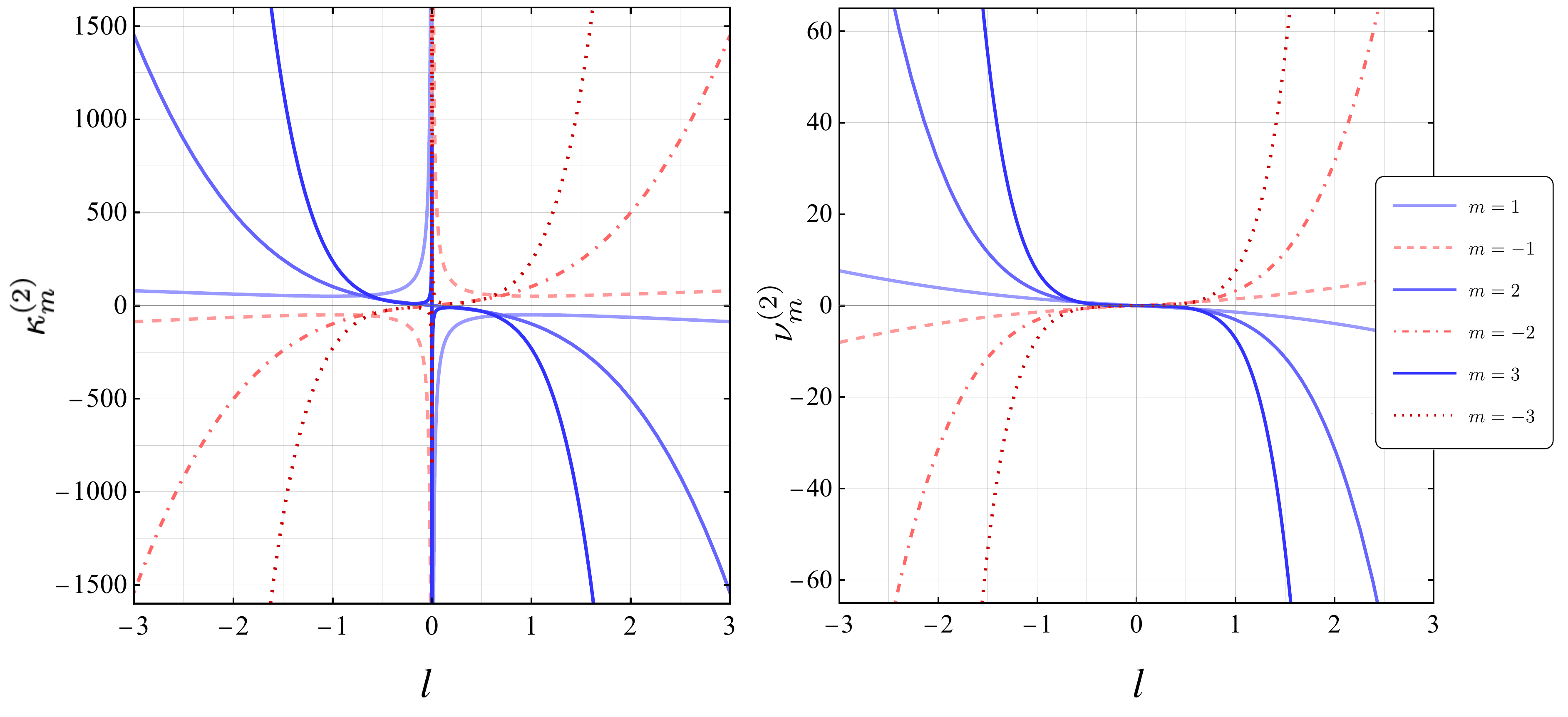}
\caption{{ The behaviour of dynamical TRC with $l$ is quite different from the static TRC in terms of the latter's aforementioned properties in Sec. \ref{zero}.} $\omega=0.01$ in the plots. $k^{(2)}_m$ and $k^{(1)}_m$ have the same expression in terms of $\alpha$ and $\beta$ (see Eqs. \eqref{abk}, \eqref{TLN1} and \eqref{DC1}). However, $\alpha$ and $\beta$ in $k^{(2)}_m$ [Eq. \eqref{alphabetaT2}] differ from those in $k^{(1)}_m$ [Eq. \eqref{alphabetaT}]. As $l\to \frac{\omega}{2m}$, a special case arises, because $2\alpha\to |m|$ via Eq. \eqref{alphabetaT2} (see the discussion around equations \eqref{Theformula3}-\eqref{nustar}). In this limit, (Left) $\kappa^{(2)}_{m}$ diverge [Eq. \eqref{TRCN}], and  (Right) $\nu^{(2)}_{m}$ take finite values [Eq. \eqref{TRCN}]. The natural symmetry $(l,m)\to (-l,-m)$ is reflected in all the curves.}
\label{fig5}
\end{figure}
One can check, by casting Eq. \eqref{Teukolsky5} into the Schr\"odinger form in \eqref{Sch} through \eqref{H_R} as illustrated in Appendix \ref{hofr}, the corresponding effective potential becomes, $V_{eff}(r)=\left(  \frac{\omega^2-2\omega ml}{r^2}+\frac{m^2 l^2}{r^4}\right)-\left(\frac{r^2 -1}{r^2}\right)\left[\frac{5}{4r^4}+\frac{1}{r^2}\left(m^2-\frac{1}{4}\right)\right].$ The value of the effective potential at $r=1$, $V_{eff}(r=1)=\tilde{\omega}^2$ provides the correct frequency of the modes at the event horizon. However, in the asymptotic region, $V_{eff}(r)$ as $r\to \infty$ is zero, providing a nonoscillatory solution ${R}^{(2)}_{\omega, m}$ in contrast with the oscillatory $R_{\omega,m}$, as seen in Fig. \ref{HOS}.  In the notation of $\tilde{R}^{(2)}_{\omega, m}$, the symbol `tilde' is used to indicate the {near-horizon} frequency correction, since we have denoted the event horizon frequency of the modes with `tilde'. Superscript `${(2)}$' indicates that $\tilde{R}^{(2)}_{\omega, m}$ is derived from the second-order correction in $\omega$ in the near-horizon region because the term $\omega^2(r^2-1)\ll 1$ in \eqref{Teukolsky4} for all values of $\omega$ in the vicinity of the event horizon.
This improves {the analysis significantly upon the first-order approximation} in Sec. \ref{1storderT}, since the latter requires $\omega \ll 1$ for $R^{(1)}_{\omega,m}\approx R_{\omega,m}$ in the near zone. Far away from the event horizon, the term  $\omega^2(r^2-1)$ becomes significant, and therefore, $\tilde{R}^{(2)}_{\omega, m}$ deviates from $R_{\omega,m}$, as shown in the bottom right panel of Fig. \ref{HOS}. In the near zone, the approximate solution $\tilde{R}^{(2)}_{\omega, m}$ matches better with the numerical $R_{\omega,m}$ than the zeroth- and {the first-order} approximations, shown in Fig. \ref{HOS}.
In the same manner as in the Appendix. \ref{hy1}, we can obtain the modified $\alpha$ and $\beta$ from \eqref{Teukolsky5},
\begin{equation}\label{alphabetaT2}
(\alpha,\beta)=\left(\frac{|m|}{2}\sqrt{1+\frac{2\omega l}{m}-\frac{\omega^2}{m^2}},-\frac{i}{2}(\omega-ml)\right).
\end{equation}
{ The procedure for finding TRCs is the same as in Sec. \ref{1storderT}, except for the updated values of $\alpha$ and $\beta$ in Eq. \eqref{alphabetaT2}.}
%With such a modification, $R_1=\sqrt{\frac{1}{\omega^2}+1}$, $R_2$ is no longer relevant.
Unlike {the} {first-order} approximate solution $R^{(1)}_{\omega,m}$ in Eq. \eqref{lVV} that does not capture the dynamical effect for $l=0$ , now the dynamical effect for $l=0$ appears through the second order term $\frac{\omega ^2}{m^2}$ in $\alpha$ of Eq. \eqref{alphabetaT2}. In order to achieve the tidal effect, we require $\alpha$ to be real. If we want the near zone to extend beyond the stationary limit surface, then, due to the modified near-zone limit, we must have $\omega \ll 1/|l|$. Thus, the set $S$ in  \eqref{lmS} that determines $\omega_{\rm max}$, is now modified providing bigger $\omega_{\rm max}$,
\begin{align}
 S=\left\{lm+|m|\sqrt{l^2+1}, \frac{1}{|l|}\right\}.&\label{lmS2}
\end{align}
The quantity $lm+|m|\sqrt{l^2+1}$ in \eqref{lmS2} provides $\omega <|m|$ for the tidal effect. $L$ and $L'$ in \eqref{lVV} are now modified providing bigger range in $|l|$,
\begin{align}
& L=\emptyset,\qquad L'=\left\{ \frac{1}{\omega}\right\}~{\rm for}~lm>0,~ {\rm and} ~L'=\left\{ \frac{1}{\omega},\frac{m^2-\omega^2}{2\omega|m|}\right\}~{\rm for}~lm<0. & \label{lV2}
\end{align}
Since $L$ is a null set, we can evaluate the TRC for the nonrotating ($l=0$) and slowly rotating ABH/AWH in Fig. \ref{fig5}, which is not possible in the {first-order} approximation in Eq. \eqref{lVV}.

%Teukolsky equation linearized in $\omega$ has the following limitations:

%(i) It does not capture $l\to 0$ limit, as described by \eqref{lVV}

%(ii) $\omega$ is restricted to small values, as described by \eqref{lmSV}.

%\end{comment}

%\bibliography{DTLN_DSLEE_v2}

\begin{thebibliography}{100}%
	\makeatletter
	\providecommand \@ifxundefined [1]{%
		\@ifx{#1\undefined}
	}%
	\providecommand \@ifnum [1]{%
		\ifnum #1\expandafter \@firstoftwo
		\else \expandafter \@secondoftwo
		\fi
	}%
	\providecommand \@ifx [1]{%
		\ifx #1\expandafter \@firstoftwo
		\else \expandafter \@secondoftwo
		\fi
	}%
	\providecommand \natexlab [1]{#1}%
	\providecommand \enquote  [1]{``#1''}%
	\providecommand \bibnamefont  [1]{#1}%
	\providecommand \bibfnamefont [1]{#1}%
	\providecommand \citenamefont [1]{#1}%
	\providecommand \href@noop [0]{\@secondoftwo}%
	\providecommand \href [0]{\begingroup \@sanitize@url \@href}%
	\providecommand \@href[1]{\@@startlink{#1}\@@href}%
	\providecommand \@@href[1]{\endgroup#1\@@endlink}%
	\providecommand \@sanitize@url [0]{\catcode `\\12\catcode `\$12\catcode
		`\&12\catcode `\#12\catcode `\^12\catcode `\_12\catcode `\%12\relax}%
	\providecommand \@@startlink[1]{}%
	\providecommand \@@endlink[0]{}%
	\providecommand \url  [0]{\begingroup\@sanitize@url \@url }%
	\providecommand \@url [1]{\endgroup\@href {#1}{\urlprefix }}%
	\providecommand \urlprefix  [0]{URL }%
	\providecommand \Eprint [0]{\href }%
	\providecommand \doibase [0]{https://doi.org/}%
	\providecommand \selectlanguage [0]{\@gobble}%
	\providecommand \bibinfo  [0]{\@secondoftwo}%
	\providecommand \bibfield  [0]{\@secondoftwo}%
	\providecommand \translation [1]{[#1]}%
	\providecommand \BibitemOpen [0]{}%
	\providecommand \bibitemStop [0]{}%
	\providecommand \bibitemNoStop [0]{.\EOS\space}%
	\providecommand \EOS [0]{\spacefactor3000\relax}%
	\providecommand \BibitemShut  [1]{\csname bibitem#1\endcsname}%
	\let\auto@bib@innerbib\@empty
	%</preamble>
	\bibitem [{\citenamefont {Love}(1909)}]{AEHLove}%
	\BibitemOpen
	\bibfield  {author} {\bibinfo {author} {\bibfnamefont {A.~E.~H.}\
			\bibnamefont {Love}},\ }\bibfield  {title} {\bibinfo {title} {The yielding of
			the earth to disturbing forces},\ }\href
	{https://doi.org/10.1093/mnras/69.6.476} {\bibfield  {journal} {\bibinfo
			{journal} {Monthly Notices of the Royal Astronomical Society}\ }\textbf
		{\bibinfo {volume} {69}},\ \bibinfo {pages} {476} (\bibinfo {year}
		{1909})}\BibitemShut {NoStop}%
	\bibitem [{\citenamefont {Hinderer}(2008)}]{Hinderer_2008}%
	\BibitemOpen
	\bibfield  {author} {\bibinfo {author} {\bibfnamefont {T.}~\bibnamefont
			{Hinderer}},\ }\bibfield  {title} {\bibinfo {title} {Tidal love numbers of
			neutron stars},\ }\href {https://doi.org/10.1086/533487} {\bibfield
		{journal} {\bibinfo  {journal} {The Astrophysical Journal}\ }\textbf
		{\bibinfo {volume} {677}},\ \bibinfo {pages} {1216} (\bibinfo {year}
		{2008})}\BibitemShut {NoStop}%
	\bibitem [{\citenamefont {Flanagan}\ and\ \citenamefont
		{Hinderer}(2008)}]{Flanagan}%
	\BibitemOpen
	\bibfield  {author} {\bibinfo {author} {\bibfnamefont {E.~E.}\ \bibnamefont
			{Flanagan}}\ and\ \bibinfo {author} {\bibfnamefont {T.}~\bibnamefont
			{Hinderer}},\ }\bibfield  {title} {\bibinfo {title} {Constraining
			neutron-star tidal love numbers with gravitational-wave detectors},\ }\href
	{https://doi.org/10.1103/PhysRevD.77.021502} {\bibfield  {journal} {\bibinfo
			{journal} {Phys. Rev. D}\ }\textbf {\bibinfo {volume} {77}},\ \bibinfo
		{pages} {021502(R)} (\bibinfo {year} {2008})}\BibitemShut {NoStop}%
	\bibitem [{\citenamefont {Damour}\ and\ \citenamefont {Nagar}(2009)}]{Damour}%
	\BibitemOpen
	\bibfield  {author} {\bibinfo {author} {\bibfnamefont {T.}~\bibnamefont
			{Damour}}\ and\ \bibinfo {author} {\bibfnamefont {A.}~\bibnamefont {Nagar}},\
	}\bibfield  {title} {\bibinfo {title} {Relativistic tidal properties of
			neutron stars},\ }\href {https://doi.org/10.1103/PhysRevD.80.084035}
	{\bibfield  {journal} {\bibinfo  {journal} {Phys. Rev. D}\ }\textbf {\bibinfo
			{volume} {80}},\ \bibinfo {pages} {084035} (\bibinfo {year}
		{2009})}\BibitemShut {NoStop}%
	\bibitem [{\citenamefont {Binnington}\ and\ \citenamefont
		{Poisson}(2009)}]{Binnington}%
	\BibitemOpen
	\bibfield  {author} {\bibinfo {author} {\bibfnamefont {T.}~\bibnamefont
			{Binnington}}\ and\ \bibinfo {author} {\bibfnamefont {E.}~\bibnamefont
			{Poisson}},\ }\bibfield  {title} {\bibinfo {title} {Relativistic theory of
			tidal love numbers},\ }\href {https://doi.org/10.1103/PhysRevD.80.084018}
	{\bibfield  {journal} {\bibinfo  {journal} {Phys. Rev. D}\ }\textbf {\bibinfo
			{volume} {80}},\ \bibinfo {pages} {084018} (\bibinfo {year}
		{2009})}\BibitemShut {NoStop}%
	\bibitem [{\citenamefont {Kol}\ and\ \citenamefont {Smolkin}(2012)}]{Kol2012}%
	\BibitemOpen
	\bibfield  {author} {\bibinfo {author} {\bibfnamefont {B.}~\bibnamefont
			{Kol}}\ and\ \bibinfo {author} {\bibfnamefont {M.}~\bibnamefont {Smolkin}},\
	}\bibfield  {title} {\bibinfo {title} {Black hole stereotyping: induced
			gravito-static polarization},\ }\href
	{https://doi.org/10.1007/JHEP02(2012)010} {\bibfield  {journal} {\bibinfo
			{journal} {Journal of High Energy Physics}\ }\textbf {\bibinfo {volume}
			{2012}},\ \bibinfo {pages} {10} (\bibinfo {year} {2012})}\BibitemShut
	{NoStop}%
	\bibitem [{\citenamefont {Hui}\ \emph {et~al.}(2021)\citenamefont {Hui},
		\citenamefont {Joyce}, \citenamefont {Penco}, \citenamefont {Santoni},\ and\
		\citenamefont {Solomon}}]{hui2021static}%
	\BibitemOpen
	\bibfield  {author} {\bibinfo {author} {\bibfnamefont {L.}~\bibnamefont
			{Hui}}, \bibinfo {author} {\bibfnamefont {A.}~\bibnamefont {Joyce}}, \bibinfo
		{author} {\bibfnamefont {R.}~\bibnamefont {Penco}}, \bibinfo {author}
		{\bibfnamefont {L.}~\bibnamefont {Santoni}},\ and\ \bibinfo {author}
		{\bibfnamefont {A.~R.}\ \bibnamefont {Solomon}},\ }\bibfield  {title}
	{\bibinfo {title} {Static response and love numbers of schwarzschild black
			holes},\ }\href {https://doi.org/10.1088/1475-7516/2021/04/052} {\bibfield
		{journal} {\bibinfo  {journal} {Journal of Cosmology and Astroparticle
				Physics}\ }\textbf {\bibinfo {volume} {2021}}\bibinfo  {number} { (04)},\
		\bibinfo {pages} {052}}\BibitemShut {NoStop}%
	\bibitem [{\citenamefont {Chia}(2021)}]{Chia}%
	\BibitemOpen
	\bibfield  {number} {  }\bibfield  {author} {\bibinfo {author} {\bibfnamefont
			{H.~S.}\ \bibnamefont {Chia}},\ }\bibfield  {title} {\bibinfo {title} {Tidal
			deformation and dissipation of rotating black holes},\ }\href
	{https://doi.org/10.1103/PhysRevD.104.024013} {\bibfield  {journal} {\bibinfo
			{journal} {Phys. Rev. D}\ }\textbf {\bibinfo {volume} {104}},\ \bibinfo
		{pages} {024013} (\bibinfo {year} {2021})}\BibitemShut {NoStop}%
	\bibitem [{\citenamefont {Goldberger}\ \emph {et~al.}(2021)\citenamefont
		{Goldberger}, \citenamefont {Li},\ and\ \citenamefont
		{Rothstein}}]{Goldberger2021}%
	\BibitemOpen
	\bibfield  {author} {\bibinfo {author} {\bibfnamefont {W.~D.}\ \bibnamefont
			{Goldberger}}, \bibinfo {author} {\bibfnamefont {J.}~\bibnamefont {Li}},\
		and\ \bibinfo {author} {\bibfnamefont {I.~Z.}\ \bibnamefont {Rothstein}},\
	}\bibfield  {title} {\bibinfo {title} {Non-conservative effects on spinning
			black holes from world-line effective field theory},\ }\href
	{https://doi.org/10.1007/JHEP06(2021)053} {\bibfield  {journal} {\bibinfo
			{journal} {Journal of High Energy Physics}\ }\textbf {\bibinfo {volume}
			{2021}},\ \bibinfo {pages} {53} (\bibinfo {year} {2021})}\BibitemShut
	{NoStop}%
	\bibitem [{\citenamefont {Charalambous}\ \emph
		{et~al.}(2021{\natexlab{a}})\citenamefont {Charalambous}, \citenamefont
		{Dubovsky},\ and\ \citenamefont {Ivanov}}]{Charalambous2021}%
	\BibitemOpen
	\bibfield  {author} {\bibinfo {author} {\bibfnamefont {P.}~\bibnamefont
			{Charalambous}}, \bibinfo {author} {\bibfnamefont {S.}~\bibnamefont
			{Dubovsky}},\ and\ \bibinfo {author} {\bibfnamefont {M.~M.}\ \bibnamefont
			{Ivanov}},\ }\bibfield  {title} {\bibinfo {title} {On the vanishing of love
			numbers for kerr black holes},\ }\href
	{https://doi.org/10.1007/JHEP05(2021)038} {\bibfield  {journal} {\bibinfo
			{journal} {Journal of High Energy Physics}\ }\textbf {\bibinfo {volume}
			{2021}},\ \bibinfo {pages} {38} (\bibinfo {year}
		{2021}{\natexlab{a}})}\BibitemShut {NoStop}%
	\bibitem [{\citenamefont {Le~Tiec}\ \emph {et~al.}(2021)\citenamefont
		{Le~Tiec}, \citenamefont {Casals},\ and\ \citenamefont {Franzin}}]{TiecKerr}%
	\BibitemOpen
	\bibfield  {author} {\bibinfo {author} {\bibfnamefont {A.}~\bibnamefont
			{Le~Tiec}}, \bibinfo {author} {\bibfnamefont {M.}~\bibnamefont {Casals}},\
		and\ \bibinfo {author} {\bibfnamefont {E.}~\bibnamefont {Franzin}},\
	}\bibfield  {title} {\bibinfo {title} {Tidal love numbers of kerr black
			holes},\ }\href {https://doi.org/10.1103/PhysRevD.103.084021} {\bibfield
		{journal} {\bibinfo  {journal} {Phys. Rev. D}\ }\textbf {\bibinfo {volume}
			{103}},\ \bibinfo {pages} {084021} (\bibinfo {year} {2021})}\BibitemShut
	{NoStop}%
	\bibitem [{\citenamefont {Abbott}\ \emph {et~al.}(2017)\citenamefont {Abbott}
		\emph {et~al.}}]{GW170817}%
	\BibitemOpen
	\bibfield  {author} {\bibinfo {author} {\bibfnamefont {B.~P.}\ \bibnamefont
			{Abbott}} \emph {et~al.} (\bibinfo {collaboration} {LIGO Scientific
			Collaboration and Virgo Collaboration}),\ }\bibfield  {title} {\bibinfo
		{title} {Gw170817: Observation of gravitational waves from a binary neutron
			star inspiral},\ }\href {https://doi.org/10.1103/PhysRevLett.119.161101}
	{\bibfield  {journal} {\bibinfo  {journal} {Phys. Rev. Lett.}\ }\textbf
		{\bibinfo {volume} {119}},\ \bibinfo {pages} {161101} (\bibinfo {year}
		{2017})}\BibitemShut {NoStop}%
	\bibitem [{\citenamefont {Guerra~Chaves}\ and\ \citenamefont
		{Hinderer}(2019)}]{Chaves_2019}%
	\BibitemOpen
	\bibfield  {author} {\bibinfo {author} {\bibfnamefont {A.}~\bibnamefont
			{Guerra~Chaves}}\ and\ \bibinfo {author} {\bibfnamefont {T.}~\bibnamefont
			{Hinderer}},\ }\bibfield  {title} {\bibinfo {title} {Probing the equation of
			state of neutron star matter with gravitational waves from binary inspirals
			in light of gw170817: a brief review},\ }\href
	{https://doi.org/10.1088/1361-6471/ab45be} {\bibfield  {journal} {\bibinfo
			{journal} {Journal of Physics G: Nuclear and Particle Physics}\ }\textbf
		{\bibinfo {volume} {46}},\ \bibinfo {pages} {123002} (\bibinfo {year}
		{2019})}\BibitemShut {NoStop}%
	\bibitem [{\citenamefont {Yunes}\ \emph {et~al.}(2022)\citenamefont {Yunes},
		\citenamefont {Miller},\ and\ \citenamefont {Yagi}}]{Yunes2022}%
	\BibitemOpen
	\bibfield  {author} {\bibinfo {author} {\bibfnamefont {N.}~\bibnamefont
			{Yunes}}, \bibinfo {author} {\bibfnamefont {M.~C.}\ \bibnamefont {Miller}},\
		and\ \bibinfo {author} {\bibfnamefont {K.}~\bibnamefont {Yagi}},\ }\bibfield
	{title} {\bibinfo {title} {Gravitational-wave and x-ray probes of the neutron
			star equation of state},\ }\href {https://doi.org/10.1038/s42254-022-00420-y}
	{\bibfield  {journal} {\bibinfo  {journal} {Nature Reviews Physics}\ }\textbf
		{\bibinfo {volume} {4}},\ \bibinfo {pages} {237} (\bibinfo {year}
		{2022})}\BibitemShut {NoStop}%
	\bibitem [{\citenamefont {Fang}\ and\ \citenamefont
		{Lovelace}(2005)}]{FangLovalace}%
	\BibitemOpen
	\bibfield  {author} {\bibinfo {author} {\bibfnamefont {H.}~\bibnamefont
			{Fang}}\ and\ \bibinfo {author} {\bibfnamefont {G.}~\bibnamefont
			{Lovelace}},\ }\bibfield  {title} {\bibinfo {title} {Tidal coupling of a
			schwarzschild black hole and circularly orbiting moon},\ }\href
	{https://doi.org/10.1103/PhysRevD.72.124016} {\bibfield  {journal} {\bibinfo
			{journal} {Phys. Rev. D}\ }\textbf {\bibinfo {volume} {72}},\ \bibinfo
		{pages} {124016} (\bibinfo {year} {2005})}\BibitemShut {NoStop}%
	\bibitem [{\citenamefont {Charalambous}\ \emph
		{et~al.}(2021{\natexlab{b}})\citenamefont {Charalambous}, \citenamefont
		{Dubovsky},\ and\ \citenamefont {Ivanov}}]{hiddensymmetryl0}%
	\BibitemOpen
	\bibfield  {author} {\bibinfo {author} {\bibfnamefont {P.}~\bibnamefont
			{Charalambous}}, \bibinfo {author} {\bibfnamefont {S.}~\bibnamefont
			{Dubovsky}},\ and\ \bibinfo {author} {\bibfnamefont {M.~M.}\ \bibnamefont
			{Ivanov}},\ }\bibfield  {title} {\bibinfo {title} {Hidden symmetry of
			vanishing love numbers},\ }\href
	{https://doi.org/10.1103/PhysRevLett.127.101101} {\bibfield  {journal}
		{\bibinfo  {journal} {Phys. Rev. Lett.}\ }\textbf {\bibinfo {volume} {127}},\
		\bibinfo {pages} {101101} (\bibinfo {year} {2021}{\natexlab{b}})}\BibitemShut
	{NoStop}%
	\bibitem [{\citenamefont {Achour}\ \emph {et~al.}(2022)\citenamefont {Achour},
		\citenamefont {Livine}, \citenamefont {Mukohyama},\ and\ \citenamefont
		{Uzan}}]{Achour2022}%
	\BibitemOpen
	\bibfield  {author} {\bibinfo {author} {\bibfnamefont {J.~B.}\ \bibnamefont
			{Achour}}, \bibinfo {author} {\bibfnamefont {E.~R.}\ \bibnamefont {Livine}},
		\bibinfo {author} {\bibfnamefont {S.}~\bibnamefont {Mukohyama}},\ and\
		\bibinfo {author} {\bibfnamefont {J.-P.}\ \bibnamefont {Uzan}},\ }\bibfield
	{title} {\bibinfo {title} {Hidden symmetry of the static response of black
			holes: applications to love numbers},\ }\href
	{https://doi.org/10.1007/JHEP07(2022)112} {\bibfield  {journal} {\bibinfo
			{journal} {Journal of High Energy Physics}\ }\textbf {\bibinfo {volume}
			{2022}},\ \bibinfo {pages} {112} (\bibinfo {year} {2022})}\BibitemShut
	{NoStop}%
	\bibitem [{\citenamefont {Pere\~niguez}\ and\ \citenamefont
		{Cardoso}(2022)}]{RSVitor}%
	\BibitemOpen
	\bibfield  {author} {\bibinfo {author} {\bibfnamefont {D.}~\bibnamefont
			{Pere\~niguez}}\ and\ \bibinfo {author} {\bibfnamefont {V.}~\bibnamefont
			{Cardoso}},\ }\bibfield  {title} {\bibinfo {title} {Love numbers and magnetic
			susceptibility of charged black holes},\ }\href
	{https://doi.org/10.1103/PhysRevD.105.044026} {\bibfield  {journal} {\bibinfo
			{journal} {Phys. Rev. D}\ }\textbf {\bibinfo {volume} {105}},\ \bibinfo
		{pages} {044026} (\bibinfo {year} {2022})}\BibitemShut {NoStop}%
	\bibitem [{\citenamefont {Charalambous}\ \emph {et~al.}(2022)\citenamefont
		{Charalambous}, \citenamefont {Dubovsky},\ and\ \citenamefont
		{Ivanov}}]{Lovesymmetry}%
	\BibitemOpen
	\bibfield  {author} {\bibinfo {author} {\bibfnamefont {P.}~\bibnamefont
			{Charalambous}}, \bibinfo {author} {\bibfnamefont {S.}~\bibnamefont
			{Dubovsky}},\ and\ \bibinfo {author} {\bibfnamefont {M.~M.}\ \bibnamefont
			{Ivanov}},\ }\bibfield  {title} {\bibinfo {title} {Love symmetry},\ }\href
	{https://doi.org/10.1007/JHEP10(2022)175} {\bibfield  {journal} {\bibinfo
			{journal} {Journal of High Energy Physics}\ }\textbf {\bibinfo {volume}
			{2022}},\ \bibinfo {pages} {175} (\bibinfo {year} {2022})}\BibitemShut
	{NoStop}%
	\bibitem [{\citenamefont {Katagiri}\ \emph {et~al.}(2023)\citenamefont
		{Katagiri}, \citenamefont {Kimura}, \citenamefont {Nakano},\ and\
		\citenamefont {Omukai}}]{valinishingtlnKS}%
	\BibitemOpen
	\bibfield  {author} {\bibinfo {author} {\bibfnamefont {T.}~\bibnamefont
			{Katagiri}}, \bibinfo {author} {\bibfnamefont {M.}~\bibnamefont {Kimura}},
		\bibinfo {author} {\bibfnamefont {H.}~\bibnamefont {Nakano}},\ and\ \bibinfo
		{author} {\bibfnamefont {K.}~\bibnamefont {Omukai}},\ }\bibfield  {title}
	{\bibinfo {title} {Vanishing love numbers of black holes in general
			relativity: From spacetime conformal symmetry of a two-dimensional reduced
			geometry},\ }\href {https://doi.org/10.1103/PhysRevD.107.124030} {\bibfield
		{journal} {\bibinfo  {journal} {Phys. Rev. D}\ }\textbf {\bibinfo {volume}
			{107}},\ \bibinfo {pages} {124030} (\bibinfo {year} {2023})}\BibitemShut
	{NoStop}%
	\bibitem [{\citenamefont {Ivanov}\ and\ \citenamefont
		{Zhou}(2023)}]{PhysRevLett.130.091403}%
	\BibitemOpen
	\bibfield  {author} {\bibinfo {author} {\bibfnamefont {M.~M.}\ \bibnamefont
			{Ivanov}}\ and\ \bibinfo {author} {\bibfnamefont {Z.}~\bibnamefont {Zhou}},\
	}\bibfield  {title} {\bibinfo {title} {Vanishing of black hole tidal love
			numbers from scattering amplitudes},\ }\href
	{https://doi.org/10.1103/PhysRevLett.130.091403} {\bibfield  {journal}
		{\bibinfo  {journal} {Phys. Rev. Lett.}\ }\textbf {\bibinfo {volume} {130}},\
		\bibinfo {pages} {091403} (\bibinfo {year} {2023})}\BibitemShut {NoStop}%
	\bibitem [{\citenamefont {Rai}\ and\ \citenamefont {Santoni}(2024)}]{Rai2024}%
	\BibitemOpen
	\bibfield  {author} {\bibinfo {author} {\bibfnamefont {M.}~\bibnamefont
			{Rai}}\ and\ \bibinfo {author} {\bibfnamefont {L.}~\bibnamefont {Santoni}},\
	}\bibfield  {title} {\bibinfo {title} {Ladder symmetries and love numbers of
			reissner-nordstr{\"o}m black holes},\ }\href
	{https://doi.org/10.1007/JHEP07(2024)098} {\bibfield  {journal} {\bibinfo
			{journal} {Journal of High Energy Physics}\ }\textbf {\bibinfo {volume}
			{2024}},\ \bibinfo {pages} {98} (\bibinfo {year} {2024})}\BibitemShut
	{NoStop}%
	\bibitem [{\citenamefont {Sharma}\ \emph {et~al.}(2024)\citenamefont {Sharma},
		\citenamefont {Ghosh},\ and\ \citenamefont {Sarkar}}]{Sharma}%
	\BibitemOpen
	\bibfield  {author} {\bibinfo {author} {\bibfnamefont {C.}~\bibnamefont
			{Sharma}}, \bibinfo {author} {\bibfnamefont {R.}~\bibnamefont {Ghosh}},\ and\
		\bibinfo {author} {\bibfnamefont {S.}~\bibnamefont {Sarkar}},\ }\bibfield
	{title} {\bibinfo {title} {Exploring ladder symmetry and love numbers for
			static and rotating black holes},\ }\href
	{https://doi.org/10.1103/PhysRevD.109.L041505} {\bibfield  {journal}
		{\bibinfo  {journal} {Phys. Rev. D}\ }\textbf {\bibinfo {volume} {109}},\
		\bibinfo {pages} {L041505} (\bibinfo {year} {2024})}\BibitemShut {NoStop}%
	\bibitem [{\citenamefont {Bonelli}\ \emph {et~al.}(2022)\citenamefont
		{Bonelli}, \citenamefont {Iossa}, \citenamefont {Lichtig},\ and\
		\citenamefont {Tanzini}}]{Bonelli}%
	\BibitemOpen
	\bibfield  {author} {\bibinfo {author} {\bibfnamefont {G.}~\bibnamefont
			{Bonelli}}, \bibinfo {author} {\bibfnamefont {C.}~\bibnamefont {Iossa}},
		\bibinfo {author} {\bibfnamefont {D.~P.}\ \bibnamefont {Lichtig}},\ and\
		\bibinfo {author} {\bibfnamefont {A.}~\bibnamefont {Tanzini}},\ }\bibfield
	{title} {\bibinfo {title} {Exact solution of kerr black hole perturbations
			via ${\mathrm{cft}}_{2}$ and instanton counting: Greybody factor, quasinormal
			modes, and love numbers},\ }\href
	{https://doi.org/10.1103/PhysRevD.105.044047} {\bibfield  {journal} {\bibinfo
			{journal} {Phys. Rev. D}\ }\textbf {\bibinfo {volume} {105}},\ \bibinfo
		{pages} {044047} (\bibinfo {year} {2022})}\BibitemShut {NoStop}%
	\bibitem [{\citenamefont {Saketh}\ \emph {et~al.}(2024)\citenamefont {Saketh},
		\citenamefont {Zhou},\ and\ \citenamefont {Ivanov}}]{Scattering2}%
	\BibitemOpen
	\bibfield  {author} {\bibinfo {author} {\bibfnamefont {M.~V.~S.}\
			\bibnamefont {Saketh}}, \bibinfo {author} {\bibfnamefont {Z.}~\bibnamefont
			{Zhou}},\ and\ \bibinfo {author} {\bibfnamefont {M.~M.}\ \bibnamefont
			{Ivanov}},\ }\bibfield  {title} {\bibinfo {title} {Dynamical tidal response
			of kerr black holes from scattering amplitudes},\ }\href
	{https://doi.org/10.1103/PhysRevD.109.064058} {\bibfield  {journal} {\bibinfo
			{journal} {Phys. Rev. D}\ }\textbf {\bibinfo {volume} {109}},\ \bibinfo
		{pages} {064058} (\bibinfo {year} {2024})}\BibitemShut {NoStop}%
	\bibitem [{\citenamefont {Chakraborty}\ \emph {et~al.}(2024)\citenamefont
		{Chakraborty}, \citenamefont {Maggio}, \citenamefont {Silvestrini},\ and\
		\citenamefont {Pani}}]{Sumanta}%
	\BibitemOpen
	\bibfield  {author} {\bibinfo {author} {\bibfnamefont {S.}~\bibnamefont
			{Chakraborty}}, \bibinfo {author} {\bibfnamefont {E.}~\bibnamefont {Maggio}},
		\bibinfo {author} {\bibfnamefont {M.}~\bibnamefont {Silvestrini}},\ and\
		\bibinfo {author} {\bibfnamefont {P.}~\bibnamefont {Pani}},\ }\bibfield
	{title} {\bibinfo {title} {Dynamical tidal love numbers of kerr-like compact
			objects},\ }\href {https://doi.org/10.1103/PhysRevD.110.084042} {\bibfield
		{journal} {\bibinfo  {journal} {Phys. Rev. D}\ }\textbf {\bibinfo {volume}
			{110}},\ \bibinfo {pages} {084042} (\bibinfo {year} {2024})}\BibitemShut
	{NoStop}%
	\bibitem [{\citenamefont {Perry}\ and\ \citenamefont
		{Rodriguez}(2023)}]{perry}%
	\BibitemOpen
	\bibfield  {author} {\bibinfo {author} {\bibfnamefont {M.}~\bibnamefont
			{Perry}}\ and\ \bibinfo {author} {\bibfnamefont {M.~J.}\ \bibnamefont
			{Rodriguez}},\ }\href {https://arxiv.org/abs/2310.03660} {\bibinfo {title}
		{Dynamical love numbers for kerr black holes}} (\bibinfo {year} {2023}),\
	\Eprint {https://arxiv.org/abs/2310.03660} {arXiv:2310.03660 [gr-qc]}
	\BibitemShut {NoStop}%
	\bibitem [{\citenamefont {Chakraborty}\ \emph {et~al.}(2025)\citenamefont
		{Chakraborty}, \citenamefont {Luca}, \citenamefont {Gualtieri},\ and\
		\citenamefont {Pani}}]{chakraborty2025dynamicallovenumbersblack}%
	\BibitemOpen
	\bibfield  {author} {\bibinfo {author} {\bibfnamefont {S.}~\bibnamefont
			{Chakraborty}}, \bibinfo {author} {\bibfnamefont {V.~D.}\ \bibnamefont
			{Luca}}, \bibinfo {author} {\bibfnamefont {L.}~\bibnamefont {Gualtieri}},\
		and\ \bibinfo {author} {\bibfnamefont {P.}~\bibnamefont {Pani}},\ }\href
	{https://arxiv.org/abs/2507.22994} {\bibinfo {title} {Dynamical love numbers
			of black holes: theory and gravitational waveforms}} (\bibinfo {year}
	{2025}),\ \Eprint {https://arxiv.org/abs/2507.22994} {arXiv:2507.22994
		[gr-qc]} \BibitemShut {NoStop}%
	\bibitem [{\citenamefont {Combaluzier-Szteinsznaider}\ \emph
		{et~al.}(2026)\citenamefont {Combaluzier-Szteinsznaider}, \citenamefont
		{Glazer}, \citenamefont {Joyce}, \citenamefont {Rodriguez},\ and\
		\citenamefont {Santoni}}]{Maria}%
	\BibitemOpen
	\bibfield  {author} {\bibinfo {author} {\bibfnamefont {O.}~\bibnamefont
			{Combaluzier-Szteinsznaider}}, \bibinfo {author} {\bibfnamefont
			{D.}~\bibnamefont {Glazer}}, \bibinfo {author} {\bibfnamefont
			{A.}~\bibnamefont {Joyce}}, \bibinfo {author} {\bibfnamefont {M.~J.}\
			\bibnamefont {Rodriguez}},\ and\ \bibinfo {author} {\bibfnamefont
			{L.}~\bibnamefont {Santoni}},\ }\href {https://arxiv.org/abs/2511.02372}
	{\bibinfo {title} {Dynamical tidal response of schwarzschild black holes}}
	(\bibinfo {year} {2026}),\ \Eprint {https://arxiv.org/abs/2511.02372}
	{arXiv:2511.02372 [gr-qc]} \BibitemShut {NoStop}%
	\bibitem [{\citenamefont {Chakraborty}\ \emph
		{et~al.}(2026{\natexlab{a}})\citenamefont {Chakraborty}, \citenamefont
		{Saketh}, \citenamefont {Hinderer},\ and\ \citenamefont
		{Steinhoff}}]{chakraborty2026dynamical}%
	\BibitemOpen
	\bibfield  {author} {\bibinfo {author} {\bibfnamefont {S.}~\bibnamefont
			{Chakraborty}}, \bibinfo {author} {\bibfnamefont {M.~V.~S.}\ \bibnamefont
			{Saketh}}, \bibinfo {author} {\bibfnamefont {T.}~\bibnamefont {Hinderer}},\
		and\ \bibinfo {author} {\bibfnamefont {J.}~\bibnamefont {Steinhoff}},\ }\href
	{https://arxiv.org/abs/2605.00693} {\bibinfo {title} {Dynamical tidal love
			numbers of black holes under generic perturbations: Connecting black hole
			perturbation theory with effective field theory}} (\bibinfo {year}
	{2026}{\natexlab{a}}),\ \Eprint {https://arxiv.org/abs/2605.00693}
	{arXiv:2605.00693 [gr-qc]} \BibitemShut {NoStop}%
	\bibitem [{\citenamefont {Teukolsky}(1972)}]{Teukolsky}%
	\BibitemOpen
	\bibfield  {author} {\bibinfo {author} {\bibfnamefont {S.~A.}\ \bibnamefont
			{Teukolsky}},\ }\bibfield  {title} {\bibinfo {title} {Rotating black holes:
			Separable wave equations for gravitational and electromagnetic
			perturbations},\ }\href {https://doi.org/10.1103/PhysRevLett.29.1114}
	{\bibfield  {journal} {\bibinfo  {journal} {Phys. Rev. Lett.}\ }\textbf
		{\bibinfo {volume} {29}},\ \bibinfo {pages} {1114} (\bibinfo {year}
		{1972})}\BibitemShut {NoStop}%
	\bibitem [{\citenamefont {Chakraborty}\ \emph
		{et~al.}(2026{\natexlab{b}})\citenamefont {Chakraborty}, \citenamefont
		{Heidmann},\ and\ \citenamefont {Pani}}]{2yr1-9ymw}%
	\BibitemOpen
	\bibfield  {author} {\bibinfo {author} {\bibfnamefont {S.}~\bibnamefont
			{Chakraborty}}, \bibinfo {author} {\bibfnamefont {P.}~\bibnamefont
			{Heidmann}},\ and\ \bibinfo {author} {\bibfnamefont {P.}~\bibnamefont
			{Pani}},\ }\bibfield  {title} {\bibinfo {title} {Fermionic response of black
			holes in general relativity},\ }\href {https://doi.org/10.1103/2yr1-9ymw}
	{\bibfield  {journal} {\bibinfo  {journal} {Phys. Rev. D}\ }\textbf {\bibinfo
			{volume} {113}},\ \bibinfo {pages} {L061503} (\bibinfo {year}
		{2026}{\natexlab{b}})}\BibitemShut {NoStop}%
	\bibitem [{\citenamefont {Pang}\ \emph {et~al.}(2026)\citenamefont {Pang},
		\citenamefont {Tian}, \citenamefont {Zhang},\ and\ \citenamefont
		{Jiang}}]{pang2026}%
	\BibitemOpen
	\bibfield  {author} {\bibinfo {author} {\bibfnamefont {X.}~\bibnamefont
			{Pang}}, \bibinfo {author} {\bibfnamefont {Y.}~\bibnamefont {Tian}}, \bibinfo
		{author} {\bibfnamefont {H.}~\bibnamefont {Zhang}},\ and\ \bibinfo {author}
		{\bibfnamefont {Q.}~\bibnamefont {Jiang}},\ }\href
	{https://arxiv.org/abs/2510.10036} {\bibinfo {title} {Fermionic love number
			of reissner-nordstr\"om black holes}} (\bibinfo {year} {2026}),\ \Eprint
	{https://arxiv.org/abs/2510.10036} {arXiv:2510.10036 [gr-qc]} \BibitemShut
	{NoStop}%
	\bibitem [{\citenamefont {Castro}\ \emph {et~al.}(2010)\citenamefont {Castro},
		\citenamefont {Maloney},\ and\ \citenamefont {Strominger}}]{Castro}%
	\BibitemOpen
	\bibfield  {author} {\bibinfo {author} {\bibfnamefont {A.}~\bibnamefont
			{Castro}}, \bibinfo {author} {\bibfnamefont {A.}~\bibnamefont {Maloney}},\
		and\ \bibinfo {author} {\bibfnamefont {A.}~\bibnamefont {Strominger}},\
	}\bibfield  {title} {\bibinfo {title} {Hidden conformal symmetry of the kerr
			black hole},\ }\href {https://doi.org/10.1103/PhysRevD.82.024008} {\bibfield
		{journal} {\bibinfo  {journal} {Phys. Rev. D}\ }\textbf {\bibinfo {volume}
			{82}},\ \bibinfo {pages} {024008} (\bibinfo {year} {2010})}\BibitemShut
	{NoStop}%
	\bibitem [{\citenamefont {Rodríguez}\ \emph {et~al.}(2026)\citenamefont
		{Rodríguez}, \citenamefont {Santoni},\ and\ \citenamefont
		{Solomon}}]{rodriguez2026lovenumbersblackholes}%
	\BibitemOpen
	\bibfield  {author} {\bibinfo {author} {\bibfnamefont {M.~J.}\ \bibnamefont
			{Rodríguez}}, \bibinfo {author} {\bibfnamefont {L.}~\bibnamefont
			{Santoni}},\ and\ \bibinfo {author} {\bibfnamefont {A.~R.}\ \bibnamefont
			{Solomon}},\ }\href {https://arxiv.org/abs/2604.08653} {\bibinfo {title}
		{Love numbers of black holes and compact objects}} (\bibinfo {year} {2026}),\
	\Eprint {https://arxiv.org/abs/2604.08653} {arXiv:2604.08653 [gr-qc]}
	\BibitemShut {NoStop}%
	\bibitem [{\citenamefont {Cardoso}\ \emph {et~al.}(2017)\citenamefont
		{Cardoso}, \citenamefont {Franzin}, \citenamefont {Maselli}, \citenamefont
		{Pani},\ and\ \citenamefont {Raposo}}]{Vitor2}%
	\BibitemOpen
	\bibfield  {author} {\bibinfo {author} {\bibfnamefont {V.}~\bibnamefont
			{Cardoso}}, \bibinfo {author} {\bibfnamefont {E.}~\bibnamefont {Franzin}},
		\bibinfo {author} {\bibfnamefont {A.}~\bibnamefont {Maselli}}, \bibinfo
		{author} {\bibfnamefont {P.}~\bibnamefont {Pani}},\ and\ \bibinfo {author}
		{\bibfnamefont {G.}~\bibnamefont {Raposo}},\ }\bibfield  {title} {\bibinfo
		{title} {Testing strong-field gravity with tidal love numbers},\ }\href
	{https://doi.org/10.1103/PhysRevD.95.084014} {\bibfield  {journal} {\bibinfo
			{journal} {Phys. Rev. D}\ }\textbf {\bibinfo {volume} {95}},\ \bibinfo
		{pages} {084014} (\bibinfo {year} {2017})}\BibitemShut {NoStop}%
	\bibitem [{\citenamefont {Cardoso}\ \emph {et~al.}(2018)\citenamefont
		{Cardoso}, \citenamefont {Kimura}, \citenamefont {Maselli},\ and\
		\citenamefont {Senatore}}]{Vitor3}%
	\BibitemOpen
	\bibfield  {author} {\bibinfo {author} {\bibfnamefont {V.}~\bibnamefont
			{Cardoso}}, \bibinfo {author} {\bibfnamefont {M.}~\bibnamefont {Kimura}},
		\bibinfo {author} {\bibfnamefont {A.}~\bibnamefont {Maselli}},\ and\ \bibinfo
		{author} {\bibfnamefont {L.}~\bibnamefont {Senatore}},\ }\bibfield  {title}
	{\bibinfo {title} {Black holes in an effective field theory extension of
			general relativity},\ }\href {https://doi.org/10.1103/PhysRevLett.121.251105}
	{\bibfield  {journal} {\bibinfo  {journal} {Phys. Rev. Lett.}\ }\textbf
		{\bibinfo {volume} {121}},\ \bibinfo {pages} {251105} (\bibinfo {year}
		{2018})}\BibitemShut {NoStop}%
	\bibitem [{\citenamefont {Maselli}\ \emph {et~al.}(2019)\citenamefont
		{Maselli}, \citenamefont {Pani}, \citenamefont {Cardoso}, \citenamefont
		{Abdelsalhin}, \citenamefont {Gualtieri},\ and\ \citenamefont
		{Ferrari}}]{Maselli_2019}%
	\BibitemOpen
	\bibfield  {author} {\bibinfo {author} {\bibfnamefont {A.}~\bibnamefont
			{Maselli}}, \bibinfo {author} {\bibfnamefont {P.}~\bibnamefont {Pani}},
		\bibinfo {author} {\bibfnamefont {V.}~\bibnamefont {Cardoso}}, \bibinfo
		{author} {\bibfnamefont {T.}~\bibnamefont {Abdelsalhin}}, \bibinfo {author}
		{\bibfnamefont {L.}~\bibnamefont {Gualtieri}},\ and\ \bibinfo {author}
		{\bibfnamefont {V.}~\bibnamefont {Ferrari}},\ }\bibfield  {title} {\bibinfo
		{title} {From micro to macro and back: probing near-horizon quantum
			structures with gravitational waves},\ }\href
	{https://doi.org/10.1088/1361-6382/ab30ff} {\bibfield  {journal} {\bibinfo
			{journal} {Classical and Quantum Gravity}\ }\textbf {\bibinfo {volume}
			{36}},\ \bibinfo {pages} {167001} (\bibinfo {year} {2019})}\BibitemShut
	{NoStop}%
	\bibitem [{\citenamefont {Unruh}(1981)}]{unruh}%
	\BibitemOpen
	\bibfield  {author} {\bibinfo {author} {\bibfnamefont {W.~G.}\ \bibnamefont
			{Unruh}},\ }\bibfield  {title} {\bibinfo {title} {{Experimental Black-Hole
				Evaporation?}},\ }\href {https://doi.org/10.1103/PhysRevLett.46.1351}
	{\bibfield  {journal} {\bibinfo  {journal} {Phys. Rev. Lett.}\ }\textbf
		{\bibinfo {volume} {46}},\ \bibinfo {pages} {1351} (\bibinfo {year}
		{1981})}\BibitemShut {NoStop}%
	\bibitem [{\citenamefont {Visser}(1993)}]{visser1993}%
	\BibitemOpen
	\bibfield  {author} {\bibinfo {author} {\bibfnamefont {M.}~\bibnamefont
			{Visser}},\ }\href {https://arxiv.org/abs/gr-qc/9311028} {\bibinfo {title}
		{Acoustic propagation in fluids: an unexpected example of lorentzian
			geometry}} (\bibinfo {year} {1993}),\ \Eprint
	{https://arxiv.org/abs/gr-qc/9311028} {arXiv:gr-qc/9311028 [gr-qc]}
	\BibitemShut {NoStop}%
	\bibitem [{\citenamefont {Unruh}(1995)}]{Unruh95}%
	\BibitemOpen
	\bibfield  {author} {\bibinfo {author} {\bibfnamefont {W.~G.}\ \bibnamefont
			{Unruh}},\ }\bibfield  {title} {\bibinfo {title} {{Sonic analogue of black
				holes and the effects of high frequencies on black hole evaporation}},\
	}\href {https://doi.org/10.1103/PhysRevD.51.2827} {\bibfield  {journal}
		{\bibinfo  {journal} {Phys. Rev. D}\ }\textbf {\bibinfo {volume} {51}},\
		\bibinfo {pages} {2827} (\bibinfo {year} {1995})}\BibitemShut {NoStop}%
	\bibitem [{\citenamefont {Barcel{\'o}}\ \emph {et~al.}(2011)\citenamefont
		{Barcel{\'o}}, \citenamefont {Liberati},\ and\ \citenamefont {Visser}}]{BLV}%
	\BibitemOpen
	\bibfield  {author} {\bibinfo {author} {\bibfnamefont {C.}~\bibnamefont
			{Barcel{\'o}}}, \bibinfo {author} {\bibfnamefont {S.}~\bibnamefont
			{Liberati}},\ and\ \bibinfo {author} {\bibfnamefont {M.}~\bibnamefont
			{Visser}},\ }\bibfield  {title} {\bibinfo {title} {{Analogue Gravity}},\
	}\href {https://doi.org/10.12942/lrr-2011-3} {\bibfield  {journal} {\bibinfo
			{journal} {Living Reviews in Relativity}\ }\textbf {\bibinfo {volume} {14}},\
		\bibinfo {pages} {3} (\bibinfo {year} {2011})}\BibitemShut {NoStop}%
	\bibitem [{\citenamefont {Visser}(1998{\natexlab{a}})}]{MattCQG}%
	\BibitemOpen
	\bibfield  {author} {\bibinfo {author} {\bibfnamefont {M.}~\bibnamefont
			{Visser}},\ }\bibfield  {title} {\bibinfo {title} {{Acoustic black holes:
				horizons, ergospheres and Hawking radiation}},\ }\href
	{http://stacks.iop.org/0264-9381/15/i=6/a=024} {\bibfield  {journal}
		{\bibinfo  {journal} {Classical and Quantum Gravity}\ }\textbf {\bibinfo
			{volume} {15}},\ \bibinfo {pages} {1767} (\bibinfo {year}
		{1998}{\natexlab{a}})}\BibitemShut {NoStop}%
	\bibitem [{\citenamefont {Visser}(1998{\natexlab{b}})}]{Visser}%
	\BibitemOpen
	\bibfield  {author} {\bibinfo {author} {\bibfnamefont {M.}~\bibnamefont
			{Visser}},\ }\bibfield  {title} {\bibinfo {title} {{Hawking Radiation without
				Black Hole Entropy}},\ }\href {https://doi.org/10.1103/PhysRevLett.80.3436}
	{\bibfield  {journal} {\bibinfo  {journal} {Phys. Rev. Lett.}\ }\textbf
		{\bibinfo {volume} {80}},\ \bibinfo {pages} {3436} (\bibinfo {year}
		{1998}{\natexlab{b}})}\BibitemShut {NoStop}%
	\bibitem [{\citenamefont {Liberati}\ \emph {et~al.}(2000)\citenamefont
		{Liberati}, \citenamefont {Sonego},\ and\ \citenamefont
		{Visser}}]{StefanoLiberati_2000}%
	\BibitemOpen
	\bibfield  {author} {\bibinfo {author} {\bibfnamefont {S.}~\bibnamefont
			{Liberati}}, \bibinfo {author} {\bibfnamefont {S.}~\bibnamefont {Sonego}},\
		and\ \bibinfo {author} {\bibfnamefont {M.}~\bibnamefont {Visser}},\
	}\bibfield  {title} {\bibinfo {title} {Unexpectedly large surface gravities
			for acoustic horizons?},\ }\href
	{https://doi.org/10.1088/0264-9381/17/15/305} {\bibfield  {journal} {\bibinfo
			{journal} {Classical and Quantum Gravity}\ }\textbf {\bibinfo {volume}
			{17}},\ \bibinfo {pages} {2903} (\bibinfo {year} {2000})}\BibitemShut
	{NoStop}%
	\bibitem [{\citenamefont {Novello}\ \emph {et~al.}(2002)\citenamefont
		{Novello}, \citenamefont {Visser},\ and\ \citenamefont {Volovik}}]{nvmv}%
	\BibitemOpen
	\bibfield  {author} {\bibinfo {author} {\bibfnamefont {M.}~\bibnamefont
			{Novello}}, \bibinfo {author} {\bibfnamefont {M.}~\bibnamefont {Visser}},\
		and\ \bibinfo {author} {\bibfnamefont {G.}~\bibnamefont {Volovik}},\ }\href
	{https://doi.org/10.1142/4861} {\emph {\bibinfo {title} {Artificial Black
				Holes}}}\ (\bibinfo  {publisher} {WORLD SCIENTIFIC},\ \bibinfo {year}
	{2002})\BibitemShut {NoStop}%
	\bibitem [{\citenamefont {Barceló}\ \emph {et~al.}(2004)\citenamefont
		{Barceló}, \citenamefont {Liberati}, \citenamefont {Sonego},\ and\
		\citenamefont {Visser}}]{Barcelo_2004}%
	\BibitemOpen
	\bibfield  {author} {\bibinfo {author} {\bibfnamefont {C.}~\bibnamefont
			{Barceló}}, \bibinfo {author} {\bibfnamefont {S.}~\bibnamefont {Liberati}},
		\bibinfo {author} {\bibfnamefont {S.}~\bibnamefont {Sonego}},\ and\ \bibinfo
		{author} {\bibfnamefont {M.}~\bibnamefont {Visser}},\ }\bibfield  {title}
	{\bibinfo {title} {{Causal structure of analogue spacetimes}},\ }\href
	{https://doi.org/10.1088/1367-2630/6/1/186} {\bibfield  {journal} {\bibinfo
			{journal} {New Journal of Physics}\ }\textbf {\bibinfo {volume} {6}},\
		\bibinfo {pages} {186} (\bibinfo {year} {2004})}\BibitemShut {NoStop}%
	\bibitem [{\citenamefont {Visser}\ and\ \citenamefont
		{Weinfurtner}(2005)}]{Visser_2005}%
	\BibitemOpen
	\bibfield  {author} {\bibinfo {author} {\bibfnamefont {M.}~\bibnamefont
			{Visser}}\ and\ \bibinfo {author} {\bibfnamefont {S.}~\bibnamefont
			{Weinfurtner}},\ }\bibfield  {title} {\bibinfo {title} {Vortex analogue for
			the equatorial geometry of the kerr black hole},\ }\href
	{https://doi.org/10.1088/0264-9381/22/12/011} {\bibfield  {journal} {\bibinfo
			{journal} {Classical and Quantum Gravity}\ }\textbf {\bibinfo {volume}
			{22}},\ \bibinfo {pages} {2493} (\bibinfo {year} {2005})}\BibitemShut
	{NoStop}%
	\bibitem [{\citenamefont {Rousseaux}\ \emph {et~al.}(2008)\citenamefont
		{Rousseaux}, \citenamefont {Mathis}, \citenamefont {Maïssa}, \citenamefont
		{Philbin},\ and\ \citenamefont {Leonhardt}}]{Rousseaux2008}%
	\BibitemOpen
	\bibfield  {author} {\bibinfo {author} {\bibfnamefont {G.}~\bibnamefont
			{Rousseaux}}, \bibinfo {author} {\bibfnamefont {C.}~\bibnamefont {Mathis}},
		\bibinfo {author} {\bibfnamefont {P.}~\bibnamefont {Maïssa}}, \bibinfo
		{author} {\bibfnamefont {T.~G.}\ \bibnamefont {Philbin}},\ and\ \bibinfo
		{author} {\bibfnamefont {U.}~\bibnamefont {Leonhardt}},\ }\bibfield  {title}
	{\bibinfo {title} {{Observation of negative-frequency waves in a water tank:
				a classical analogue to the Hawking effect?}},\ }\href
	{https://doi.org/10.1088/1367-2630/10/5/053015} {\bibfield  {journal}
		{\bibinfo  {journal} {New Journal of Physics}\ }\textbf {\bibinfo {volume}
			{10}},\ \bibinfo {pages} {053015} (\bibinfo {year} {2008})}\BibitemShut
	{NoStop}%
	\bibitem [{\citenamefont {Weinfurtner}\ \emph {et~al.}(2011)\citenamefont
		{Weinfurtner}, \citenamefont {Tedford}, \citenamefont {Penrice},
		\citenamefont {Unruh},\ and\ \citenamefont {Lawrence}}]{Weinfurtner}%
	\BibitemOpen
	\bibfield  {author} {\bibinfo {author} {\bibfnamefont {S.}~\bibnamefont
			{Weinfurtner}}, \bibinfo {author} {\bibfnamefont {E.~W.}\ \bibnamefont
			{Tedford}}, \bibinfo {author} {\bibfnamefont {M.~C.~J.}\ \bibnamefont
			{Penrice}}, \bibinfo {author} {\bibfnamefont {W.~G.}\ \bibnamefont {Unruh}},\
		and\ \bibinfo {author} {\bibfnamefont {G.~A.}\ \bibnamefont {Lawrence}},\
	}\bibfield  {title} {\bibinfo {title} {{Measurement of Stimulated Hawking
				Emission in an Analogue System}},\ }\href
	{https://doi.org/10.1103/PhysRevLett.106.021302} {\bibfield  {journal}
		{\bibinfo  {journal} {Phys. Rev. Lett.}\ }\textbf {\bibinfo {volume} {106}},\
		\bibinfo {pages} {021302} (\bibinfo {year} {2011})}\BibitemShut {NoStop}%
	\bibitem [{\citenamefont {Unruh}(2014)}]{Unruh2014}%
	\BibitemOpen
	\bibfield  {author} {\bibinfo {author} {\bibfnamefont {W.~G.}\ \bibnamefont
			{Unruh}},\ }\bibfield  {title} {\bibinfo {title} {{Has Hawking Radiation Been
				Measured?}},\ }\href {https://doi.org/10.1007/s10701-014-9778-0} {\bibfield
		{journal} {\bibinfo  {journal} {Foundations of Physics}\ }\textbf {\bibinfo
			{volume} {44}},\ \bibinfo {pages} {532} (\bibinfo {year} {2014})}\BibitemShut
	{NoStop}%
	\bibitem [{\citenamefont {Euv\'e}\ \emph {et~al.}(2016)\citenamefont {Euv\'e},
		\citenamefont {Michel}, \citenamefont {Parentani}, \citenamefont {Philbin},\
		and\ \citenamefont {Rousseaux}}]{Euve}%
	\BibitemOpen
	\bibfield  {author} {\bibinfo {author} {\bibfnamefont {L.-P.}\ \bibnamefont
			{Euv\'e}}, \bibinfo {author} {\bibfnamefont {F.}~\bibnamefont {Michel}},
		\bibinfo {author} {\bibfnamefont {R.}~\bibnamefont {Parentani}}, \bibinfo
		{author} {\bibfnamefont {T.~G.}\ \bibnamefont {Philbin}},\ and\ \bibinfo
		{author} {\bibfnamefont {G.}~\bibnamefont {Rousseaux}},\ }\bibfield  {title}
	{\bibinfo {title} {{Observation of Noise Correlated by the Hawking Effect in
				a Water Tank}},\ }\href {https://doi.org/10.1103/PhysRevLett.117.121301}
	{\bibfield  {journal} {\bibinfo  {journal} {Phys. Rev. Lett.}\ }\textbf
		{\bibinfo {volume} {117}},\ \bibinfo {pages} {121301} (\bibinfo {year}
		{2016})}\BibitemShut {NoStop}%
	\bibitem [{\citenamefont {Garay}\ \emph {et~al.}(2000)\citenamefont {Garay},
		\citenamefont {Anglin}, \citenamefont {Cirac},\ and\ \citenamefont
		{Zoller}}]{PhysRevLett85.4643}%
	\BibitemOpen
	\bibfield  {author} {\bibinfo {author} {\bibfnamefont {L.~J.}\ \bibnamefont
			{Garay}}, \bibinfo {author} {\bibfnamefont {J.~R.}\ \bibnamefont {Anglin}},
		\bibinfo {author} {\bibfnamefont {J.~I.}\ \bibnamefont {Cirac}},\ and\
		\bibinfo {author} {\bibfnamefont {P.}~\bibnamefont {Zoller}},\ }\bibfield
	{title} {\bibinfo {title} {{Sonic Analog of Gravitational Black Holes in
				Bose-Einstein Condensates}},\ }\href
	{https://doi.org/10.1103/PhysRevLett.85.4643} {\bibfield  {journal} {\bibinfo
			{journal} {Phys. Rev. Lett.}\ }\textbf {\bibinfo {volume} {85}},\ \bibinfo
		{pages} {4643} (\bibinfo {year} {2000})}\BibitemShut {NoStop}%
	\bibitem [{\citenamefont {Barcel{\'{o}}}\ \emph {et~al.}(2001)\citenamefont
		{Barcel{\'{o}}}, \citenamefont {Liberati},\ and\ \citenamefont
		{Visser}}]{barcelo2001analogue}%
	\BibitemOpen
	\bibfield  {author} {\bibinfo {author} {\bibfnamefont {C.}~\bibnamefont
			{Barcel{\'{o}}}}, \bibinfo {author} {\bibfnamefont {S.}~\bibnamefont
			{Liberati}},\ and\ \bibinfo {author} {\bibfnamefont {M.}~\bibnamefont
			{Visser}},\ }\bibfield  {title} {\bibinfo {title} {{Analogue gravity from
				Bose-Einstein condensates}},\ }\href
	{https://doi.org/10.1088/0264-9381/18/6/312} {\bibfield  {journal} {\bibinfo
			{journal} {Classical and Quantum Gravity}\ }\textbf {\bibinfo {volume}
			{18}},\ \bibinfo {pages} {1137} (\bibinfo {year} {2001})}\BibitemShut
	{NoStop}%
	\bibitem [{\citenamefont {Basak}\ and\ \citenamefont
		{Majumdar}(2003)}]{Basak_2003}%
	\BibitemOpen
	\bibfield  {author} {\bibinfo {author} {\bibfnamefont {S.}~\bibnamefont
			{Basak}}\ and\ \bibinfo {author} {\bibfnamefont {P.}~\bibnamefont
			{Majumdar}},\ }\bibfield  {title} {\bibinfo {title} {‘superresonance’
			from a rotating acoustic black hole},\ }\href
	{https://doi.org/10.1088/0264-9381/20/18/304} {\bibfield  {journal} {\bibinfo
			{journal} {Classical and Quantum Gravity}\ }\textbf {\bibinfo {volume}
			{20}},\ \bibinfo {pages} {3907} (\bibinfo {year} {2003})}\BibitemShut
	{NoStop}%
	\bibitem [{\citenamefont {Volovik}(1999)}]{Volovik_1999}%
	\BibitemOpen
	\bibfield  {author} {\bibinfo {author} {\bibfnamefont {G.~E.}\ \bibnamefont
			{Volovik}},\ }\bibfield  {title} {\bibinfo {title} {{Simulation of a
				Painlev{\'{e} }-Gullstrand black hole in a thin 3He-A film}},\ }\href
	{https://doi.org/10.1134/1.568079} {\bibfield  {journal} {\bibinfo  {journal}
			{Journal of Experimental and Theoretical Physics Letters}\ }\textbf {\bibinfo
			{volume} {69}},\ \bibinfo {pages} {705} (\bibinfo {year} {1999})}\BibitemShut
	{NoStop}%
	\bibitem [{\citenamefont {Volovik}(2005)}]{Volovik2005}%
	\BibitemOpen
	\bibfield  {author} {\bibinfo {author} {\bibfnamefont {G.~E.}\ \bibnamefont
			{Volovik}},\ }\bibfield  {title} {\bibinfo {title} {Hydraulic jump as a white
			hole},\ }\href {https://doi.org/10.1134/1.2166908} {\bibfield  {journal}
		{\bibinfo  {journal} {Journal of Experimental and Theoretical Physics
				Letters}\ }\textbf {\bibinfo {volume} {82}},\ \bibinfo {pages} {624}
		(\bibinfo {year} {2005})}\BibitemShut {NoStop}%
	\bibitem [{\citenamefont {Volovik}(2006)}]{Volovik2006}%
	\BibitemOpen
	\bibfield  {author} {\bibinfo {author} {\bibfnamefont {G.~E.}\ \bibnamefont
			{Volovik}},\ }\bibfield  {title} {\bibinfo {title} {{Horizons and Ergoregions
				in Superfluids}},\ }\href {https://doi.org/10.1007/s10909-006-9248-y}
	{\bibfield  {journal} {\bibinfo  {journal} {Journal of Low Temperature
				Physics}\ }\textbf {\bibinfo {volume} {145}},\ \bibinfo {pages} {337}
		(\bibinfo {year} {2006})}\BibitemShut {NoStop}%
	\bibitem [{\citenamefont {Marino}(2008)}]{Marino}%
	\BibitemOpen
	\bibfield  {author} {\bibinfo {author} {\bibfnamefont {F.}~\bibnamefont
			{Marino}},\ }\bibfield  {title} {\bibinfo {title} {{Acoustic black holes in a
				two-dimensional ``photon fluid''}},\ }\href
	{https://doi.org/10.1103/PhysRevA.78.063804} {\bibfield  {journal} {\bibinfo
			{journal} {Phys. Rev. A}\ }\textbf {\bibinfo {volume} {78}},\ \bibinfo
		{pages} {063804} (\bibinfo {year} {2008})}\BibitemShut {NoStop}%
	\bibitem [{\citenamefont {Nguyen}\ \emph {et~al.}(2015)\citenamefont {Nguyen},
		\citenamefont {Gerace}, \citenamefont {Carusotto}, \citenamefont {Sanvitto},
		\citenamefont {Galopin}, \citenamefont {Lema\^{\i}tre}, \citenamefont
		{Sagnes}, \citenamefont {Bloch},\ and\ \citenamefont {Amo}}]{Nguyen}%
	\BibitemOpen
	\bibfield  {author} {\bibinfo {author} {\bibfnamefont {H.~S.}\ \bibnamefont
			{Nguyen}}, \bibinfo {author} {\bibfnamefont {D.}~\bibnamefont {Gerace}},
		\bibinfo {author} {\bibfnamefont {I.}~\bibnamefont {Carusotto}}, \bibinfo
		{author} {\bibfnamefont {D.}~\bibnamefont {Sanvitto}}, \bibinfo {author}
		{\bibfnamefont {E.}~\bibnamefont {Galopin}}, \bibinfo {author} {\bibfnamefont
			{A.}~\bibnamefont {Lema\^{\i}tre}}, \bibinfo {author} {\bibfnamefont
			{I.}~\bibnamefont {Sagnes}}, \bibinfo {author} {\bibfnamefont
			{J.}~\bibnamefont {Bloch}},\ and\ \bibinfo {author} {\bibfnamefont
			{A.}~\bibnamefont {Amo}},\ }\bibfield  {title} {\bibinfo {title} {{Acoustic
				Black Hole in a Stationary Hydrodynamic Flow of Microcavity Polaritons}},\
	}\href {https://doi.org/10.1103/PhysRevLett.114.036402} {\bibfield  {journal}
		{\bibinfo  {journal} {Phys. Rev. Lett.}\ }\textbf {\bibinfo {volume} {114}},\
		\bibinfo {pages} {036402} (\bibinfo {year} {2015})}\BibitemShut {NoStop}%
	\bibitem [{\citenamefont {Gerace}\ and\ \citenamefont
		{Carusotto}(2012)}]{PhysRevB.86.144505}%
	\BibitemOpen
	\bibfield  {author} {\bibinfo {author} {\bibfnamefont {D.}~\bibnamefont
			{Gerace}}\ and\ \bibinfo {author} {\bibfnamefont {I.}~\bibnamefont
			{Carusotto}},\ }\bibfield  {title} {\bibinfo {title} {{Analog Hawking
				radiation from an acoustic black hole in a flowing polariton superfluid}},\
	}\href {https://doi.org/10.1103/PhysRevB.86.144505} {\bibfield  {journal}
		{\bibinfo  {journal} {Phys. Rev. B}\ }\textbf {\bibinfo {volume} {86}},\
		\bibinfo {pages} {144505} (\bibinfo {year} {2012})}\BibitemShut {NoStop}%
	\bibitem [{\citenamefont {Carusotto}\ \emph {et~al.}(2008)\citenamefont
		{Carusotto}, \citenamefont {Fagnocchi}, \citenamefont {Recati}, \citenamefont
		{Balbinot},\ and\ \citenamefont {Fabbri}}]{Carusotto_2008}%
	\BibitemOpen
	\bibfield  {author} {\bibinfo {author} {\bibfnamefont {I.}~\bibnamefont
			{Carusotto}}, \bibinfo {author} {\bibfnamefont {S.}~\bibnamefont
			{Fagnocchi}}, \bibinfo {author} {\bibfnamefont {A.}~\bibnamefont {Recati}},
		\bibinfo {author} {\bibfnamefont {R.}~\bibnamefont {Balbinot}},\ and\
		\bibinfo {author} {\bibfnamefont {A.}~\bibnamefont {Fabbri}},\ }\bibfield
	{title} {\bibinfo {title} {{Numerical observation of Hawking radiation from
				acoustic black holes in atomic Bose{\textendash}Einstein condensates}},\
	}\href {https://doi.org/10.1088/1367-2630/10/10/103001} {\bibfield  {journal}
		{\bibinfo  {journal} {New Journal of Physics}\ }\textbf {\bibinfo {volume}
			{10}},\ \bibinfo {pages} {103001} (\bibinfo {year} {2008})}\BibitemShut
	{NoStop}%
	\bibitem [{\citenamefont {Sch\"utzhold}\ and\ \citenamefont
		{Unruh}(2002)}]{RalfBill}%
	\BibitemOpen
	\bibfield  {author} {\bibinfo {author} {\bibfnamefont {R.}~\bibnamefont
			{Sch\"utzhold}}\ and\ \bibinfo {author} {\bibfnamefont {W.~G.}\ \bibnamefont
			{Unruh}},\ }\bibfield  {title} {\bibinfo {title} {Gravity wave analogues of
			black holes},\ }\href {https://doi.org/10.1103/PhysRevD.66.044019} {\bibfield
		{journal} {\bibinfo  {journal} {Phys. Rev. D}\ }\textbf {\bibinfo {volume}
			{66}},\ \bibinfo {pages} {044019} (\bibinfo {year} {2002})}\BibitemShut
	{NoStop}%
	\bibitem [{\citenamefont {Jannes}\ \emph {et~al.}(2011)\citenamefont {Jannes},
		\citenamefont {Piquet}, \citenamefont {Ma\"{\i}ssa}, \citenamefont {Mathis},\
		and\ \citenamefont {Rousseaux}}]{ros2011}%
	\BibitemOpen
	\bibfield  {author} {\bibinfo {author} {\bibfnamefont {G.}~\bibnamefont
			{Jannes}}, \bibinfo {author} {\bibfnamefont {R.}~\bibnamefont {Piquet}},
		\bibinfo {author} {\bibfnamefont {P.}~\bibnamefont {Ma\"{\i}ssa}}, \bibinfo
		{author} {\bibfnamefont {C.}~\bibnamefont {Mathis}},\ and\ \bibinfo {author}
		{\bibfnamefont {G.}~\bibnamefont {Rousseaux}},\ }\bibfield  {title} {\bibinfo
		{title} {{Experimental demonstration of the supersonic-subsonic bifurcation
				in the circular jump: A hydrodynamic white hole}},\ }\href
	{https://doi.org/10.1103/PhysRevE.83.056312} {\bibfield  {journal} {\bibinfo
			{journal} {Phys. Rev. E}\ }\textbf {\bibinfo {volume} {83}},\ \bibinfo
		{pages} {056312} (\bibinfo {year} {2011})}\BibitemShut {NoStop}%
	\bibitem [{\citenamefont {Dolan}\ and\ \citenamefont
		{Oliveira}(2013)}]{PhysRevD.87.124038}%
	\BibitemOpen
	\bibfield  {author} {\bibinfo {author} {\bibfnamefont {S.~R.}\ \bibnamefont
			{Dolan}}\ and\ \bibinfo {author} {\bibfnamefont {E.~S.}\ \bibnamefont
			{Oliveira}},\ }\bibfield  {title} {\bibinfo {title} {Scattering by a draining
			bathtub vortex},\ }\href {https://doi.org/10.1103/PhysRevD.87.124038}
	{\bibfield  {journal} {\bibinfo  {journal} {Phys. Rev. D}\ }\textbf {\bibinfo
			{volume} {87}},\ \bibinfo {pages} {124038} (\bibinfo {year}
		{2013})}\BibitemShut {NoStop}%
	\bibitem [{\citenamefont {Steinhauer}(2016)}]{Steinhauer16}%
	\BibitemOpen
	\bibfield  {author} {\bibinfo {author} {\bibfnamefont {J.}~\bibnamefont
			{Steinhauer}},\ }\bibfield  {title} {\bibinfo {title} {{Observation of
				quantum Hawking radiation and its entanglement in an analogue black hole}},\
	}\href {http://dx.doi.org/10.1038/nphys3863} {\bibfield  {journal} {\bibinfo
			{journal} {Nat. Phys.}\ }\textbf {\bibinfo {volume} {12}},\ \bibinfo {pages}
		{959} (\bibinfo {year} {2016})}\BibitemShut {NoStop}%
	\bibitem [{\citenamefont {Mu{\~{n}}oz~de Nova}\ \emph
		{et~al.}(2019)\citenamefont {Mu{\~{n}}oz~de Nova}, \citenamefont {Golubkov},
		\citenamefont {Kolobov},\ and\ \citenamefont {Steinhauer}}]{Munoz}%
	\BibitemOpen
	\bibfield  {author} {\bibinfo {author} {\bibfnamefont {J.~R.}\ \bibnamefont
			{Mu{\~{n}}oz~de Nova}}, \bibinfo {author} {\bibfnamefont {K.}~\bibnamefont
			{Golubkov}}, \bibinfo {author} {\bibfnamefont {V.~I.}\ \bibnamefont
			{Kolobov}},\ and\ \bibinfo {author} {\bibfnamefont {J.}~\bibnamefont
			{Steinhauer}},\ }\bibfield  {title} {\bibinfo {title} {Observation of thermal
			hawking radiation and its temperature in an analogue black hole},\ }\href
	{https://doi.org/10.1038/s41586-019-1241-0} {\bibfield  {journal} {\bibinfo
			{journal} {Nature}\ }\textbf {\bibinfo {volume} {569}},\ \bibinfo {pages}
		{688} (\bibinfo {year} {2019})}\BibitemShut {NoStop}%
	\bibitem [{\citenamefont {Peloquin}\ \emph {et~al.}(2016)\citenamefont
		{Peloquin}, \citenamefont {Euv\'e}, \citenamefont {Philbin},\ and\
		\citenamefont {Rousseaux}}]{analogueWormholes}%
	\BibitemOpen
	\bibfield  {author} {\bibinfo {author} {\bibfnamefont {C.}~\bibnamefont
			{Peloquin}}, \bibinfo {author} {\bibfnamefont {L.-P.}\ \bibnamefont
			{Euv\'e}}, \bibinfo {author} {\bibfnamefont {T.}~\bibnamefont {Philbin}},\
		and\ \bibinfo {author} {\bibfnamefont {G.}~\bibnamefont {Rousseaux}},\
	}\bibfield  {title} {\bibinfo {title} {Analog wormholes and black hole laser
			effects in hydrodynamics},\ }\href
	{https://doi.org/10.1103/PhysRevD.93.084032} {\bibfield  {journal} {\bibinfo
			{journal} {Phys. Rev. D}\ }\textbf {\bibinfo {volume} {93}},\ \bibinfo
		{pages} {084032} (\bibinfo {year} {2016})}\BibitemShut {NoStop}%
	\bibitem [{\citenamefont {Richartz}\ \emph {et~al.}(2015)\citenamefont
		{Richartz}, \citenamefont {Prain}, \citenamefont {Liberati},\ and\
		\citenamefont {Weinfurtner}}]{PhysRevD.91.124018}%
	\BibitemOpen
	\bibfield  {author} {\bibinfo {author} {\bibfnamefont {M.}~\bibnamefont
			{Richartz}}, \bibinfo {author} {\bibfnamefont {A.}~\bibnamefont {Prain}},
		\bibinfo {author} {\bibfnamefont {S.}~\bibnamefont {Liberati}},\ and\
		\bibinfo {author} {\bibfnamefont {S.}~\bibnamefont {Weinfurtner}},\
	}\bibfield  {title} {\bibinfo {title} {Rotating black holes in a draining
			bathtub: Superradiant scattering of gravity waves},\ }\href
	{https://doi.org/10.1103/PhysRevD.91.124018} {\bibfield  {journal} {\bibinfo
			{journal} {Phys. Rev. D}\ }\textbf {\bibinfo {volume} {91}},\ \bibinfo
		{pages} {124018} (\bibinfo {year} {2015})}\BibitemShut {NoStop}%
	\bibitem [{\citenamefont {Torres}\ \emph {et~al.}(2017)\citenamefont {Torres},
		\citenamefont {Patrick}, \citenamefont {Coutant}, \citenamefont {Richartz},
		\citenamefont {Tedford},\ and\ \citenamefont {Weinfurtner}}]{Torres2017}%
	\BibitemOpen
	\bibfield  {author} {\bibinfo {author} {\bibfnamefont {T.}~\bibnamefont
			{Torres}}, \bibinfo {author} {\bibfnamefont {S.}~\bibnamefont {Patrick}},
		\bibinfo {author} {\bibfnamefont {A.}~\bibnamefont {Coutant}}, \bibinfo
		{author} {\bibfnamefont {M.}~\bibnamefont {Richartz}}, \bibinfo {author}
		{\bibfnamefont {E.~W.}\ \bibnamefont {Tedford}},\ and\ \bibinfo {author}
		{\bibfnamefont {S.}~\bibnamefont {Weinfurtner}},\ }\bibfield  {title}
	{\bibinfo {title} {Rotational superradiant scattering in a vortex flow},\
	}\href {https://doi.org/10.1038/nphys4151} {\bibfield  {journal} {\bibinfo
			{journal} {Nature Physics}\ }\textbf {\bibinfo {volume} {13}},\ \bibinfo
		{pages} {833} (\bibinfo {year} {2017})}\BibitemShut {NoStop}%
	\bibitem [{\citenamefont {Datta}\ \emph {et~al.}(2018)\citenamefont {Datta},
		\citenamefont {Shaikh},\ and\ \citenamefont {Das}}]{Datta_2018}%
	\BibitemOpen
	\bibfield  {author} {\bibinfo {author} {\bibfnamefont {S.}~\bibnamefont
			{Datta}}, \bibinfo {author} {\bibfnamefont {M.~A.}\ \bibnamefont {Shaikh}},\
		and\ \bibinfo {author} {\bibfnamefont {T.~K.}\ \bibnamefont {Das}},\
	}\bibfield  {title} {\bibinfo {title} {{Acoustic geometry obtained through
				the perturbation of Bernoulli’s constant}},\ }\href
	{https://doi.org/10.1016/j.newast.2018.03.003} {\bibfield  {journal}
		{\bibinfo  {journal} {New Astronomy}\ }\textbf {\bibinfo {volume} {63}},\
		\bibinfo {pages} {65–74} (\bibinfo {year} {2018})}\BibitemShut {NoStop}%
	\bibitem [{\citenamefont {Euv\'e}\ \emph {et~al.}(2020)\citenamefont {Euv\'e},
		\citenamefont {Robertson}, \citenamefont {James}, \citenamefont {Fabbri},\
		and\ \citenamefont {Rousseaux}}]{EuveII}%
	\BibitemOpen
	\bibfield  {author} {\bibinfo {author} {\bibfnamefont {L.-P.}\ \bibnamefont
			{Euv\'e}}, \bibinfo {author} {\bibfnamefont {S.}~\bibnamefont {Robertson}},
		\bibinfo {author} {\bibfnamefont {N.}~\bibnamefont {James}}, \bibinfo
		{author} {\bibfnamefont {A.}~\bibnamefont {Fabbri}},\ and\ \bibinfo {author}
		{\bibfnamefont {G.}~\bibnamefont {Rousseaux}},\ }\bibfield  {title} {\bibinfo
		{title} {{Scattering of Co-Current Surface Waves on an Analogue Black
				Hole}},\ }\href {https://doi.org/10.1103/PhysRevLett.124.141101} {\bibfield
		{journal} {\bibinfo  {journal} {Phys. Rev. Lett.}\ }\textbf {\bibinfo
			{volume} {124}},\ \bibinfo {pages} {141101} (\bibinfo {year}
		{2020})}\BibitemShut {NoStop}%
	\bibitem [{\citenamefont {Bhattacharjee}\ and\ \citenamefont
		{Ray}(2021)}]{JKBRoyWH}%
	\BibitemOpen
	\bibfield  {author} {\bibinfo {author} {\bibfnamefont {J.~K.}\ \bibnamefont
			{Bhattacharjee}}\ and\ \bibinfo {author} {\bibfnamefont {A.~K.}\ \bibnamefont
			{Ray}},\ }\bibfield  {title} {\bibinfo {title} {Surface tension and
			instability in the hydrodynamic white hole of a circular hydraulic jump},\
	}\href {https://doi.org/10.1103/PhysRevFluids.6.104801} {\bibfield  {journal}
		{\bibinfo  {journal} {Phys. Rev. Fluids}\ }\textbf {\bibinfo {volume} {6}},\
		\bibinfo {pages} {104801} (\bibinfo {year} {2021})}\BibitemShut {NoStop}%
	\bibitem [{\citenamefont {Kolobov}\ \emph {et~al.}(2021)\citenamefont
		{Kolobov}, \citenamefont {Golubkov}, \citenamefont {Mu{\~n}oz~de Nova},\ and\
		\citenamefont {Steinhauer}}]{Kolobov2021}%
	\BibitemOpen
	\bibfield  {author} {\bibinfo {author} {\bibfnamefont {V.~I.}\ \bibnamefont
			{Kolobov}}, \bibinfo {author} {\bibfnamefont {K.}~\bibnamefont {Golubkov}},
		\bibinfo {author} {\bibfnamefont {J.~R.}\ \bibnamefont {Mu{\~n}oz~de Nova}},\
		and\ \bibinfo {author} {\bibfnamefont {J.}~\bibnamefont {Steinhauer}},\
	}\bibfield  {title} {\bibinfo {title} {{Observation of stationary spontaneous
				Hawking radiation and the time evolution of an analogue black hole}},\ }\href
	{https://doi.org/10.1038/s41567-020-01076-0} {\bibfield  {journal} {\bibinfo
			{journal} {Nature Physics}\ }\textbf {\bibinfo {volume} {17}},\ \bibinfo
		{pages} {362} (\bibinfo {year} {2021})}\BibitemShut {NoStop}%
	\bibitem [{\citenamefont {Ribeiro}\ \emph {et~al.}(2022)\citenamefont
		{Ribeiro}, \citenamefont {Baak},\ and\ \citenamefont
		{Fischer}}]{ESangshinCaio2022}%
	\BibitemOpen
	\bibfield  {author} {\bibinfo {author} {\bibfnamefont {C.~C.~H.}\
			\bibnamefont {Ribeiro}}, \bibinfo {author} {\bibfnamefont {S.-S.}\
			\bibnamefont {Baak}},\ and\ \bibinfo {author} {\bibfnamefont {U.~R.}\
			\bibnamefont {Fischer}},\ }\bibfield  {title} {\bibinfo {title} {Existence of
			steady-state black hole analogs in finite quasi-one-dimensional bose-einstein
			condensates},\ }\href {https://doi.org/10.1103/PhysRevD.105.124066}
	{\bibfield  {journal} {\bibinfo  {journal} {Phys. Rev. D}\ }\textbf {\bibinfo
			{volume} {105}},\ \bibinfo {pages} {124066} (\bibinfo {year}
		{2022})}\BibitemShut {NoStop}%
	\bibitem [{\citenamefont {Fourdrinoy}\ \emph {et~al.}(2022)\citenamefont
		{Fourdrinoy}, \citenamefont {Robertson}, \citenamefont {James}, \citenamefont
		{Fabbri},\ and\ \citenamefont {Rousseaux}}]{PhysRevD.105.085022}%
	\BibitemOpen
	\bibfield  {author} {\bibinfo {author} {\bibfnamefont {J.}~\bibnamefont
			{Fourdrinoy}}, \bibinfo {author} {\bibfnamefont {S.}~\bibnamefont
			{Robertson}}, \bibinfo {author} {\bibfnamefont {N.}~\bibnamefont {James}},
		\bibinfo {author} {\bibfnamefont {A.}~\bibnamefont {Fabbri}},\ and\ \bibinfo
		{author} {\bibfnamefont {G.}~\bibnamefont {Rousseaux}},\ }\bibfield  {title}
	{\bibinfo {title} {{Correlations on weakly time-dependent transcritical
				white-hole flows}},\ }\href {https://doi.org/10.1103/PhysRevD.105.085022}
	{\bibfield  {journal} {\bibinfo  {journal} {Phys. Rev. D}\ }\textbf {\bibinfo
			{volume} {105}},\ \bibinfo {pages} {085022} (\bibinfo {year}
		{2022})}\BibitemShut {NoStop}%
	\bibitem [{\citenamefont {Syu}\ \emph {et~al.}(2022)\citenamefont {Syu},
		\citenamefont {Lee},\ and\ \citenamefont {Lin}}]{SyuLee2022}%
	\BibitemOpen
	\bibfield  {author} {\bibinfo {author} {\bibfnamefont {W.-C.}\ \bibnamefont
			{Syu}}, \bibinfo {author} {\bibfnamefont {D.-S.}\ \bibnamefont {Lee}},\ and\
		\bibinfo {author} {\bibfnamefont {C.-Y.}\ \bibnamefont {Lin}},\ }\bibfield
	{title} {\bibinfo {title} {Analogous hawking radiation and quantum
			entanglement in two-component bose-einstein condensates: The gapped
			excitations},\ }\href {https://doi.org/10.1103/PhysRevD.106.044016}
	{\bibfield  {journal} {\bibinfo  {journal} {Phys. Rev. D}\ }\textbf {\bibinfo
			{volume} {106}},\ \bibinfo {pages} {044016} (\bibinfo {year}
		{2022})}\BibitemShut {NoStop}%
	\bibitem [{\citenamefont {Syu}\ and\ \citenamefont {Lee}(2023)}]{SyuLee2023}%
	\BibitemOpen
	\bibfield  {author} {\bibinfo {author} {\bibfnamefont {W.-C.}\ \bibnamefont
			{Syu}}\ and\ \bibinfo {author} {\bibfnamefont {D.-S.}\ \bibnamefont {Lee}},\
	}\bibfield  {title} {\bibinfo {title} {Analogous hawking radiation from
			gapped excitations in a transonic flow of binary bose-einstein condensates},\
	}\href {https://doi.org/10.1103/PhysRevD.107.084049} {\bibfield  {journal}
		{\bibinfo  {journal} {Phys. Rev. D}\ }\textbf {\bibinfo {volume} {107}},\
		\bibinfo {pages} {084049} (\bibinfo {year} {2023})}\BibitemShut {NoStop}%
	\bibitem [{\citenamefont {Syu}\ and\ \citenamefont
		{Lee}(2024)}]{PhysRevD.110.044017}%
	\BibitemOpen
	\bibfield  {author} {\bibinfo {author} {\bibfnamefont {W.-C.}\ \bibnamefont
			{Syu}}\ and\ \bibinfo {author} {\bibfnamefont {D.-S.}\ \bibnamefont {Lee}},\
	}\bibfield  {title} {\bibinfo {title} {Acoustic quasibound states and
			tachyonic instabilities from binary bose-einstein condensates},\ }\href
	{https://doi.org/10.1103/PhysRevD.110.044017} {\bibfield  {journal} {\bibinfo
			{journal} {Phys. Rev. D}\ }\textbf {\bibinfo {volume} {110}},\ \bibinfo
		{pages} {044017} (\bibinfo {year} {2024})}\BibitemShut {NoStop}%
	\bibitem [{\citenamefont {Bossard}\ \emph {et~al.}(2023)\citenamefont
		{Bossard}, \citenamefont {James}, \citenamefont {Aucouturier}, \citenamefont
		{Fourdrinoy}, \citenamefont {Robertson},\ and\ \citenamefont
		{Rousseaux}}]{bossard2023createanalogueblackhole}%
	\BibitemOpen
	\bibfield  {author} {\bibinfo {author} {\bibfnamefont {A.}~\bibnamefont
			{Bossard}}, \bibinfo {author} {\bibfnamefont {N.}~\bibnamefont {James}},
		\bibinfo {author} {\bibfnamefont {C.}~\bibnamefont {Aucouturier}}, \bibinfo
		{author} {\bibfnamefont {J.}~\bibnamefont {Fourdrinoy}}, \bibinfo {author}
		{\bibfnamefont {S.}~\bibnamefont {Robertson}},\ and\ \bibinfo {author}
		{\bibfnamefont {G.}~\bibnamefont {Rousseaux}},\ }\href
	{https://arxiv.org/abs/2307.11022} {\bibinfo {title} {How to create analogue
			black hole or white fountain horizons and laser cavities in experimental free
			surface hydrodynamics?}} (\bibinfo {year} {2023}),\ \Eprint
	{https://arxiv.org/abs/2307.11022} {arXiv:2307.11022 [physics.flu-dyn]}
	\BibitemShut {NoStop}%
	\bibitem [{\citenamefont {Datta}\ and\ \citenamefont
		{Fischer}(2025)}]{Datta_2025}%
	\BibitemOpen
	\bibfield  {author} {\bibinfo {author} {\bibfnamefont {S.}~\bibnamefont
			{Datta}}\ and\ \bibinfo {author} {\bibfnamefont {U.~R.}\ \bibnamefont
			{Fischer}},\ }\bibfield  {title} {\bibinfo {title} {Probing penrose-type
			singularities inside sonic black holes},\ }\href
	{https://doi.org/10.1088/1361-6382/ae2412} {\bibfield  {journal} {\bibinfo
			{journal} {Classical and Quantum Gravity}\ }\textbf {\bibinfo {volume}
			{42}},\ \bibinfo {pages} {245009} (\bibinfo {year} {2025})}\BibitemShut
	{NoStop}%
	\bibitem [{\citenamefont {De~Luca}\ \emph {et~al.}(2025)\citenamefont
		{De~Luca}, \citenamefont {Khek}, \citenamefont {Khoury},\ and\ \citenamefont
		{Trodden}}]{PRDLuca}%
	\BibitemOpen
	\bibfield  {author} {\bibinfo {author} {\bibfnamefont {V.}~\bibnamefont
			{De~Luca}}, \bibinfo {author} {\bibfnamefont {B.}~\bibnamefont {Khek}},
		\bibinfo {author} {\bibfnamefont {J.}~\bibnamefont {Khoury}},\ and\ \bibinfo
		{author} {\bibfnamefont {M.}~\bibnamefont {Trodden}},\ }\bibfield  {title}
	{\bibinfo {title} {Tidal love numbers of analog black holes},\ }\href
	{https://doi.org/10.1103/PhysRevD.111.044069} {\bibfield  {journal} {\bibinfo
			{journal} {Phys. Rev. D}\ }\textbf {\bibinfo {volume} {111}},\ \bibinfo
		{pages} {044069} (\bibinfo {year} {2025})}\BibitemShut {NoStop}%
	\bibitem [{\citenamefont {Poisson}\ and\ \citenamefont
		{Will}(2014)}]{poisson2014gravity}%
	\BibitemOpen
	\bibfield  {author} {\bibinfo {author} {\bibfnamefont {E.}~\bibnamefont
			{Poisson}}\ and\ \bibinfo {author} {\bibfnamefont {C.}~\bibnamefont {Will}},\
	}\href {https://books.google.com.tw/books?id=PZ5cAwAAQBAJ} {\emph {\bibinfo
			{title} {Gravity: Newtonian, Post-Newtonian, Relativistic}}}\ (\bibinfo
	{publisher} {Cambridge University Press},\ \bibinfo {year}
	{2014})\BibitemShut {NoStop}%
	\bibitem [{\citenamefont {Chakraborty}\ and\ \citenamefont
		{Pani}(2026)}]{chakraborty2026tidalresponsecompactobjects}%
	\BibitemOpen
	\bibfield  {author} {\bibinfo {author} {\bibfnamefont {S.}~\bibnamefont
			{Chakraborty}}\ and\ \bibinfo {author} {\bibfnamefont {P.}~\bibnamefont
			{Pani}},\ }\href {https://arxiv.org/abs/2604.08679} {\bibinfo {title} {Tidal
			response of compact objects}} (\bibinfo {year} {2026}),\ \Eprint
	{https://arxiv.org/abs/2604.08679} {arXiv:2604.08679 [gr-qc]} \BibitemShut
	{NoStop}%
	\bibitem [{\citenamefont {Landau}\ and\ \citenamefont
		{Lifshitz}(1987)}]{Landau1987Fluid}%
	\BibitemOpen
	\bibfield  {author} {\bibinfo {author} {\bibfnamefont {L.~D.}\ \bibnamefont
			{Landau}}\ and\ \bibinfo {author} {\bibfnamefont {E.~M.}\ \bibnamefont
			{Lifshitz}},\ }\href {http://www.worldcat.org/isbn/0750627670} {\emph
		{\bibinfo {title} {Fluid Mechanics, Second Edition: Volume 6 (Course of
				Theoretical Physics)}}},\ \bibinfo {edition} {2nd}\ ed.\ (\bibinfo
	{publisher} {Butterworth-Heinemann},\ \bibinfo {year} {1987})\BibitemShut
	{NoStop}%
	\bibitem [{\citenamefont {Clarke}\ and\ \citenamefont
		{Carswell}(2007)}]{clarke2007principles}%
	\BibitemOpen
	\bibfield  {author} {\bibinfo {author} {\bibfnamefont {C.}~\bibnamefont
			{Clarke}}\ and\ \bibinfo {author} {\bibfnamefont {B.}~\bibnamefont
			{Carswell}},\ }\href {https://books.google.com.tw/books?id=odiiDgAAQBAJ}
	{\emph {\bibinfo {title} {Principles of Astrophysical Fluid Dynamics}}},\
	Principles of Astrophysical Fluid Dynamics\ (\bibinfo  {publisher} {Cambridge
		University Press},\ \bibinfo {year} {2007})\BibitemShut {NoStop}%
	\bibitem [{\citenamefont {Carroll}(2019)}]{carroll_2019}%
	\BibitemOpen
	\bibfield  {author} {\bibinfo {author} {\bibfnamefont {S.~M.}\ \bibnamefont
			{Carroll}},\ }\href {https://doi.org/10.1017/9781108770385} {\emph {\bibinfo
			{title} {Spacetime and Geometry: An Introduction to General Relativity}}}\
	(\bibinfo  {publisher} {Cambridge University Press},\ \bibinfo {year}
	{2019})\BibitemShut {NoStop}%
	\bibitem [{\citenamefont {Penrose}(1965)}]{Penrose65PRL}%
	\BibitemOpen
	\bibfield  {author} {\bibinfo {author} {\bibfnamefont {R.}~\bibnamefont
			{Penrose}},\ }\bibfield  {title} {\bibinfo {title} {{Gravitational Collapse
				and Space-Time Singularities}},\ }\href
	{https://doi.org/10.1103/PhysRevLett.14.57} {\bibfield  {journal} {\bibinfo
			{journal} {Phys. Rev. Lett.}\ }\textbf {\bibinfo {volume} {14}},\ \bibinfo
		{pages} {57} (\bibinfo {year} {1965})}\BibitemShut {NoStop}%
	\bibitem [{\citenamefont {Hawking}(1975)}]{Hawking1975}%
	\BibitemOpen
	\bibfield  {author} {\bibinfo {author} {\bibfnamefont {S.~W.}\ \bibnamefont
			{Hawking}},\ }\bibfield  {title} {\bibinfo {title} {Particle creation by
			black holes},\ }\href {https://doi.org/10.1007/BF02345020} {\bibfield
		{journal} {\bibinfo  {journal} {Communications in Mathematical Physics}\
		}\textbf {\bibinfo {volume} {43}},\ \bibinfo {pages} {199} (\bibinfo {year}
		{1975})}\BibitemShut {NoStop}%
	\bibitem [{\citenamefont {Starobinskii}(1973)}]{Starobinskii:1973vzb}%
	\BibitemOpen
	\bibfield  {author} {\bibinfo {author} {\bibfnamefont {A.~A.}\ \bibnamefont
			{Starobinskii}},\ }\bibfield  {title} {\bibinfo {title} {{Amplification of
				waves during reflection from a rotating ''black hole''}},\ }\href@noop {}
	{\bibfield  {journal} {\bibinfo  {journal} {Sov. Phys. JETP}\ }\textbf
		{\bibinfo {volume} {37}},\ \bibinfo {pages} {28} (\bibinfo {year}
		{1973})}\BibitemShut {NoStop}%
	\bibitem [{\citenamefont {Stein}\ and\ \citenamefont
		{Shakarchi}(2010)}]{stein2010complex}%
	\BibitemOpen
	\bibfield  {author} {\bibinfo {author} {\bibfnamefont {E.}~\bibnamefont
			{Stein}}\ and\ \bibinfo {author} {\bibfnamefont {R.}~\bibnamefont
			{Shakarchi}},\ }\href {https://books.google.com.tw/books?id=0ECHh9tjPUAC}
	{\emph {\bibinfo {title} {Complex Analysis}}},\ Princeton lectures in
	analysis\ (\bibinfo  {publisher} {Princeton University Press},\ \bibinfo
	{year} {2010})\BibitemShut {NoStop}%
	\bibitem [{\citenamefont {Beals}\ and\ \citenamefont
		{Wong}(2010)}]{Beals_Wong_2010}%
	\BibitemOpen
	\bibfield  {author} {\bibinfo {author} {\bibfnamefont {R.}~\bibnamefont
			{Beals}}\ and\ \bibinfo {author} {\bibfnamefont {R.}~\bibnamefont {Wong}},\
	}\href
	{https://www.cambridge.org/core/books/special-functions/A0E697B24555D5E56B0FDC1CD9101DE1}
	{\emph {\bibinfo {title} {Special Functions: A Graduate Text}}},\ Cambridge
	Studies in Advanced Mathematics\ (\bibinfo  {publisher} {Cambridge University
		Press},\ \bibinfo {year} {2010})\BibitemShut {NoStop}%
	\bibitem [{\citenamefont {Cannizzaro}\ \emph {et~al.}(2026)\citenamefont
		{Cannizzaro}, \citenamefont {Luca},\ and\ \citenamefont
		{Pani}}]{tidaldeformabilitydisk}%
	\BibitemOpen
	\bibfield  {author} {\bibinfo {author} {\bibfnamefont {E.}~\bibnamefont
			{Cannizzaro}}, \bibinfo {author} {\bibfnamefont {V.~D.}\ \bibnamefont
			{Luca}},\ and\ \bibinfo {author} {\bibfnamefont {P.}~\bibnamefont {Pani}},\
	}\href {https://arxiv.org/abs/2408.14208} {\bibinfo {title} {Tidal
			deformability of black holes surrounded by thin accretion disks}} (\bibinfo
	{year} {2026}),\ \Eprint {https://arxiv.org/abs/2408.14208} {arXiv:2408.14208
		[astro-ph.HE]} \BibitemShut {NoStop}%
	\bibitem [{\citenamefont {Bhatt}\ \emph {et~al.}(2025)\citenamefont {Bhatt},
		\citenamefont {Chakraborty},\ and\ \citenamefont {Bose}}]{Bhatt2025}%
	\BibitemOpen
	\bibfield  {author} {\bibinfo {author} {\bibfnamefont {R.~P.}\ \bibnamefont
			{Bhatt}}, \bibinfo {author} {\bibfnamefont {S.}~\bibnamefont {Chakraborty}},\
		and\ \bibinfo {author} {\bibfnamefont {S.}~\bibnamefont {Bose}},\ }\bibfield
	{title} {\bibinfo {title} {Rotating black holes experience dynamical tides},\
	}\href {https://doi.org/10.1103/PhysRevD.111.L041504} {\bibfield  {journal}
		{\bibinfo  {journal} {Phys. Rev. D}\ }\textbf {\bibinfo {volume} {111}},\
		\bibinfo {pages} {L041504} (\bibinfo {year} {2025})}\BibitemShut {NoStop}%
	\bibitem [{\citenamefont {Datta}\ and\ \citenamefont
		{Fischer}(2022)}]{Datta_2022}%
	\BibitemOpen
	\bibfield  {author} {\bibinfo {author} {\bibfnamefont {S.}~\bibnamefont
			{Datta}}\ and\ \bibinfo {author} {\bibfnamefont {U.~R.}\ \bibnamefont
			{Fischer}},\ }\bibfield  {title} {\bibinfo {title} {{Analogue gravitational
				field from nonlinear fluid dynamics}},\ }\href
	{https://doi.org/10.1088/1361-6382/ac4828} {\bibfield  {journal} {\bibinfo
			{journal} {Classical and Quantum Gravity}\ }\textbf {\bibinfo {volume}
			{39}},\ \bibinfo {pages} {075018} (\bibinfo {year} {2022})}\BibitemShut
	{NoStop}%
	\bibitem [{\citenamefont {Biondi}\ \emph {et~al.}(2024)\citenamefont {Biondi},
		\citenamefont {Robertson},\ and\ \citenamefont {Rousseaux}}]{bionditwolayer}%
	\BibitemOpen
	\bibfield  {author} {\bibinfo {author} {\bibfnamefont {A.}~\bibnamefont
			{Biondi}}, \bibinfo {author} {\bibfnamefont {S.}~\bibnamefont {Robertson}},\
		and\ \bibinfo {author} {\bibfnamefont {G.}~\bibnamefont {Rousseaux}},\ }\href
	{https://arxiv.org/abs/2409.16864} {\bibinfo {title} {Vortical scattering
			channel in an aquatic space-time}} (\bibinfo {year} {2024}),\ \Eprint
	{https://arxiv.org/abs/2409.16864} {arXiv:2409.16864 [gr-qc]} \BibitemShut
	{NoStop}%
	\bibitem [{\citenamefont {Tamura}\ \emph {et~al.}(2025)\citenamefont {Tamura},
		\citenamefont {Khlebnikov}, \citenamefont {Chen},\ and\ \citenamefont
		{Hung}}]{2dsupersonicBEC}%
	\BibitemOpen
	\bibfield  {author} {\bibinfo {author} {\bibfnamefont {H.}~\bibnamefont
			{Tamura}}, \bibinfo {author} {\bibfnamefont {S.}~\bibnamefont {Khlebnikov}},
		\bibinfo {author} {\bibfnamefont {C.-A.}\ \bibnamefont {Chen}},\ and\
		\bibinfo {author} {\bibfnamefont {C.-L.}\ \bibnamefont {Hung}},\ }\bibfield
	{title} {\bibinfo {title} {Observation of self-oscillating supersonic flow
			across an acoustic horizon in two dimensions},\ }\href
	{https://doi.org/10.1103/t2sn-kx99} {\bibfield  {journal} {\bibinfo
			{journal} {Phys. Rev. A}\ }\textbf {\bibinfo {volume} {112}},\ \bibinfo
		{pages} {L031301} (\bibinfo {year} {2025})}\BibitemShut {NoStop}%
	\bibitem [{\citenamefont {Andrews}\ \emph {et~al.}(1997)\citenamefont
		{Andrews}, \citenamefont {Kurn}, \citenamefont {Miesner}, \citenamefont
		{Durfee}, \citenamefont {Townsend}, \citenamefont {Inouye},\ and\
		\citenamefont {Ketterle}}]{Andrews}%
	\BibitemOpen
	\bibfield  {author} {\bibinfo {author} {\bibfnamefont {M.~R.}\ \bibnamefont
			{Andrews}}, \bibinfo {author} {\bibfnamefont {D.~M.}\ \bibnamefont {Kurn}},
		\bibinfo {author} {\bibfnamefont {H.-J.}\ \bibnamefont {Miesner}}, \bibinfo
		{author} {\bibfnamefont {D.~S.}\ \bibnamefont {Durfee}}, \bibinfo {author}
		{\bibfnamefont {C.~G.}\ \bibnamefont {Townsend}}, \bibinfo {author}
		{\bibfnamefont {S.}~\bibnamefont {Inouye}},\ and\ \bibinfo {author}
		{\bibfnamefont {W.}~\bibnamefont {Ketterle}},\ }\bibfield  {title} {\bibinfo
		{title} {{Propagation of Sound in a Bose-Einstein Condensate}},\ }\href
	{https://doi.org/10.1103/PhysRevLett.79.553} {\bibfield  {journal} {\bibinfo
			{journal} {Phys. Rev. Lett.}\ }\textbf {\bibinfo {volume} {79}},\ \bibinfo
		{pages} {553} (\bibinfo {year} {1997})}\BibitemShut {NoStop}%
	\bibitem [{\citenamefont {Jacobson}(2013)}]{Jacobson2013}%
	\BibitemOpen
	\bibfield  {author} {\bibinfo {author} {\bibfnamefont {T.}~\bibnamefont
			{Jacobson}},\ }\bibinfo {title} {Black holes and hawking radiation in
		spacetime and its analogues},\ in\ \href
	{https://doi.org/10.1007/978-3-319-00266-8_1} {\emph {\bibinfo {booktitle}
			{Analogue Gravity Phenomenology: Analogue Spacetimes and Horizons, from
				Theory to Experiment}}},\ \bibinfo {editor} {edited by\ \bibinfo {editor}
		{\bibfnamefont {D.}~\bibnamefont {Faccio}}, \bibinfo {editor} {\bibfnamefont
			{F.}~\bibnamefont {Belgiorno}}, \bibinfo {editor} {\bibfnamefont
			{S.}~\bibnamefont {Cacciatori}}, \bibinfo {editor} {\bibfnamefont
			{V.}~\bibnamefont {Gorini}}, \bibinfo {editor} {\bibfnamefont
			{S.}~\bibnamefont {Liberati}},\ and\ \bibinfo {editor} {\bibfnamefont
			{U.}~\bibnamefont {Moschella}}}\ (\bibinfo  {publisher} {Springer
		International Publishing},\ \bibinfo {address} {Cham},\ \bibinfo {year}
	{2013})\ pp.\ \bibinfo {pages} {1--29}\BibitemShut {NoStop}%
	\bibitem [{\citenamefont {Fischer}\ and\ \citenamefont
		{Datta}(2023)}]{Datta_2023}%
	\BibitemOpen
	\bibfield  {author} {\bibinfo {author} {\bibfnamefont {U.~R.}\ \bibnamefont
			{Fischer}}\ and\ \bibinfo {author} {\bibfnamefont {S.}~\bibnamefont
			{Datta}},\ }\bibfield  {title} {\bibinfo {title} {Dispersive censor of
			acoustic spacetimes with a shock-wave singularity},\ }\href
	{https://doi.org/10.1103/PhysRevD.107.084023} {\bibfield  {journal} {\bibinfo
			{journal} {Phys. Rev. D}\ }\textbf {\bibinfo {volume} {107}},\ \bibinfo
		{pages} {084023} (\bibinfo {year} {2023})}\BibitemShut {NoStop}%
		\bibitem [{\citenamefont {Berti}\ \emph {et~al.}(2004)\citenamefont {Berti},
  \citenamefont {Cardoso},\ and\ \citenamefont {Lemos}}]{QNMBertiE}%
  \BibitemOpen
  \bibfield  {author} {\bibinfo {author} {\bibfnamefont {E.}~\bibnamefont
  {Berti}}, \bibinfo {author} {\bibfnamefont {V.}~\bibnamefont {Cardoso}},\
  and\ \bibinfo {author} {\bibfnamefont {J.~P.~S.}\ \bibnamefont {Lemos}},\
  }\bibfield  {title} {\bibinfo {title} {Quasinormal modes and classical wave
  propagation in analogue black holes},\ }\href
  {https://doi.org/10.1103/PhysRevD.70.124006} {\bibfield  {journal} {\bibinfo
  {journal} {Phys. Rev. D}\ }\textbf {\bibinfo {volume} {70}},\ \bibinfo
  {pages} {124006} (\bibinfo {year} {2004})}\BibitemShut {NoStop}%
  \bibitem [{\citenamefont {Abramowitz}\ and\ \citenamefont
  {Stegun}(1965)}]{handbook}%
  \BibitemOpen
  \bibfield  {author} {\bibinfo {author} {\bibfnamefont {M.}~\bibnamefont
  {Abramowitz}}\ and\ \bibinfo {author} {\bibfnamefont {I.}~\bibnamefont
  {Stegun}},\ }\href {https://books.google.com.tw/books?id=MtU8uP7XMvoC} {\emph
  {\bibinfo {title} {Handbook of Mathematical Functions: With Formulas, Graphs,
  and Mathematical Tables}}},\ Applied mathematics series\ (\bibinfo
  {publisher} {Dover Publications},\ \bibinfo {year} {1965})\BibitemShut
  {NoStop}%
\end{thebibliography}
\end{document}